\documentclass{aa}  
\bibpunct{(}{)}{;}{a}{}{,}
\usepackage{graphicx}
\usepackage{txfonts}
\usepackage{enumitem}
\usepackage{physics}
\usepackage[squaren,Gray]{SIunits}
\usepackage{textcomp}
\usepackage{natbib}
\usepackage{subfigure}
\usepackage[colorlinks=false]{hyperref}
\usepackage{bm}
\usepackage[fleqn]{cases}
\usepackage{animate}
\usepackage{xcolor}
\usepackage{dsfont}
\usepackage{makecell}
\usepackage{multirow}
\usepackage{diagbox}

\DeclareMathOperator{\sign}{sign}

\newcommand{\ft}{\widetilde{f}}

\newcommand{\mA}{\mathcal{A}}
\newcommand{\mB}{\mathcal{B}}

\newcommand{\mE}{\mathcal{E}}

\newcommand{\kx}{k_{x}}

\newcommand{\Up}{U'}
\newcommand{\Um}{U}
\newcommand{\rom}{\rho_0}
\newcommand{\yo}{y_0}
\newcommand{\rc}{r_{\mathrm{c}}}

\newcommand{\Ro}{R_\mathrm{o}}
\DeclareMathOperator{\e}{e}

\renewcommand{\sf}{\sigma_{f}}
\newcommand{\bo}[1]{{\bm{#1}}}
\newcommand{\bn}{\bo\nabla}
\newcommand{\bO}{\bo n}
\newcommand{\df}[2]{\frac{\mathrm{d}#1}{\mathrm{d}#2}}
\newcommand{\dfd}[2]{\frac{\mathrm{d}^2#1}{\mathrm{d}#2^2}}
\newcommand{\wg}{V_\mathrm{g}}
\newcommand{\Od}{R}
\newcommand{\od}{\Od_{\theta_0=0,\,\pi}}
\newcommand{\Ri}{\mathrm{Ri}}

\newcommand{\sn}{\sigma_3}
\newcommand{\ak}{\alpha_k}
\renewcommand{\tt}[1]{\,10^{#1}}
\newcommand{\ron}{\rho_\mathrm{T}}
\newcommand{\jp}[1]{\textcolor{red}{#1}}

\definecolor{vin}{HTML}{842121}
\newcommand{\aava}[1]{\textcolor{black}{#1}}
\definecolor{rf}{HTML}{871515}

\makeatletter
\renewcommand*\aa@pageof{, page \thepage{} of \pageref*{LastPage}}
\makeatother
\newenvironment{my_enumerate}{
\begin{enumerate}
     \setlength{\itemsep}{1pt}
     \setlength{\parskip}{0pt}
     \setlength{\parsep}{0pt}}
{\end{enumerate}}

\begin{document}

   \title{The complex interplay between tidal inertial waves and zonal flows in differentially rotating stellar and planetary convective regions\\
   I. Free waves
   }
   \titlerunning{Astoul et al.: The complex interplay between tidal inertial waves and zonal flows.}

   \subtitle{}

   \author{Aurélie Astoul
          \inst{1,2}
          \and
	Junho Park\inst{1,3}
	\and
          Stéphane Mathis\inst{1}
	\and
	Clément Baruteau\inst{4}
	\and
	Florian Gallet \inst{5}
          }

   \institute{Laboratoire AIM Paris-Saclay, CEA/DRF - CNRS - Université Paris Diderot, IRFU/DAp Centre de Saclay, 91191 Gif-sur-Yvette, France
	\and Department of Applied Mathematics, School of Mathematics, University of Leeds, Leeds, LS2 9JT, UK
	\and
	\aava{Fluid and Complex Systems Research Centre, Coventry University, Coventry CV1 5FB, UK}
	\and
	IRAP, Observatoire Midi-Pyrénées, Université de Toulouse, 14 avenue Edouard Belin, 31400 Toulouse, France
	\and
	Université Grenoble Alpes, CNRS, IPAG, 38000 Grenoble, France
}
   \date{\jp{
   }}
 
  \abstract
   {Quantifying tidal interactions in close-in two-body systems is of prime interest since they have a crucial impact on the architecture and on the rotational history of the bodies. Various studies have shown that the dissipation of tides in \aava{either} body is very sensitive to its structure and to its dynamics. Furthermore, solar-like stars and giant gaseous planets in our solar system are the seat of differential rotation in their outer convective envelope. 
   In this respect, numerical simulations of tidal interactions in these objects have shown that the propagation and dissipation properties of tidally-excited inertial waves can be strongly modified in the presence of differential rotation.}
   {In particular, tidal inertial waves may strongly interact with zonal flows at the so-called corotation resonances, where the wave’s Doppler-shifted frequency cancels out. 
   The energy dissipation at such resonances could deeply modify the orbital and spin evolutions of tidally interacting systems. In this context, we aim to provide a deep physical understanding of the dynamics of tidal waves at corotation resonances, in the presence of differential rotation profiles that are typical of low-mass stars and giant planets.}
   {In this work, we have developed an analytical local model of an inclined shearing box describing a small patch of the differentially rotating convective zone of a star or a planet.  We investigate the propagation and the transmission of free inertial waves at corotation, and more generally at critical levels, which are singularities in the governing wave differential equation. Through the construction of an invariant  called the wave action flux, we identify different regimes of wave transmission at critical levels, which are confirmed with a one-dimensional three-layer numerical model.}
   {We find that inertial waves can be either fully transmitted, strongly damped, or even  amplified after crossing a critical level. The occurrence of these regimes depends on the assumed profile of differential rotation, on the nature as well as the latitude of the critical level, and on wave parameters such as the inertial frequency and the longitudinal and vertical wavenumbers.
   Waves can thus either deposit their action flux to the fluid when damped at critical levels, or they can extract action flux to the fluid when amplified at critical levels. Both situations could lead to significant angular momentum exchange between the tidally interacting bodies.
}
   {}


   \maketitle
\section{Introduction}
%
%
Tidal interactions are known to drive the late evolution of short-period planetary systems, like Hot-Jupiters orbiting around their host star, and in our solar system the satellites around Jupiter and Saturn \citep[e.g.,][]{O2014,M2019}. In particular, the dissipation of tides in the convective envelope of low-mass host stars and giant planets can modify the spin of the tidally pertubed body,  the orbital period and the spin-orbit angle of the perturber \citep[e.g.][]{H1980,FR2006,L2012,BM2016,DM2018}. 
Inertial waves, which are driven by tidal forcing and restored by the Coriolis acceleration, are an important source of tidal dissipation in stellar \citep{OL2007,BO2009,BM2016} and planetary convective zones \citep{OL2004}, where the action of turbulent motions on tidal flows is most often modelled as an effective frictional force or a viscous force with an effective viscosity that is much larger than the molecular viscosity \citep[e.g.,][]{Z1966a,Z1977,DB2020}. 
For coplanar and circular systems, inertial waves are excited so long as the companion orbits beyond half its corotation radius (the orbit where the host’s rotation frequency is equal to the mean motion). 
Low-mass stars from K to F spectral type and giant gaseous planets both harbour a convective envelope surrounding a radiative and a solid (or diluted) core, respectively \citep[e.g.][]{K2012,DC2019}. 
In these objects, inertial waves then propagate in a spherical shell and do not form regular normal modes of oscillation as in spherical \aava{and ellipsoidal geometries \citep[respectively]{G1969,B1889}}. In contrast, they can focus on limit cycles also called attractors of characteristics \citep{ML1995} that are confined within the convective envelope \citep[see also][]{RV1997}.
With a non-zero viscosity, attractors take the form of shear layers where the tidal wave’s energy and angular momentum can be deposited by viscous dissipation \citep{RG2001}. Besides, viscous dissipation across shear layers can be more important as viscosity is weaker, as demonstrated notably by \cite{OL2004}, and \cite{AM2015}. In that respect, tidal dissipation of inertial waves can compete with the dissipation of gravito-inertial waves in the radiative core or be greater by several orders of magnitude than the dissipation of equilibrium tidal flows in the convective zone \citep[i.e., the non-wave like fluid's response; see, e.g.,][]{OL2007}. 
The dissipation of tidally-forced waves can have a great impact on the orbital and rotational evolution of the system \citep{AL2014,BM2016,GB2018,BR2019}.
Moreover, the dissipation of the stellar dynamical and equilibrium tides varies significantly along the evolution of the star, and is highly
dependent on stellar parameters like the mass, the angular velocity, and the metallicity of stars \citep{M2015,GB2017,BG2017}. This makes desirable the inclusion of all stellar processes on tidal interaction, in particular differential rotation.

The frequency-averaged tidal dissipation is often used to quantify the response of a body subject to tidal perturbations \citep{OL2004,JG2008}. Yet, the dissipation of a tidally-forced inertial wave is strongly correlated with the presence of an attractor at a specific eigenfrequency of the spherical shell \citep[see][]{O2009,RV2010}. Tidal dissipation at a given frequency may then alter differently each orbital and spin elements of the two-body systems as postulated for instance by \cite{L2012} to explain the survival of hot-Jupiters with completely damped spin-orbit angle, and revisited by \cite{DM2018} with an improved treatment of dynamical tides in the convective region. In addition in the context of Jupiter and Saturn moon systems, \cite{FL2016} and \cite{LF2018} also investigated the dependence in frequency of tidal dissipation to explain rapid outward migration of the moons,  through resonant locking of tidally-forced internal modes in the giant gaseous planets. This concept could for example explain the high dissipation observed in Saturn as derived from astrometric measurements at the frequency of Rhea \citep{LJ2017}, and at the frequency of Titan \citep{LC2020}. 

Furthermore, the fact that all layers in a star or a planet do not rotate at the same speed, i.e. differential rotation,  is rarely taken into account in the determination of tidal dissipation. Yet, differential rotation seems ubiquitous in low-mass stars and giant gaseous planets. The Sun's surface is rotating in $\sim25$ days at the equator versus $\sim35$ days near the poles, and a latitude-dependent rotational gradient has also been observed in the Sun's convective envelope thanks to helioseismology \citep{SA1998,TC2003}. Through asteroseismology,  latitudinal shears have been found to be comparable to that of the Sun for Sun analogs \citep{BB2019}, and can be even larger for solar-like stars \citep{BB2018}. Essentially, differential rotation in low-mass stars depends on the effective temperature \citep{BCC2005,BJ2017}, and seems to be more important as the convective envelope is thinner. 
Solar-like and anti-solar-like (with faster poles and slower equator) rotation profiles are expected for G and K-type stars based on 3D numerical simulations \citep[see in particular][]{BS2017,BS2018}, while cylindrical rotation profile is expected for fast rotators \citep{GW2013}.
Regarding giant gaseous planets in our solar system, the extent of zonal winds, which are visible on their surface as running lengthwise bands, has been recently constrained by the probes Cassini and Juno. They extend to $3000\,\kilo\meter$ depth for Jupiter \citep{KG2017}, while they penetrate down to 9000 km in Saturn \citep{GK2019}. Thus, the outermost molecular convective envelopes \citep{MW2019,DC2019} are the seat of cylindrical differential rotation.

The study of the impact of differential rotation on the propagation and dissipation properties of inertial modes of oscillation began with the work of \cite{BR2013}. They examined the impact of either a shellular (radial) or a cylindrical rotation profile on free inertial waves in an incompressible background, by means of a Wentzel-Kramers-Brillouin-Jeffreys (WKBJ) linear analysis for an inviscid fluid and by solving the linearised hydrodynamics equations for a viscous fluid via a spectral code.
Their linear analysis highlighted major differences compared to the case of solid-body rotation. Two regimes of propagation have been found, in which inertial modes of oscillation can develop along curved paths of characteristic in the entire convective shell (which the authors named D modes), or in a restricted region of the convective shell, encompassed between a turning surface and one of the shell's boundaries (DT modes). Compared to solid-body rotation, the frequency range of propagation of inertial modes is broader. \cite{BR2013} also pointed out strong dissipation of wave energy at corotation resonances where the Doppler-shifted wave frequency vanishes within the fluid. All these new properties have been retrieved by \cite{GB2016}, who in turn examined a conical (latitudinal) rotation profile, which is typical of low-mass (F- to K-type) stars. They also confirmed the existence of unstable inertial modes (i.e., modes with positive growth rate) at corotation resonances, which were found only for shellular rotation in \cite{BR2013}. Tidal forcing of inertial waves with conical rotation has been introduced by \cite{GM2016} within a linear numerical exploration, which also underlined the strong dissipation of inertial waves at corotation resonances, particularly at low viscosities. \cite{FB2014} also studied tidally-forced inertial waves, but through non-linear numerical simulations. Differential rotation was triggered in their simulations by tidal waves depositing energy and angular momentum in an initially uniformly rotating spherical shell. In some cases, they observed hydrodynamical shear instabilities when the Ekman number (the ratio between the viscous and Coriolis accelerations) is sufficiently small.

Understanding how inertial waves interact with corotation resonances is thus a key issue in quantifying tidal dissipation, especially since waves may deeply interact with the background flow at this  particular location, which in turn may alter the background flow \citep[as it was proposed first by ][for terrestrial mountain waves]{EP1961}. In binary systems and for late-type stars, \cite{GN1989} have shown that the angular momentum transported by gravity waves and exchanged at corotation can lead to the  successive synchronisation of the layers, from the base to the top of the radiative envelope.
 More generally, a body of work in various domains from astrophysical disks \citep[e.g.][]{GT1979,BM2008,LB2009,TL2009} to geophysical fluid dynamics \citep[e.g.][]{B1966,YT1984} has tried to understand the properties of wave propagation and dissipation around corotation, and more generally at all special locations in fluids that correspond to singularities in the linear wave propagation equation. We will refer to them as critical levels in the following \citep{M1986}, or to critical layers in the case of a viscous medium. This distinction is analogous to that between shear layers and attractors of characteristics that are kind of singularities for the governing equation of inertial waves in a spherical shell. The aforementioned singularities can act very differently, with either severe absorption at the critical level \citep[like in ][for stratified vertical shear flows]{BB1967}, or no attenuation if the wave propagates in a peculiar direction \citep[for stratified vertical shear flows with rotation and magnetism]{J1967,A1972,G1975}. In other cases, a critical level may even give rise to wave amplification under certain conditions related to the first and second derivatives of the mean flow velocity \citep[for barotropic and stratified shear flows, respectively]{LT1978,LB1985}. These studies have in common the use of an invariant quantity (the Reynolds stress or the wave action for rotating or magnetic flows) as a diagnostic tool to interpret the role of the critical level in terms of energy transmission and to quantify  exchanges between the wave and the mean flow \citep{EP1961,B1966}. 
 
In light of these various studies, it is necessary to consider carefully corotation in differentially rotating convective zones. 
A local model can notably allow us a detailed understanding of physical processes at critical levels.
 While the propagation through a critical level of gravito-inertial waves in stratified shear flows and of Rossby waves in baroclinic and barotropic flows has been largely studied in the past decades, the behaviour of inertial waves in a latitudinal sheared flow with critical levels has been poorly investigated so far \citep[e.g.][for a review]{L1988}. This is why we develop in this work a local Cartesian shearing box model to understand the complex interplay between tidal waves and zonal flows near critical levels. The concept of a shearing box \aava{for tidal flows} has been introduced by \cite{OL2012} to investigate the interactions between \aava{large-scale tidal perturbations} and convective motions. In our model, we focus on latitudinal differential rotation of the mean flow, varying the box orientation to model either cylindrical or conical rotation. The behaviour of free inertial waves in this framework is then examined near critical levels using both analytical and numerical approaches. 
  
 This paper is organised as follows. 
 In Sect. \ref{sec2}, we describe the local shear model with its main assumptions and the system of governing equations. 
 In Sect. \ref{sec3}, we establish a second-order ordinary differential equation (ODE) for the latitudinal perturbed velocity, and we derive the propagation properties of inertial waves for an inviscid fluid. 
 This \aava{ODE} is solved near each critical level for both conical  and cylindrical rotation profiles, and we interpret energy flux exchanges between the waves and the mean flow. We use in Sect. \ref{secnum} a three-layer numerical model to test our analytical predictions at critical levels. Viscosity is included and non-linear mean flow profiles are also used.
 Astrophysical applications are discussed in Sect. \ref{astrodi} with  implications for low-mass stars hosting close exoplanets and giant gaseous planets in our solar system. In sect. \ref{conclu}, we summarise the main results of the paper, and discuss some perspectives and caveats.
\section{Local Cartesian model including differential rotation}
\label{sec2}
\begin{figure}
\centering
\includegraphics[width=0.3\textwidth]{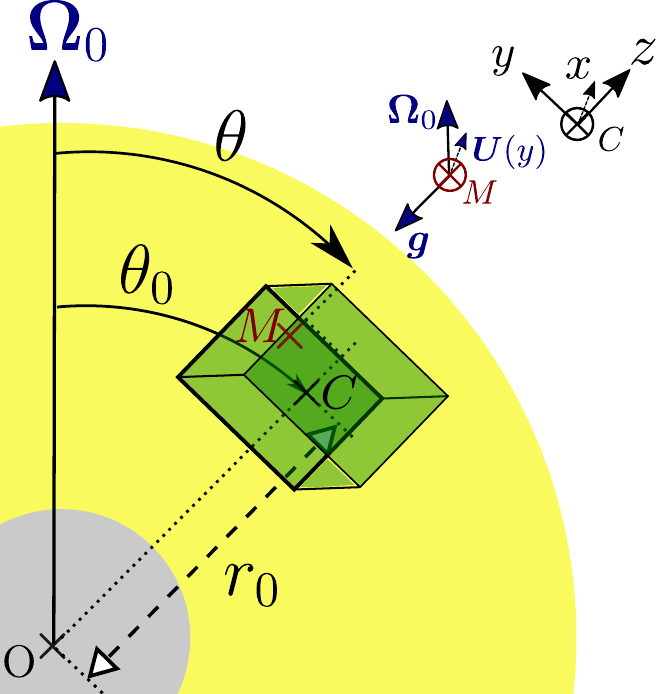}
\caption{Sketch of the local Cartesian box in the convective region of a low-mass star or giant planet. Global spherical coordinates like the depth $r_0$, the inclination of the box $\theta_0$, and the co-latitude  $\theta$ of a point of interest $M$ inside the box, are shown to ease the analogy between the spherical and the Cartesian geometries.}
\label{box}
\end{figure}
\subsection{Presentation of the model}
The local model takes the form of an inclined sheared box, centred at a point $C$ of a convective shell, as illustrated in Fig. \ref{box}. 
The inclined box model has already been used by \citet{AM2015} to characterise analytically the properties of tidal gravito-inertial waves in the presence of viscous and thermal diffusion in stably stratified or convective regions, and by \citet{AB2017} in layered semi-convective regions in giant planets interiors \citep[see also][for two-dimensional numerical simulations of inertial wave attractors]{JO2014}. The local coordinate system $(x,y,z)$ corresponds to the local azimuthal, latitudinal and radial direction of global spherical coordinates, respectively, as presented in Table \ref{tab_corres}. 
The mean flow velocity $\bo U$ is directed along the local azimuthal axis ${\bo e}_x$ (we neglect possible meridional flows) and differential rotation is embodied by a latitudinal shear $\partial_yU$. As the box is tilted by an angle $\theta_0$ relative to the rotation axis, the rotation vector in the local coordinate system is
\begin{equation}
2\bo\Omega_0=(0,2\Omega_0\sin\theta_0,2\Omega_0\cos\theta_0)=2\Omega_0\,(0,\ft,f),
\end{equation}
where $\Omega_{0}$ is the rotation frequency of the star at the pole, $\ft$ and $f$ are the normalised horizontal and vertical Coriolis components, respectively. Note that the inclusion of both these components means that we go beyond the traditional $f$-plane approximation \citep[see also,][]{GZ2008}.  
Furthermore, we make several hypotheses to model wave propagation in a latitudinal shear flow. 
The buoyancy acceleration is kept in the fluid equations for the background flow. \aava{The effective gravity acceleration $\bo g$ also includes the centrifugal acceleration, the fluid's angular velocity being assumed small compared to the critical angular velocity $\sqrt{GM/R^3}$ where $G$, $M$ and $R$ are the gravitational constant, the mass and the radius of the body, respectively. Thus, the geometry of the body is close to spherical. Furthermore,} the vector $\bo g$ is supposed to be uniform and constant in the whole box. This requires that the typical length of the box $L$ \aava{satisfies} $L\ll H_p$, where $H_p=-p(\mathrm{d}z/\mathrm{d}p)$ is the vertical pressure scale height, with $p$ the pressure. We can assume this because tidally excited waves are expected to have small-scale structures \citep{OL2004,RV2010,AB2017}. Moreover, the dimensions of the box are chosen to be small compared to the depth of the convective envelope so as to remove curvature effects. 
\subsection{Mean flow profile}
\label{mean_flow_sect}
In global spherical geometry, the mean flow based on a conical rotation profile $\Omega(\theta)$ is written \citep[e.g. in][]{GB2016}:
\begin{equation}
\bo u=\boldsymbol\Omega\times\bo r=r\sin\theta\Omega(\theta)\bo e_\varphi,
\label{mean_flow_g}
\end{equation}
where $\bo e_\varphi$ is the azimuthal unit vector, $r$ and $\theta$ are the radius and colatitude, respectively. We introduce $\bo u_0=r\sin\theta\Omega_0\bo e_\varphi$ the mean flow at a point $M$ inside the box (see Fig. \ref{box}) without differential rotation, where we remind that $\Omega_0$ is the spin frequency at the pole. We shall also use the shear contrast $\delta\Omega=\Omega(\theta)-\Omega_0$, i.e. the difference between the angular frequency at colatitude $\theta$ and at the pole. The shear contrast is positive for the Sun since the equator rotates faster than the pole, and negative for anti-solar-like rotating stars.
Using the notations of Fig. \ref{box}, the centre $C$ of the box is located at a distance $r_0\sin\theta_0$ from the rotation axis. Accordingly, the \aava{latitudinal coordinate} of the point $M$ in the local frame is
\begin{equation}
y=r\sin(\theta_0-\theta).
\end{equation}
It should be noted that the radial coordinate $r$ of the point $M$ in spherical geometry can be written as $r=r_0+z$. Nevertheless, we neglect vertical displacements in the expression of the local shear, because we are interested in how the (one-dimensional) horizontal shear affects the wave dynamics while a lot of studies on differential rotation in stars have focused on the vertical shear \citep[e.g.][]{MP2004,DM2009,AM2013,MP2018}.
Since $y/r$ and so $\theta_0-\theta$ are small, we have written in Table \ref{tab_corres} the correspondences in terms of mean flows and shears between the two geometries. 
\begin{table}
\begin{tabular}{c|cc}
Geometry &Local Cartesian &Global Spherical  \\\hline\hline
\rule[-1ex]{0pt}{4ex} Basis &$(\bo e_x,\bo e_y, \bo e_z)$ & $(\bo e_\varphi,-\bo e_\theta, \bo e_r)$ \\
\rule[-1ex]{0pt}{4ex} \begin{tabular}{c} Conical coordinate \end{tabular} & $y/r_0$ & $\theta_0-\theta$\\
\rule[-1ex]{0pt}{4ex} Mean flow & $\bo\Um$ & $\bo u-\bo u_0$ \\
\rule[-1ex]{0pt}{4ex}Conical shear & $\partial_y\Um$ & $-\partial_\theta(\sin\theta\delta\Omega)$ \\
\end{tabular}
\caption{Correspondence between local and global coordinate systems.}
\label{tab_corres}
\end{table}

As an example, the shear contrast from solid-body rotation used by \cite{GB2016} was:
\begin{equation}
\delta\Omega(\theta)=\Omega_0\chi\sin^2\theta,
\label{law}
\end{equation}
where $\chi$ is the magnitude of the shear between the equator and the pole. Performing a second-order Taylor expansion around a fixed colatitude $\theta_0$, such that $\theta=\theta_0-y/r_0$ and at a specified depth $r_0$ inside the convective region, the local mean flow $\Um$ can be recast as 
\begin{equation}
\begin{aligned}
u-u_0& =r_0\Omega_0\chi\sin^3\theta\\
&\simeq \Omega_0\chi\left[r_0\sin^3\theta_0-3\cos\theta_0\sin^2\theta_0 y\textcolor{white}{\frac{3}{2}}\right.\\
&\hspace{1cm}\left.+\frac{y^2}{r_0}\left(3\cos\theta^2_0\sin\theta_0-\frac{3}{2}\sin^3\theta_0\right)+O(y^3)\right].
\end{aligned}
\label{U_conic}
\end{equation}
We point out that the Taylor expansion 
must be pushed further at the pole $\theta_{0}=0$ (and at the pole $\theta_{0}=\pi$ with an opposite sign):
\begin{equation}
u-u_0\simeq-\frac{\Omega_0\chi}{r_0^2}y^3+O(y^4).
\label{cubicpro}
\end{equation}
Accordingly, we can approximate a conical shear as a linear mean flow at the first order when the box is tilted.
We recall that conical shear has been observed in the solar convective zone and is expected in slowly and in moderately rotating solar-like stars  \citep[we refer the reader to sect. \ref{appli_lat} for a detailed discussion; see also][]{BG2015,BS2018,BB2018,BB2019}. When the box is at the pole, $y$ becomes the distance from the rotation axis (called hereafter axial distance). Thus, the mean flow mimics a cylindrical differential rotation that can be modelled using a cubic profile in $y$ given Eq. (\ref{cubicpro}). This rotation profile is found in Jupiter and Saturn, as well as in rapidly rotating stars as demonstrated for instance by \cite{GW2013} and \cite{BG2015}.
\subsection{System of equations}
To derive the system of governing equations for tidal waves in the local reference frame, we made several hypotheses.
Stratification terms, which usually drive the propagation of internal gravity waves, have been kept for clarity sake and will be carefully kept or removed after applying the Boussinesq approximation and setting the equations for inertial waves. 
Moreover, we assume that the action of turbulence can be modelled as a Rayleigh friction term in the momentum equation with an effective \aava{frictional damping rate} $\sf$. This \aava{simplifies the analytical solution} of the fluid equations compared to the usual modelling of turbulence as an effective viscous force \citep[see in particular][]{O2009}. 
 The momentum, continuity and thermodynamic equations for tidal waves in a differentially rotating Cartesian framework thus are:
\begin{align}
\df{\bo u} t+2\bo\Omega_0\times\bo u&= -\frac{\bn p}{\rho}+\bo g-\sf\bo u+\bo f \label{eq:1},\\
\frac{\partial\rho}{\partial t}+\bn\cdot(\rho\bo u) &=0\label{eq:2},\\
\df\rho t -\frac{1}{c_\mathrm{s}^2}\df p t&=0\label{eq:3},
\end{align} 
where $\bo u$, $p$, $\rho$, and $\bo f$ denote the velocity, pressure, density and volumetric tidal forcing, respectively. We have also introduced $c_\mathrm{s}$ the sound speed and \aava{$\df{}{t}=\frac{\partial}{\mathrm{d} t}+\bo u\cdot\bo\nabla$} is the total derivative operator. 

All variables are then linearised at first order: zero-order terms correspond to background equilibrium quantities while first-order terms represent the leading perturbation. The local velocity, density and pressure are therefore written:
\begin{equation}
\left\{\begin{aligned}
\bo u&=\Um(y)\,\bo e_x + \bo u'\\\
\rho&=\rom+\epsilon\rho'\\
p&=p_0+\epsilon p'
\end{aligned}\right.,
\end{equation}
 where $\bo u=(u,v,w)$ in the local Cartesian basis.
 We have introduced the dimensionless parameter $\epsilon$
\begin{equation}
\epsilon=\frac{(2\Omega_0)^2 L}{g},
\end{equation}
where we have used $1/(2\Omega_0)$ a characteristic time scale and $L$ a characteristic length scale of the mean flow. These notations are based on those of \cite{G1975}, and adapted to our model.
In the following, we will work with dimensionless variables using the above scaling, including $2\Omega_{\aava{0}} L$ to scale velocity and $\ron gL$ to scale pressure, with $\ron$ the reference density.
The dimensionless momentum equation of the mean flow is:
\begin{equation}
\epsilon\,\bO\times\bo{\Um}=-\frac{\bn p_0}{\rom}-\bo e_z,
\label{eq0}
\end{equation}
with $\bO$ the unit vector parallel to the rotation axis. Projecting Eq. (\ref{eq0}) into Cartesian coordinates, one can derive:
\begin{equation}
\left\{\begin{aligned}
\partial_xp_0&=0,\\
\partial_yp_0&=-\epsilon\rom f\Um,\\
\partial_zp_0&=-\rom +\epsilon\rom\ft\Um.
\end{aligned}\right.
\end{equation}
At the leading order in $\epsilon$, one can recognise the hydrostatic balance, and at the first-order the geostrophic balance \citep[the set is akin the thermal-wind equilibrium assumption, see e.g.][]{G1975,YT1984}. We underline that tending $\epsilon$ to zero is similar to assuming the Boussinesq approximation. Indeed, all density variations are neglected, except the ones involved in the buoyancy force. 
The dimensionless Brunt-Väisälä frequency is 
\begin{equation}
N^2=\epsilon^{-1}\left(F\frac{\partial_zp_0}{\rom}-\partial_z\ln\rom\right),
\end{equation}
where we have introduced the dimensionless number $F=gL/c_\mathrm{s}^2$, which is small when filtering acoustic waves. Consequently, the curl of Eq. (\ref{eq0}) gives
\begin{equation}
\left\{\begin{aligned}
&\partial_x\rom=0, \\
&\partial_y\ln\rom=\epsilon\left(\ft\partial_y\Um-fUF\right),\\
&\partial_z\ln\rom=\epsilon\left(F\ft U-N^2\right)-F,
\end{aligned}\right.
\label{rhorel}
\end{equation}
where we neglect the second-order terms in $\epsilon$. 

Now, we make several assumptions to treat the propagation of inertial waves. As the convective motions
are essentially adiabatic, the convective zone can be assumed neutrally stratified to a first approximation. Hence, the Brunt-Väisälä frequency $N$ is cancelled out in the third density relationship Eq. (\ref{rhorel}). Moreover, we make the Boussinesq approximation, which means that we \aava{neglect} terms in $\epsilon$ and $F$ in the final set of perturbed equations. 
Thus, the dimensionless linearised momentum, continuity, and thermodynamic equations are finally:
\begin{align}
\df{\bo u'} t+v\partial_y\bo U+\bO\times\bo u'&= -\frac{\bn p'}{\rom}-\frac{\rho'}{\rho_0}\bo e_z-\sf\bo u'+\bo f \label{pert:1},\\
\bn\cdot\bo u' &=0,\label{pert:2}\\
\df{\rho'}{t}&=-\rho_0v\ft\partial_y\Um\aava{,} \label{rhop}
\end{align}
\aava{where we remind that $v$ is the latitudinal velocity perturbation}.
 We emphasise that, although vertical stratification has been filtered in the limit $N$ goes to zero, an horizontal stratification term remains in Eq. (\ref{rhop}). As a result, we consider the inertial waves propagating in the inclined shear box where the mean flow is maintained by the thermal-wind balance.
\subsection{Equilibrium state of the background flow}
\label{aside}
It is noteworthy to discuss the choice of keeping buoyancy forces in the zero-order momentum equation.
Without gravitational forces, the momentum equation for mean dimensional variables is written as a geostrophic balance:
\begin{equation}
2\rom\bo\Omega_{\aava{0}}\times\bo U=-\bn p_0.
\label{stab}
\end{equation}
This balance satisfies the Taylor-Proudman theorem 
\citep{R2015}, namely the geostrophic flow is independent of the coordinate parallel to the rotation axis.
When taking the x-axis (the only non-zero) projection of the curl of this equation, one gets the following relationship:
\begin{equation}
(2\bo\Omega_{\aava{0}}\cdot\bn)(\rom\Um)=0.
\label{relstab}
\end{equation}
Without vertical stratification embodied by the Brunt-Väisälä frequency, nor latitudinal stratification, so for an incompressible fluid, the equilibrium of \aava{a y-dependent} mean flow is not ensured. An alternative to conserve the equilibrium without stratification would be to consider a $z$-dependence of the mean flow. 
Such other possibility is not considered in this paper since we are mainly interested in latitudinal mean flow profiles. 
Furthermore, in addition to maintaining differential rotation, the latitudinal stratification can allow to  construct an invariant that is useful for studying energy transfer at critical levels: the wave action flux. This will further discussed in Sect. \ref{sec3}. Lastly, since $\ft=0$ at the poles, the latitudinal stratification term will not appear in the perturbed fluid equations (as we can see from Eq. (\ref{rhop})).
\section{Dynamics of inertial waves at critical levels: analytical predictions}
\label{sec3}
In this section, we investigate analytically the behaviour of inertial waves at critical levels in a non-dissipative fluid at various colatitudes. 
For this purpose, we consider perturbations $q$ in the normal mode
\begin{equation}
q(x,y,z,t)=q(y)\exp{i( k_x x+k_zz-\omega t)}+\mathrm{c.c.}
\label{q}
\end{equation}
with $\omega$ the complex inertial frequency, $k_x$ and $k_z$ the real streamwise and vertical wavenumbers, respectively, and c.c. the complex conjugate. 
\subsection{\aava{Wave propagation} equation \aava{in the latitudinal direction}}
Using the modal form (\ref{q}) for $\rho$, $p$ and $\bo u$,  we solve the set of hydrodynamic equations,  Eqs. (\ref{pert:1}) to (\ref{rhop}), for the latitudinal velocity $v$. Considering free inertial waves (i.e. without forcing terms), the set of perturbation equations can be recast into a single second-order ODE for $v$: 
\begin{equation}
Av''+Bv'+Cv=0, 
\label{Pc}
\end{equation}
where the prime now denotes the derivative according to $y$, and $A$, $B$, and $C$ are the coefficients that can be simplified without friction as follows:
\begin{equation}
\begin{aligned}
A=&\sigma^2-\ft^2,\\
B=&-2k_x\ft^2\frac{\Um'}{\sigma}-2ik_z\ft f,\\
C=&-k_\perp^2\sigma^2+k_z^2f(f-\Um')-2ik_xk_z\ft f\frac{U'}{\sigma}\\
&-2k_x^2\ft^2\frac{\Um'^2}{\sigma^2}+\frac{k_x\Um''}{\sigma}\left(\sigma^2-\ft^2\right),
\label{coefs}
\end{aligned}
\end{equation}
where $k_{\perp}=\sqrt{k_x^2+k_z^2}$ is the absolute wavenumber in the direction perpendicular to the $y$-direction\aava{, and $\sigma=\omega-k_x\Um$ is the (dimensionless) Doppler-shifted wave frequency}. We refer the reader to the Appendix \ref{derPc} for the detailed ODE derivation with \aava{friction} and tidal source terms. 
Eq. (\ref{Pc}) becomes singular when $A=0$ or $\sigma=0$, and these singular points are called critical levels \citep[see e.g][]{B1966,G1975}. The critical level where the Doppler-shifted frequency equals to zero (i.e. $\sigma=0$) can be met when the mean flow matches the local phase velocity, and is also known as corotation resonance \citep[e.g. in][]{GN1989,GT1979,OL2004}. When the Coriolis acceleration is not taken into account, as to treat internal gravity waves, the corotation resonance is the unique critical level \cite[see e.g.][]{BB1967}.
At colatitudes other than the poles, the critical levels come in three flavours, the corotation $\sigma=0$ and two other critical levels that are defined, in our model, by $\sigma=\pm\ft$ (where we remind that $\ft$ in the latitudinal component of the rotation vector). These critical levels were similarly reported for vertical shear flows as in the studies of \citet{J1967} and \citet{G1975} for vertical and inclined rotation vectors, respectively. In these work, the Doppler-shifted frequency at critical levels other than the corotation resonance equals to $\pm2\Omega_{v}$ where $\Omega_{v}$ is the vertical component of the rotation vector. 
\subsection{Propagation properties}
\label{propa_prop}
\subsubsection{Dispersion relation, group and phase velocities}
The 3D dispersion relation is a fourth-order equation in Doppler-shifted frequency when injecting wave-like solutions in the three directions $x,\ y,\text{ and } z$ in Eq. (\ref{Pc}). In order to understand the main properties of waves at the critical level, we make the short-wavelength approximation as in \cite{BR2013} and \cite{GB2016} in the \aava{meridional plane}. This involves keeping only the second-order derivatives in the $y$ and $z$ directions and it reduces the relation dispersion to a second-order equation when injecting plane wave-like solutions. In the local meridional plane, the \aava{differential} equation reduces to \aava{a Poincaré-like equation}:
\begin{equation}
\left(\sigma^2-\ft^2\right)\partial_{y,y}v-2\ft f\partial_{y,z}v+\left[\sigma^2+f\left(\Um'-f\right)\right]\partial_{z,z}v=0,
\end{equation}
\aava{where we recover the Poincaré equation \citep[for the propagation of inertial waves in the inviscid limit,][]{C1922} in the meridional plane when there is no shear ($U'=0$) and at the poles ($\ft=0$ and $f=1$).}
Moreover, we set $v\propto\exp{-i(k_z z -k_x x)}$ so as to write the wave dispersion relation for the Doppler-shifted frequency $\sigma$:
\begin{equation}
\sigma^2=\frac{1}{||\bo k||^2}\left[\left(\bO\cdot\bo k\right)^2-k_z^2f\Um'\right],
\end{equation}
where $||\bo k||=k_y^2+k_z^2$ is the norm of the wave vector in the \aava{meridional} plane (e.g. for fixed $\kx$) like in \citet{BR2013}. Compared to solid-body rotation \citep[see e.g.][]{R2015}, an additional term ($k_z^2f\Up$) is present, which accounts for the latitudinal shear. \aava{Assuming that $\sigma^2$ takes positive values} \citep[as in][]{BR2013,GB2016}, we therefore introduce 
\begin{equation}
\gamma=\sqrt{(\bO\cdot\bo k)^2-k_z^2f\Up}.
\end{equation}
We can then explicit the phase velocity in the \aava{meridional} plane:
\begin{equation}
\bo v_\phi=\frac{\sigma}{{||\bo k||}^2}{\bo k}=\pm \frac{\gamma}{||\bo k||^3}{\bo k}.
\label{vp}
\end{equation}
In the same way, we can derive the expression for the group velocity in the \aava{meridional} plane:
\begin{equation}
\bo v_\mathrm{g}=\bo\nabla_{\bo k}\sigma=\pm\frac{\gamma}{||\bo k||^3}\left\{-\bo k+\left(\frac{||\bo k||}{\gamma}\right)^2\left[{\bO}(\bO\cdot\bo k)-k_zf\Up \bo e_z\right]\right\}.
\label{vg}
\end{equation}
Note that without differential rotation, the group velocity reduces to its well-known expression for solid-body rotation \citep[e.g., see][]{R2015}:
\begin{equation}
{\bo v_\mathrm{g}}=\pm \frac{{\bo k}\wedge({\bO\wedge\bo k})}{||{\bo k}||^3}.
\end{equation}
Moreover, as in solid-body rotation, the group velocity (Eq.~\ref{vg}) and the phase velocity (Eq.~\ref{vp}) lie in perpendicular planes: $\bo v_\mathrm{g}\cdot\bo v_\phi=0$.

When the box is located at the North pole ($\theta_0=0$ in Fig. \ref{box}), by setting $\kappa=1-\Up$ we recover
\begin{equation}
\left\{
\begin{aligned}
\sigma^2&=\frac{k_z^2}{||\bo k||^2}\kappa^2\\
\bo v_\phi&=\pm\kappa\frac{k_z\bo k}{||\bo k||^3}\\
\bo v_\mathrm{g}&=\pm\kappa\frac{k_y}{||\bo k||^3}(-k_z\bo e_y+k_y\bo e_z)
\end{aligned}\right.
\end{equation}
as in \citet{LB2009} or \cite{BR2013}, where $\kappa$ can be identified to their epicyclic frequency and $k_y$ corresponds to the cylindrical component of the wavenumber ($k_s$). 
\subsubsection{Phase and group velocity at singularities}
We derive in this section the conditions required to meet singularities in terms of wavenumbers and shear, and examine what the implications are for the phase and group velocities.
When the box is inclined, for $\sigma\rightarrow0$ we must have: 
\begin{itemize}[label=$-$]
\item$\gamma\rightarrow0$, meaning that $\bm v_\phi\rightarrow\bm0$ while $\ |\bm v_\mathrm{g}\cdot\bm e_y|\rightarrow\infty$ and $|\bm v_\mathrm{g}\cdot\bm e_z|\rightarrow\infty$. 
\end{itemize}
\cite{GB2016} found similar results by studying the propagation of free inertial waves in a global frame with conical shear, namely when their parameter $\mathcal{B}$ (which is homogeneous to a frequency and equivalent to our $\gamma$ parameter) goes to zero, the group velocity goes to infinity while the phase velocity cancels out. According to their work, an inertial wave may propagate across the corotation.\\
Now to get $\sigma\rightarrow\pm\ft$, we either need:
\begin{itemize}[label=$-$]
\item $|k_y|\rightarrow\infty$  at fixed $k_z$, which implies $\bm v_\phi \rightarrow\bm0$ and $\bm v_\mathrm{g}\rightarrow\bm0$ and means that inertial waves cannot get through the critical level,
\item $|k_z|\rightarrow0$ at fixed $k_y$, which gives $\bm v_\phi\cdot\bm e_z \rightarrow0$ and $\bm v_\mathrm{g}\cdot\bm e_y\rightarrow0$ while $|\bm v_\phi\cdot\bm e_y|\rightarrow\ft/k_y$, and $|\bm v_\mathrm{g}\cdot\bm e_z|\rightarrow f/k_y$: the wave may then cross the critical level with some preferential direction.
\end{itemize}
Again, these conditions share some similarities with those observed for corotation in a global spherical geometry. The first aforementioned possibility (first item above) is analogous to the global phase and group velocities tending to zero when $k_s\rightarrow\infty$, with $k_s$ the axial wavenumber in cylindrical coordinates \citep{BR2013,GB2016}. This makes sense since the axial distance is $s=r\sin\theta$, and $y\sim r_0(\theta_0-\theta)$ here. However, the second aforementioned condition (second item above) is slightly different from both these previous works, in that $|k_z|\rightarrow0$ at fixed $k_s$, with $k_z$ the global vertical wavenumber, i.e. along the rotation axis unlike our local vertical wavenumber $k_z$ along the spherical radial coordinate. 

We point out that the singularities at $\sigma=\pm\ft$ arise in our model because the rotation vector is inclined with respect to the local vertical axis of the box. In the global model of \cite{GB2016}, three conditions for a wave to meet the corotation exist, and these conditions are actually quite similar to the three above conditions for waves in our model to interact either with the corotation $\sigma=0$ or the other critical levels at $\sigma=\pm\ft$.
Hence, the local critical levels at $\sigma=\pm\ft$ behave partly like the corotation in the global framework, as if we partially broke the degeneracy in the local framework of the origin of the corotation found in the global framework. 

When the box is at the North pole, the conditions to meet corotation are similar but lead to different relationships for the phase and group velocities:
\begin{itemize}[label=$-$]
\item$\kappa\rightarrow0$ (i.e. $\Up\rightarrow1$) meaning that $\bm v_\phi\rightarrow\bm0$, and $\bm v_\mathrm{g}\rightarrow\bo 0$: the wave is totally absorbed at corotation. 
\item $|k_y|\rightarrow\infty$  at fixed $k_z$, which implies $\bm v_\phi \rightarrow\bm0$ and $\bm v_\mathrm{g}\rightarrow\bm0$: same conclusion as in the previous case and analogous case as when the box is tilted,
\item $|k_z|\rightarrow0$ at fixed $k_y$, which gives $\bo v_\phi\rightarrow\bo 0$, $\bm v_\mathrm{g}\cdot\bm e_y\rightarrow0$ while $|\bm v_\mathrm{g}\cdot\bm e_z|\rightarrow \kappa/k_y$: the wave energy does not cross the corotation in the latitudinal direction \citep[equivalent to vertical paths of characteristic in global cylindrical geometry like in ][]{BR2013}.
\end{itemize}
At the north pole\footnotemark\footnotetext{Note that a similar analysis can be undertaken at the South Pole, by using $\kappa=1+\Up$.}, we actually have a perfect match with the conditions given by \cite{BR2013} when using a cylindrical rotation profile for the mean flow. 
\subsubsection{Energetical aspects}
In this section, we examine the energetic balance associated \aava{with} inertial waves in our inclined shear box model, without assuming the short-wavelength approximation.
This energetic balance does not include potential energy because of the adiabaticity of the convective region but two additional terms appear compared to the solid-body rotation case, coming from the differential rotation.
We denote by $\eta$ and $\zeta$ the displacements along the vertical and latitudinal directions, respectively.  Considering that $\omega=k_x c$, where $c$ is the longitudinal phase velocity \citep[e.g. as in][]{BB1967}, we can use the first-order definition 
\begin{equation}
\begin{aligned}
v&=\df\zeta t=(\Um-c)\zeta_x,\\
w&=\df\eta t=(\Um-c)\eta_x.
\end{aligned}
\end{equation}
It first allows us to express the perturbed density from Eq. (\ref{rhop}) as $\rho=-\rho_0\zeta\ft\Um'$\aava{, where we remind that the symbol $'$ has been dropped out for perturbed quantities}.
Then, by multiplying the momentum equation (\ref{pert:1}) by $\rom\bo u$, we can get the energy balance equation:
\begin{equation}
(U-c)\partial_xe_\mathrm{k}+{\bo\nabla\cdot}(p\bo u)=\rom\left[-v\Up u+\zeta\ft\Um' w-\sf\bo u^2+\bo f\cdot\bo u\right]
\label{balance_e},
\end{equation}
where $e_\mathrm{k}=\rom\bm u^2/2$ is the kinetic energy density, and $p{\bm u}$ the so-called acoustic flux. 
We now integrate the above energy balance equation over $x$ and $z$, and over one wave period, as the perturbed quantities have a wave-like form in these directions. Further assuming that the box is $\delta$ thick in the $y$-direction, the energetic balance yields:
\begin{equation}
\mathcal{P}_\mathrm{ext}=\mathcal{P_\mathrm{shear}}+D_\mathrm{visc}+\mathcal{P}_\mathrm{f},
\label{power}
\end{equation}
where we have introduced from left to right the  power of the external pressures at the boundaries $-\delta/2$ and $\delta/2$ on the perturbed latitudinal flow, the work of the shear, the viscous dissipation, and the forcing power, which read respectively:
\begin{equation}
\begin{aligned}
\mathrm{P}_\mathrm{ext}&=\left[\textcolor{white}{\ft}\overline{pv}\textcolor{white}{\ft}\right]_{-\delta/2}^{\delta/2},\\
\mathcal{P_\mathrm{shear}}&=-\int_{-\delta/2}^{\delta/2}\rom\Up\left(\overline{uv}-\ft\,\overline{w\zeta}\right)\,\mathrm{d}y,\\
D_\mathrm{visc}&=-\int_{-\delta/2}^{\delta/2}\rom\sf\overline{\bo u^2}\,\mathrm{d}y,\\ 
\mathcal{P}_\mathrm{f}&=\int_{-\delta/2}^{\delta/2}\rom\overline{\bo f\cdot\bo u}\,\mathrm{d}y,
\end{aligned}
\end{equation}
where the bar represents the average in the $(z,x)$ plane over one period. Note that the energy density and the acoustic flux in the $x-$ and $z-$directions drop out in Eq. (\ref{power}) when integrating, because of the wave periodicity in those directions. The quantity $\mathcal{P}_\mathrm{shear}$ can also be seen as the power transferred from the mean flow to the perturbation (or conversely) by the Reynolds stress:
\begin{equation}
\tau=-\rom(\overline{uv}-\ft\,\overline{w\zeta})=-\rom\overline{v\left(u+\ft\eta\right)},
\label{Re}
\end{equation}
where we have used partial integration and the periodicity of perturbations in the $x$ and $z$ directions.
At the pole, $\ft=0$ so we recover the definition of the Reynolds stress in \cite{M1961}, who studied the stability of a 2D stratified $y$-sheared flow, i.e. $\tau=-\rom\overline{uv}$. This quantity can also be called the latitudinal flux of horizontal (in the sense $(z,x)$ plane) momentum in reference to the vertical flux of horizontal momentum in stratified $z$-sheared flows. Moreover, we emphasise that the latitudinal flux of energy $\overline{pv}$ is not conserved even in the inviscid free-wave problem. This is due to the presence of the shear, as already stated for example by \cite{EP1961}, who studied stratified vertically sheared flows. They underline that when the mean flow varies with height, the kinetic energy of the mean motion can be converted into wave energy. 
Without \aava{friction} and forcing, the $y$-derivative of the latitudinal flux is:
\begin{equation}
\df{}{y}\overline{pv}=-\rom\Up\overline{v\left(u+\ft\eta\right)}.
\label{flux}
\end{equation}
Using the same method as \cite{B1995}, we multiply the $x$-projection of the inviscid force-free momentum equation by $\zeta$:
\begin{equation}
-p_x\zeta=\rho_0(\Um-c)\zeta u_x+\rho_0(\Um'-f)v\zeta+\rom\ft w\zeta.
\end{equation}
By multiplying by $(U-c)$, the latitudinal flux of energy can thus be written as:
\begin{equation}
\overline{pv}=-\rom(\Um-c)\overline{v\left(u+\ft\eta\right)}.
\end{equation}
By \aava{differentiating this relationship with respect to $y$}, and by equalising with Eq. (\ref{flux}), one can obtain:
\begin{equation}
-\rom(U-c)\frac{\mathrm{d}}{\mathrm{d}y}\left\{\overline{v\left(u+\ft\eta\right)}\right\}=0,
\label{rel}
\end{equation}
that is $(U-c)\cfrac{\mathrm{d}\tau}{\mathrm{d} y}=0$ with $\tau$ the Reynolds stress. Eq.~(\ref{rel}) is naturally satisfied at corotation, where $U-c=0$, or if the Reynolds stress is uniform. \cite{BB1967} have shown that the Reynolds stress is discontinuous at a critical level, highlighting  exchanges between wave energy and the mean flow.
Compared to the analysis of \cite{B1995} for 3D stratified shear flows, Eq.~(\ref{rel}) is not vectorial, because our base flow is unidirectional.
\subsubsection{Polarisation relations}
For the forthcoming analysis, it is useful to derive expressions of the perturbed projected velocities and the perturbed reduced pressure\footnotemark\footnotetext{The quantity $\Pi$ is actually the enthalpy perturbation but we will use the denomination "reduced pressure" in the following.} $\Pi=p/\rho_0$, namely the polarisation relations (see Appendix \ref{derPi} for more details). In the inviscid free-wave problem, these perturbed quantities can be written in terms of the latitudinal velocity, its derivative, and the shear:
\begin{equation}
\begin{aligned}
\Pi&=\frac{1}{\sigma k_\perp^2}\left\{iAv'+\left[ik_x\frac{\Um'}{\sigma}A+f(k_z\ft-ik_x\sigma)\right]v\right\},\\
w&=\frac{1}{\sigma k_\perp^2}\left\{\left(i\sigma k_z-k_x\ft\right)v'+k_x\left[ik_z(\Um'-f)-k_x\ft\frac{\Um'}{\sigma}\right]v\right\},\\
u&=\frac{1}{\sigma k_\perp^2}\left\{\left(i\sigma k_x+k_z\ft\right)v'+k_z\left[ik_z(f-\Um')+k_x\ft\frac{\Um'}{\sigma}\right]v\right\}.
\end{aligned}
\label{pol}
\end{equation}
Without shear and at $\theta_0=0$, we recover the polarisation relations in the solid-body rotation case \citep[see e.g.][]{R2015}.
\subsubsection{Conservation of the wave action flux}
\label{Co}
While the latitudinal flux of energy is not conserved in the whole domain, there is a conserved quantity, called the wave action flux as introduced 
in \cite{G1975}'s paper:
\begin{equation}
\mA=\frac{\overline{pv}}{\sigma}=\frac{\rom}{2}\Re\left\{\frac{\Pi v^*}{\sigma}\right\},
\label{A}
\end{equation}
which is the latitudinal flux averaged over vertical and longitudinal wavelengths divided by the Doppler-shifted frequency. A general treatment for the derivation of the wave action as a conserved quantity can be similarly found in \cite{AM1978}. The wave action flux is related to the Reynolds stress $\tau$ as $\mA=-\tau/k_x$. 
By using the expression for the perturbed reduced pressure $\Pi$ derived in the previous section, the wave action flux now reads:
\begin{equation}
\mA=\frac{1}{2}\rom\Re\left\{\frac{iAv_yv^*+f\ft k_z|v|^2}{\sigma^2k_\perp^2} \right\}.
\label{wA}
\end{equation}
Unlike the latitudinal flux of energy, but similarly to the Reynolds stress, this wave action flux is conserved along the latitudinal direction. One can demonstrate that $\mA'=0$ in the whole domain except at critical levels, by using the expression for the reduced pressure in Eqs. (\ref{pol}) and the \aava{ODE} (\ref{Pc}). Several works have shown that a properly defined (i.e., conserved) angular momentum transport parameter can be found in $z$-sheared mean flows without rotation \citep{BB1967}, with rotation under the traditional approximation \citep{J1967}, and with rotation under the non-traditional approximation \citep{G1975}.
Verifying the conservation in the whole domain except at critical levels is really important because it brings to the fore energy transfers due to the critical levels. We specify that $\mA$ is a measure of wave energy through a surface (in the $(z,x)$ plane) since $\overline{pv}$ is the energy density transported by the group velocity\footnotemark\footnotetext{Note that the velocity of energy density $\wg$ in the latitudinal direction has been named "group velocity" in the latitudinal direction for obvious physical reasons but it differs from the group velocity defined in Sect. \ref{propa_prop} that depends on latitudinal and vertical wavenumbers unlike $\wg$.
} $\wg$ in the latitudinal direction \citep[e.g.][]{BG1968,MB2012}. It should be underlined that the wave action flux has been defined in the inviscid limit and is not conserved when the \aava{friction} is taken into account.
\subsection{Inertial waves at critical levels when the box is tilted}
\label{Fro}
In this section, we analytically investigate waves passing through the various critical levels in the tilted box.
We examine the behaviour of the waves around the corotation $\sigma=0$ and the critical levels $\sigma=\pm\ft$ when the box is tilted (for the corotation when the box is at the pole, see Section \ref{polar_case}).
\subsubsection{Critical levels at $\sigma=\pm\ft$}
\label{cc_ft}
\begin{figure*}[ht]
\resizebox{\hsize/2}{!}{\includegraphics{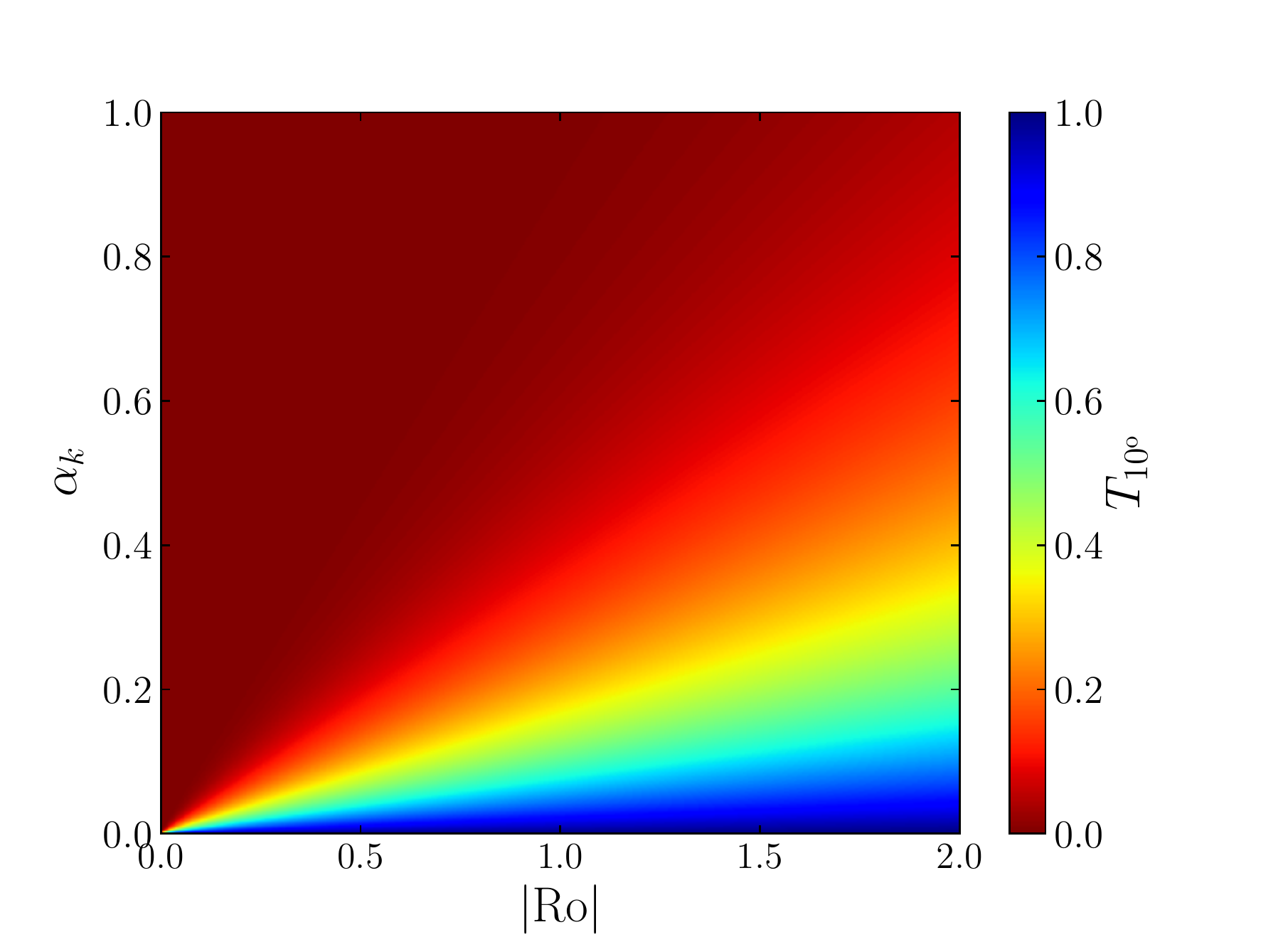}}
\resizebox{\hsize/2}{!}{\includegraphics{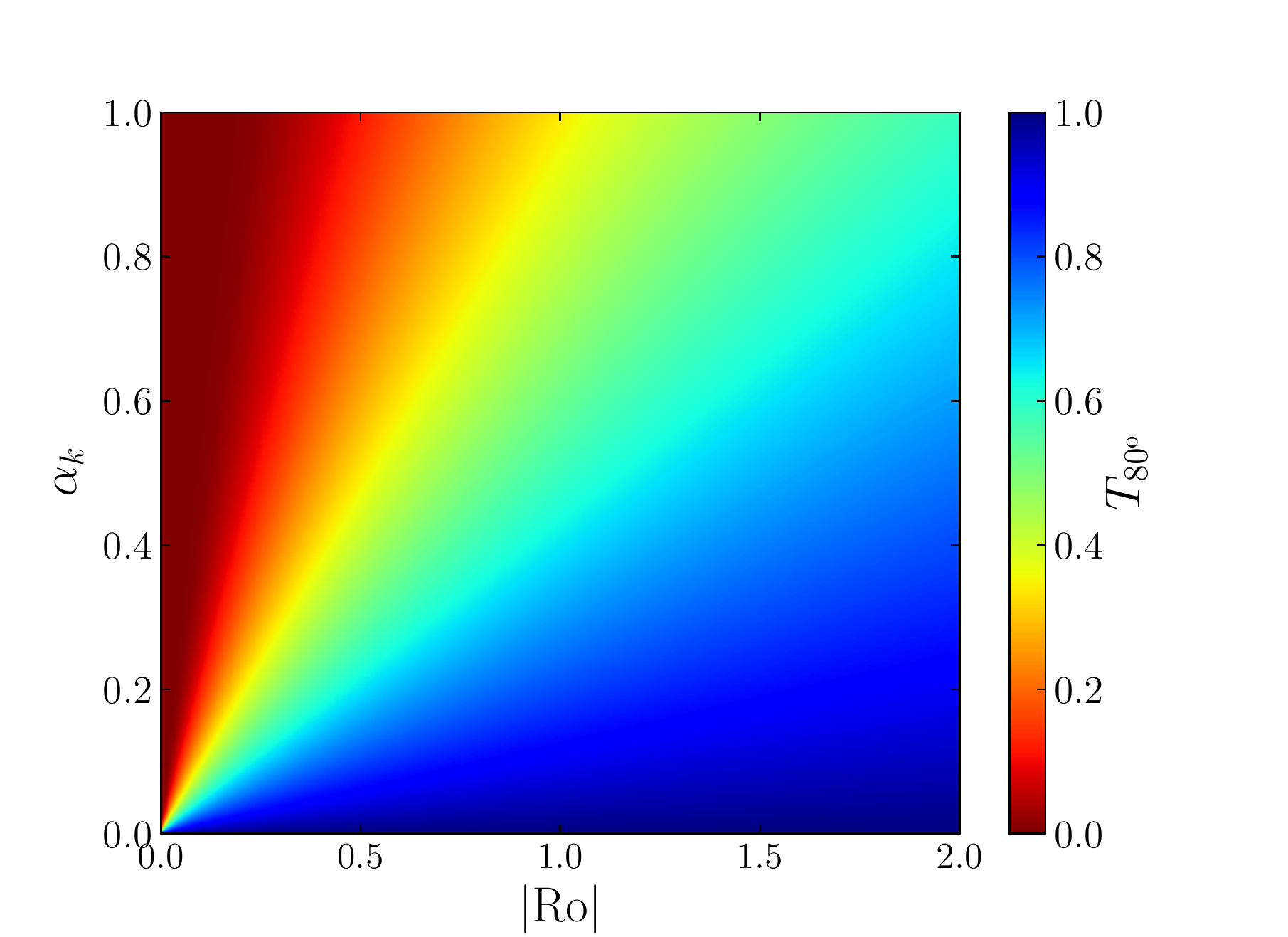}}
\caption{Transmission rate $T_{\theta_0}$ of a wave passing through any of the critical levels defined by $\sigma=\pm\ft$ as a function of the absolute value of the shear Rossby number $|\Ro|=|\Up|$ (where $\Up$ is scaled by $2\Omega_0$), the ratio of the horizontal wave numbers $\alpha_k=k_z/k_x$, and for a co-latitude of the box $\theta_0=10\degree$ (\textit{left panel}) and $\theta_0=80\degree$ (\textit{right panel}).}
\label{atnu_deg}
\end{figure*}
In this subsection, we treat both singularities $\sigma=\pm\ft$ simultaneously. Although Eq. (\ref{Pc}) does not have analytical solutions in general, it is still possible to study the behaviour of an inertial wave close to the critical levels defined by $\sigma=\pm\ft$ by approximating the ODE through its first-order Taylor expansion in the vicinity of these singularities, and then by applying the Frobenius method.
We introduce $y_\pm$, the location of the related critical level $\sigma=\pm\ft$.
For a linear mean flow profile $\Um=\Lambda y$, with $\Lambda$ a constant, $y_\pm$ are given by:
\begin{equation}
y_\pm=\frac{\omega\mp\ft}{k_x\Lambda}.
\end{equation}
Without any assumption on the mean flow profile, the first-order Taylor expansion of the ODE (\ref{Pc}) near $y_\pm$ is:
\begin{equation}
v''+\frac{\mB}{y-y_\pm} v'=0,
\label{PcDL_ft}
\end{equation}
with
\begin{equation}
\begin{aligned}
\mB=&1\pm i\frac{k_zf}{k_x\Um'_{\aava{\pm}}},\\
\end{aligned}
\end{equation}
where the symbol $\pm$ 
 refers to the regular singularities\footnotemark\footnotetext{A singular point $y_0$ of the second-order ODE $v''(y)+p(y)v'(y)+q(y)v(y)=0$ is said regular when the function $p(y)(y-y_0)$ and $q(y)(y-y_0)^2$ are analytical at $y=y_0$.} $y_+$ and $y_-$, respectively\aava{, and $\Up_\pm$ is $U'$ evaluated at these singularities.}
The Frobenius method consists in injecting the power function $(y-y_\pm)^\lambda$ in Eq. (\ref{PcDL_ft}), with $\lambda$ a constant to be determined \citep[see e.g.][]{MF1953}. The corresponding indicial equation is then:
\begin{equation}
\lambda(\lambda-1)+\mB\lambda=0,
\end{equation}
with solutions:
\begin{equation}
\lambda_\pm=\left\{0,\ \mp i\frac{k_zf}{k_x\Um_{\aava{\pm}}'}\right\}.
\end{equation}
Therefore, the two independent solutions of Eq. (\ref{PcDL_ft}) can be written as follows:
\begin{equation}
v_{1,\pm}(y)=\sum_{n=0}^{+\infty}a_n(y-y_\pm)^n\ \text{ and }\ v_{2,\pm}(y)=\sum_{n=0}^{+\infty}b_n(y-y_\pm)^{n+\lambda_\pm},
\label{sol}
\end{equation}
where $a_n$ and $b_n$ are complex constants. Both solutions are valid in the vicinity of the critical level around which they are built \aava{in the complex plane}, up to the next singularity if it exists. The coefficients $a_0$ and $b_0$ are unconstrained and depend on boundary conditions, unlike the other factors that can be determined by injecting solutions (\ref{sol}) into the \aava{linearised ODE (\ref{Pc}) around $y_\pm$ at the right order for the desired coefficients.}
 
Near the critical points $y_\pm$, the total solution $v_\pm$ is well approximated by the lowest orders of $v_{1,\pm}$ and $v_{2,\pm}$:
\begin{equation}
v_\pm(y)\simeq a_0+b_0(y-y_\pm)^{\lambda_\pm}.
\end{equation}
Owing to the existence of a branch point at $y_\pm$ (since $\lambda_\pm$ is complex), reconnecting solutions on either part of the critical levels is not straightforward. This requires both physical and mathematical arguments \citep[see in particular][]{M1961,BB1967,R1998}.
 In order to remove degeneracy of the path from positive to negative $y-y_\pm$ (i.e. choose either $\e^{+i\pi}$ or $\e^{-i\pi}$), we make use of a complex inertial frequency $\omega=\omega_\mathrm{R}+i\omega_\mathrm{I}$, assuming the radiation condition $\omega_\mathrm{I}>0$. This condition ensures a non-growing wave toward infinity. 
The Taylor expansion of the base flow at first-order in $y-y_\pm$ gives
\begin{equation}
y-y_\pm=\frac{\Um(y)-\Um_{\aava{\pm}}}{\Um'_{\aava{\pm}}},
\end{equation}
and by definition, we have
\begin{equation}
\Um_{\aava{\pm}}=\frac{\omega\mp\ft}{k_x}.
\end{equation}
Consequently, the solution below the critical level is unambiguous in terms of the above solution coefficients, and depends on
\begin{equation}
\sign[\Im\{y-y_\pm\}]=-\sign\left[k_x\Um'_{\aava{\pm}}\right].
\label{branch}
\end{equation}
In other words, when taking $y-y_\pm$ to decrease from positive to negative values, its complex argument changes continuously from $0$ to $-\sign\left[k_x\Um'_{\aava{\pm}}\right]\pi$. 
Thus, the appropriate path for determining the branch of $(y-y_\pm)^{\lambda_\pm}$ passes under (above) $y_\pm$ as long as $k_x\Um'_{\aava{\pm}}>0$ ($k_x\Um'_{\aava{\pm}}<0$) \citep[the same reasoning can be found in ][]{G1975}.
Therefore, the solution on both sides of the critical level $y_\pm$ is:
\begin{equation}
v_{\pm}(y)\simeq\left\{\begin{aligned}
&a_0+b_0|y-y_\pm|^{\mp i k_z f/k_x\Up_{\aava{\pm}}}&\text{ for }y>y_\pm\\
&a_0+b_0|y-y_\pm|^{\mp i k_z f/k_x\Up_{\aava{\pm}}}\mathrm{exp}(\mp \pi k_z f/|k_x\Up_{\aava{\pm}} |)&\text{ for }y<y_\pm
\end{aligned}\right..
\end{equation}

The remaining issue is now to know in which direction the wave is propagating. The second part of the solution can be assimilated to a wave-like solution with the varying latitudinal wavenumber $\mp (k_z f/k_x\Up_{\aava{\pm}})  \log|y-y_{\pm}|$. 
Moreover, according to Eq. (\ref{wA}), the wave action flux on either side of $y_\pm$ is:
\begin{equation}
\mA=
\frac{\rom k_zf}{2\ft k_\perp^2}\left\{\begin{aligned}
&|a_0|^2-|b_0|^2&\text{ for }y>y_\pm\\
&|a_0|^2-|b_0|^2\mathrm{exp}(\mp2\pi k_z f/|k_x\Up_{\aava{\pm}}|)&\text{ for }y<y_\pm
\end{aligned}\right..
\label{atnu}
\end{equation}
The group velocity gives the direction towards which the energy is transported, recalling that  $\wg\mE=\overline{pv}$ with $\wg$ the group velocity and $\mE$ the local energy density. By consequence, $\sign({\wg})=\sign({\mA\sigma})=\mp\sign(fk_z)$ for the solution featuring the coefficient $b_0$. If $k_zf$ is positive, this wave transports energy downward (upward) across  the critical level $\sigma=\ft$ ($\sigma=-\ft$). If $k_zf$ is negative, the wave transports energy upward (downward) across the critical level $\sigma=\ft$ ($\sigma=-\ft$). In all cases, the action flux of the wave with the
amplitude $|b_0|$ will be transmitted (in the direction given by the sign of $k_zf$ and the critical level $y_+$ or $y_-$) by a factor $T_{\theta_0}$ where
\begin{equation}
    T_{\theta_0}=\exp(-2\pi\frac{|k_zf|}{|k_x\Up_{\aava{\pm}}|})=\exp(-2\pi\frac{|\alpha_k\cos\theta_0|}{|\Ro|}),
\end{equation}
with $\ak=k_z/k_x$ and $\Ro=\Um'_{\aava{\pm}}$, after passing through the critical level. Such wave will always be attenuated since $T_{\theta_0}\leq1$.
The transmission factor $T_{\theta_0=10\degree}$ and $T_{\theta_0=80\degree}$ are displayed in Fig. \ref{atnu_deg} in terms of the absolute value of the shear Rossby number $|\Ro|$ and the ratio of wave numbers $\alpha_k$. 
The lower the amplitude of the Rossby number and the lower the inclination, the more likely the wave is to be strongly attenuated at any $\alpha_k$. We remind that a low Rossby number refers either to fast rotating stars or to low differential rotation.
At the equator, one should note that \aava{$f=0$ so there is no transmission nor exchange of wave action flux near the critical levels $y_\pm$ in the inviscid limit (see Eq. (\ref{atnu}))}. Results are the same for $\theta_0+k\pi/2$ with $k\in\{0,1,2,3\}$, and for negative Rossby numbers.
However, it has to be emphasised that the cases where the inclination satisfies $\theta_0=k\pi$ with $k\in\{0,1\}$ are not well described by the attenuated factor $T_{\theta_0}$ and require a specific treatment, as discussed in Sect. \ref{polar_case}.

It is important to note that, with fixed parameters $\{k_z, \theta_0,\Ro\}$, the attenuation of the wave action flux is specific to a single direction of wave propagation, i.e. the solution featuring the coefficient $b_0$. The solution of coefficient $a_0$ is not affected by the attenuation. It is the so-called valve effect introduced by \cite{A1972} in the context of hydromagnetic waves in a rotating fluid. It was also evidenced by \cite{G1975}, and further discussed in \cite{G1979} for magneto-gravito-inertial waves in an inviscid and compressible z-sheared fluid.
\subsubsection{Inertial wave crossing corotation}
\label{corot_inclined}
We perform the same analysis as in the previous section to treat the corotation point $\yo$ where $\sigma=0$ (i.e. $U(\yo) \aava{=U_0}=\omega/k_{x}$). \aava{The linearised \aava{ODE}~(\ref{Pc}) near the corotation using the}
 Taylor expansion of  \aava{$\sigma$ and $U$} \aava{at the lowest orders} is:
\aava{\begin{equation}
\begin{aligned}
&v''+\left(-\frac{2}{y-\yo}+2ik_z\frac{f}{\ft}-\frac{U_0''}{\Up_0}\right)v'\\&+\left[\frac{2}{(y-\yo)^2}-\frac{2ik_zf}{\ft(y-\yo)}+\frac{\Um_0''}{\Um_0'(y-\yo)}\right]v=0,
\label{Pcorot}
\end{aligned}
\end{equation}}
where $U_{\aava{0}}'$ and $U_{\aava{0}}''$ are \aava{the first and second derivatives of the mean flow profile $U$} evaluated at the critical level $\yo$.
The singularity at the corotation is a regular singularity and we can use again the Frobenius method. The indicial equation has solutions $\lambda=\{2, 1\}$. Since the difference between the two \aava{values} of the exponent \aava{$\lambda$} is an integer, one expects a second independent solution $v_2$ of Eq. (\ref{Pcorot}) \aava{ that includes a logarithmic part such as \citep[e.g.][]{SH2002}:
\begin{equation}
v_2(y)=\sum_{n=0}^{+\infty}b_n(y-\yo)^{n+1}+L\ln(y-\yo)v_1(y)
\end{equation}
 with $v_1(y)=\sum^{+\infty}_{n=0}a_n(y-\yo)^{n+2}$ the first solution, and $a_n$, $b_n$, and $L$ complex coefficients. However, when injecting $v_1+v_2$ in Eq. (\ref{Pcorot}), one finds $L=0$, meaning that a sole polynomial solution in the form
\begin{equation}
v(y)=\sum_{n=0}^{+\infty}c_n(y-\yo)^{n+1}
\label{solln}
\end{equation}
 includes all the solutions of Eq. (\ref{Pcorot}), with $c_0=b_0$ and $a_1=b_1+a_0$ determined by boundary conditions, and  $c_{n,~n\in\mathds{N}^*\backslash\{1\}}=b_n+a_{n-1}$ determined by recurrence via the expansion of Eq. (\ref{Pc}) around $\yo$. As a result, the wave action flux given by Eq. (\ref{wA}) becomes
\begin{equation}
\mA=\frac{\rom}{2k_x^2\Up^2_0k_\perp^2}\left[\ft^2\Im{c_0^*c_1}+f\ft k_z|c_0|^2\right]
\label{wAconcor}
\end{equation}
just below and above the corotation, and it is continuous there, similarly as in \citet{G1975}, but here without being restricted to a linear mean flow profile. Hence, no transfer of wave action flux is expected at corotation in the inviscid limit when the box is inclined relative to the rotation axis (i.e. for conical differential rotation), regardless of the mean flow profile. This result also holds true when the box is located at the equator.}

Like in the works of \cite{G1975} and \cite{J1967}, it is tempting to investigate the asymptotic behaviour of a wave when $y\rightarrow\infty$, in order to better constrain the propagation of waves through one or multiple critical levels. Nevertheless, the term $-\sigma^2k_\perp^2$ in the ODE (Eq. (\ref{Pc})), which can not be overlooked like in the aforementioned studies since we do not have vertical stratification, makes the singularity $y=\infty$ an essential (or irregular) singularity and the Frobenius method can not be applied. This term also prevents us to apply an analysis like the WKBJ approximation because even far from critical levels the coefficients $C$ of the ODE (in Eq. (\ref{Pc})) still have a strong dependence on the latitudinal coordinate when the box is tilted.
\subsection{Inertial waves when the box is at the poles}
\label{polar_case}
When the box is located at the North or the South pole, $\ft=0$ and the \aava{ODE} (Eq. (\ref{Pc}))  is greatly simplified.
For $\theta_0 = 0$, the dimensionless \aava{wave propagation} equation becomes indeed
\begin{equation}
v''+\left[\frac{k_z^2(1-\Up)}{\sigma^2}+\frac{k_x\Um''}{\sigma}-k_\perp^2\right]v=0.
\label{Pcpole}
\end{equation}
At the South pole (i.e. $\theta_0=\pi$), the term $1-\Up$ in Eq.~(\ref{Pcpole}) is replaced by $1+\Up$.
Note that this equation is reminiscent of the differential equation for Rossby waves in the $\beta$-plane, i.e. $2\bo\Omega =(0,\ft,f)$ and constant $\mathrm{d}f/\mathrm{d}y=\beta$, with $f$ the Coriolis parameter \cite[e.g.][]{M1961,G1975b,GH2020}. However, we can not make a direct comparison at corotation, because the singularity in the equations for Rossby waves and inertial waves is not of the same order.
 We have a second-order pole around the corotation while only first-order poles are found in the aforementioned studies. In fact, Eq. (\ref{Pcpole}) is similar to the wave equation in stratified z-sheared flows \citep[e.g.][]{J1968}.

In our polar configuration the y-coordinate is now the axial distance, and it means that the mean flow has a cylindrical profile. Such a rotation profile is expected in giant planets such as Jupiter and Saturn \citep[][respectively]{KG2017,GK2019} as a natural outcome of the Proudman-Taylor theorem for fast-rotating bodies. The propagation and dissipation of inertial modes of oscillations in the presence of critical levels for this kind of mean flow have been investigated by \cite{BR2013} in a spherical shell.
\subsubsection{Analytical solutions with constant shear}
\label{ana_pole}
Analytical solutions of the ODE Eq. (\ref{Pcpole}) are difficult to find for general profiles of the mean flow, e.g. a quadratic mean-flow profile. A linear mean-flow profile, on the other hand, has  analytic solutions, that is why we use in this section such a profile, i.e. $\Um=\Ro y$, with $\Ro$ the shear Rossby number which is taken constant here. Eq. (\ref{Pcpole}) then becomes:
\begin{equation}
v''+\left[\frac{\alpha_k^2(1-\Ro)/\Ro^2}{(y-\yo)^2}-k_\perp^2\right]v=0,
\label{pole_cstUp}
\end{equation}
where $\yo=\omega/(k_x\Ro)$ and $\ak=k_z/k_x$ the vertical to longitudinal wave number ratio. When the box is located at the South pole, the left-hand term in the bracket is $\alpha_k^2(1+\Ro)/\Ro^2$ in the numerator.
 This equation takes the form of Whittaker's equation \citep[see][]{AS1972} and solutions can be written in terms of the Whittaker functions $M$:
\begin{equation}
v(y)=AM_{0,\,\mu\,}(\tilde y)+BM_{0,\,-\mu\,}(\tilde y),
\end{equation}
 with $\tilde y=2k_\perp(y-\yo)$, 
\begin{equation}
\mu=\sqrt{\frac{1}{4}-\frac{\ak^2(1-\Ro)}{\Ro^2}},
\end{equation}
and $A$ and $B$ are complex constants given by boundary conditions. 
The Whittaker function $M_{0,-\mu}$ allows quite straightforward analytic continuation:
\begin{equation}
M_{0,\,\mu\,}(\e^{-i\pi}\tilde y)=-i\e^{-i\pi\mu}M_{0,\,\mu\,}(\tilde y).
\end{equation}
By consequence, the solution below the critical point $y=\yo$ is:
\begin{equation}
v_\mathrm{W}(y)=-i\varsigma A\e^{- i\varsigma\pi\mu}M_{0,\,\mu\,}(-\tilde y)- i\varsigma B\e^{i\varsigma\pi\mu}M_{0,\,-\mu\,}(-\tilde y).
\end{equation}
Although the Whittaker functions do not feature precisely as wave-like forms, we can already have a good idea of the attenuation factor thanks to analytic continuation as will be shown in the following section.

It is important to point out that $\mu$ can be real or complex depending on the value of 
\begin{equation}
\od=\ak^2(1\mp\Ro)/\Ro^2,
\end{equation}
which we will simply denote by $\Od$ in the following. This can drastically change the behaviour of a wave passing through the corotation.
 A necessary, but not sufficient condition to find an instability is that $\Od<1/4$ as we will demonstrate in Sect. \ref{inst}. This condition is similar to the Miles-Howard theorem for stratified $z$-sheared flow \citep{MH1964,L1988}. In these studies, the prerequisite for instability is that $\Ri<1/4$ where $\Ri$ is the Richardson number, i.e. the squared ratio of the Brunt-Väisälä frequency and the vertical \citep[or radial \aava{in global spherical geometry},][]{AM2013} shear.  In our model, unlike cases where the box is tilted, a WKBJ analysis can be performed for a linear mean flow, mainly provided that $|R|\gg1/4$, in line with the condition of stability derived in the coming sections, and detailed in Appendix \ref{wkbj}. 

\begin{figure}
\includegraphics[width=\hsize]{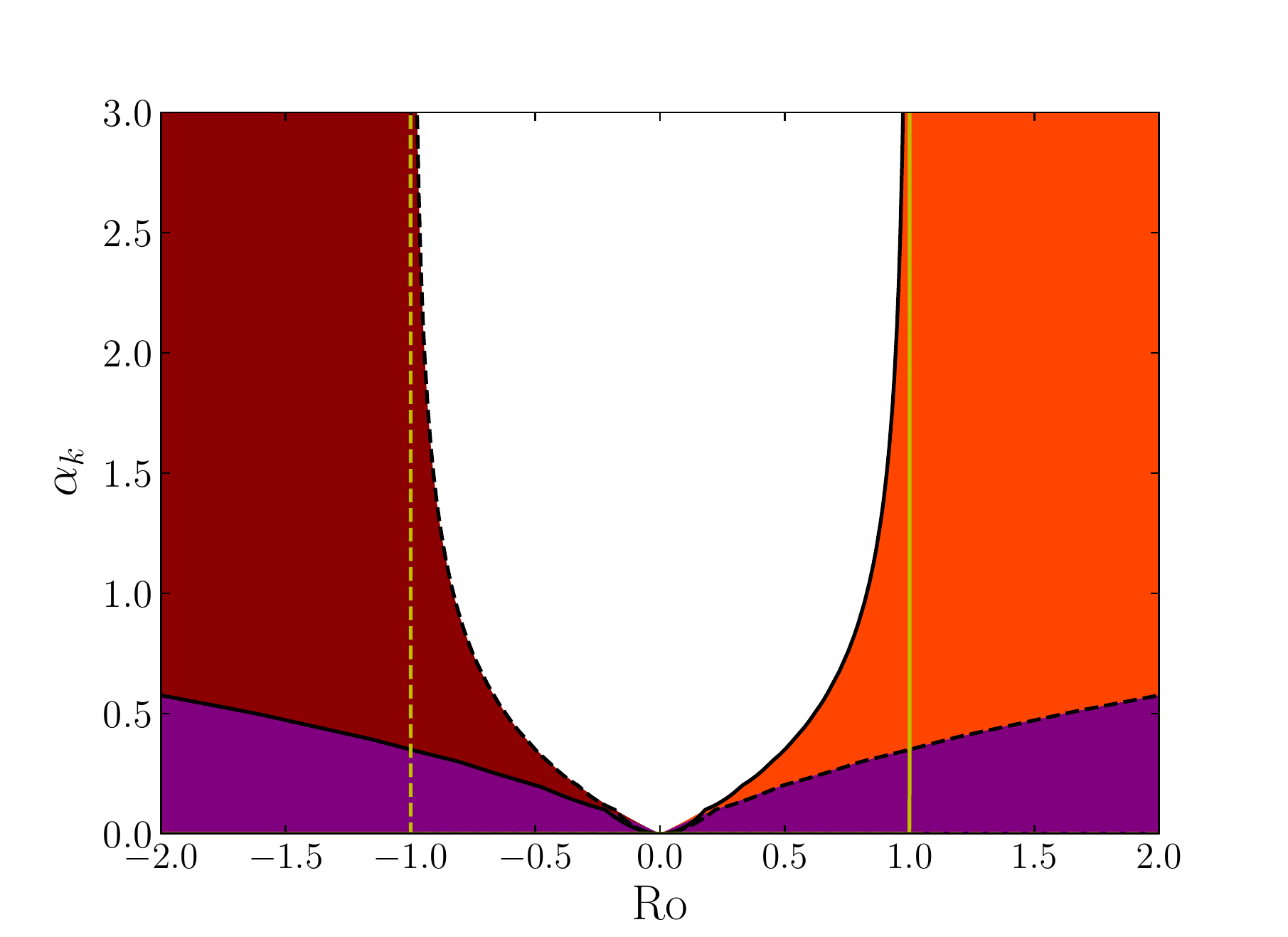}
\caption{Diagram showing how $\od \equiv \Od$ compares to $1/4$ for two positions of the box ($\theta_0=0\text{ and }\pi$), and for a range of Rossby numbers $\Ro$ and of the ratio of the vertical and longitudinal wave numbers $\alpha_k$. The solid and dashed black lines mark where $\od=1/4$ at the north and south pole, respectively. The purple domain show where $\od<1/4$ and the white region where $\od>1/4$, whether the box is at the north or the south pole. In the dark red region, $\Od_{\theta_0=0}>1/4$ and $\Od_{\theta_0=\pi}<1/4$ and vice versa in the orange region.
}
\label{O2}
\end{figure}

These various situations regarding the value of $\od$ at the North and South poles are illustrated in Fig. \ref{O2}. We stress the particular case where $\Ro=1$ ($\Ro=-1$) at the North (South) pole, and where the differential equation and its solutions take a quite simple form:
\begin{equation}
v''-k_\perp^2v=0,\ \text{ with }\ v\propto\e^{\pm k_\perp y}.
\label{eva}
\end{equation}
Solutions are then fully evanescent for such shears. 
One can notice that Eq. (\ref{eva}) is the same far from corotation, for any mean flow. 

Finally, it is clear from Fig. \ref{O2} that wave propagation is the same at the North or South pole provided a Rossby number of opposite sign. 
As a result, only the equations at the  North pole will be treated in the following, and  the word "pole" now refers to the North pole.

\subsubsection{Frobenius method at the pole}
\label{Frobpol}
Though analytic solutions are known, it is still useful to determine Frobenius solutions near corotation for two main reasons. First, these solutions can be derived for any mean flow profile near the corotation. Close to corotation, the mean flow is  approximated by a Taylor expansion at the first-order $\Um=U'_{\aava{0}}(y-\yo)$. Secondly, Frobenius solutions may feature wave-like forms, which is helpful for physical interpretation. 
Therefore, Eq. (\ref{Pcpole}) can be written near  corotation:
\begin{equation}
v''+\frac{\alpha_k^2(1-\Ro)/\Ro^2}{(y-\yo)^2}
v=0,
\label{Fropoleeq}
\end{equation}
where $\Ro=\Up_{\aava{0}}$ and $\Ro'=\Up'_{\aava{0}}$. 
The indicial equation gives:
\begin{equation}
\lambda=\frac{1}{2}\pm\mu.
\end{equation}
In the two next subsections, we examine both cases where $\mu$ is imaginary or real.
\subsubsection{Theoretical stable regime ($R>1/4$)}
\label{stab_pole}
We address here the case where $\Od>1/4$.
The same analysis as in Sect. \ref{Fro} can be carried out to determine how a wave behaves upon crossing the corotation. 
The solutions of the indicial equation can be recast as
\begin{equation}
\lambda=\frac{1}{2}\pm i|\mu|.
\end{equation}
The first-order solutions to Eq. (\ref{Fropoleeq}) in the vicinity of $\yo$ are :
\begin{equation}
\left\{\begin{aligned}
&a_0(y-\yo)^{\frac{1}{2}+i|\mu|}+b_0(y-\yo)^{\frac{1}{2}-i|\mu|},&y>\yo\\
&-i\varsigma\left[a_0(\yo-y)^{\frac{1}{2}+i|\mu|}e^{\varsigma\pi|\mu|}+b_0(\yo-y)^{\frac{1}{2}-i|\mu|}e^{-\varsigma\pi|\mu|}\right],&y<\yo\end{aligned}\right.
\label{solpol}
\end{equation}
for $a_0,\, b_0 \in \mathbb{C}$. One can recover the same form of Frobenius solutions as in \cite{AM2013}, who examined radially stratified mean flows in spherical geometry.
As $\ft=0$, the wave action flux Eq. (\ref{wA}) reduces to
\begin{equation}
\mA=-\frac{\rom}{2k_\perp^2}\Im{v_yv^*},
\label{wApol}
\end{equation}
that is, injecting the solutions on both sides of the critical level:
\begin{equation}
\mathcal{A}=-\frac{\rom}{2k_\perp^2}|\mu|\left\{
\begin{aligned}
&|a_0|^2-|b_0|^2,&y>\yo\\
&-|a_0|^2\e^{2\varsigma\pi|\mu|}+|b_0|^2\e^{-2\varsigma\pi|\mu|},&y<\yo
\end{aligned}\right..
\label{Apol_st}
\end{equation}
This formulation is quite similar to the expression of the Reynolds stress ($\tau$) in vertically stratified mean flows,  which can be found in \cite{BB1967} in Cartesian geometry. We recall indeed that $\tau=-k_x\mA$. Moreover, given Eq. (\ref{Re}), the Reynolds stress in our model reads $\tau=-\rom \overline{uv}$. Using the polarisation relations for $u$, we recover the wave action flux in Eq. (\ref{wApol}).

\begin{figure}[t!]
\includegraphics[width=\hsize]{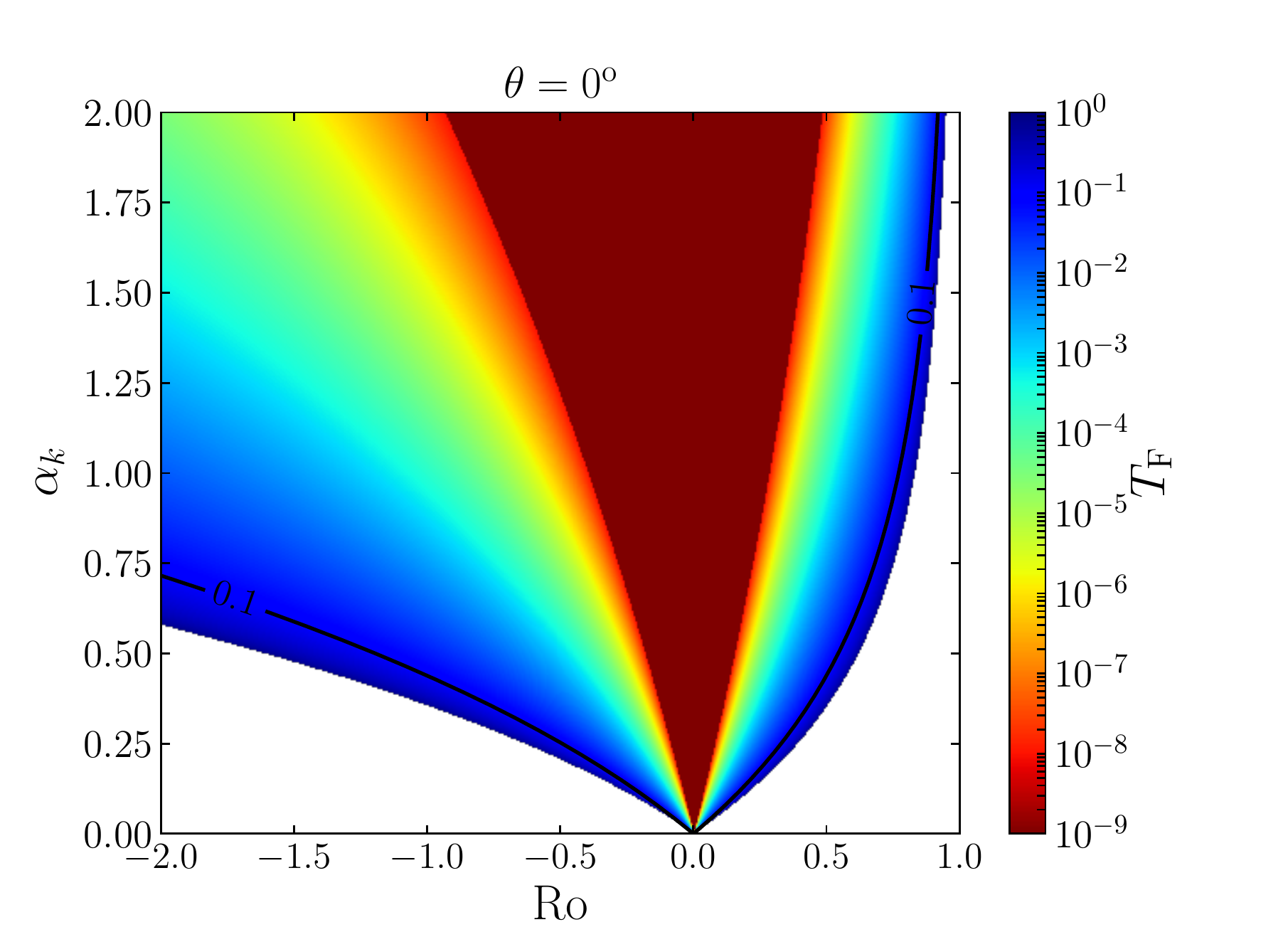}
\caption{Transmission rate $T_\mathrm{F}$ of the wave action flux across the corotation when the box is located at the pole for $\Od>1/4$. It is displayed against the Rossby number $\Ro$ and the ratio of the horizontal wave numbers $\alpha_k=k_z/k_x$. The forbidden region where $\Od<1/4$ is painted in white, and the dark red cone corresponds to values of $T_\mathrm{F}$ lower than $10^{-9}$.}
\label{atnuO2}
\end{figure}
The pre-factor $i$ in the solutions (\ref{solpol}) below the corotation does not affect the energy flow and simply indicates that the wave undergoes a phase shift of $\pi/2$ through the critical level \citep[see also][]{AM2013}.
Above the critical level, the normalised Doppler-shifted frequency satisfies $\sign(\sigma)=-\varsigma$ as for the corotation in the inclined case. The sign is reversed below the critical level. Thus, the first solution of main amplitude $|a_0|$  carries its latitudinal flux of energy upward (downward) for $\varsigma=+1$ ($\varsigma=-1$), while the second solution transfers its energy in the opposite direction in the various cases.
Therefore,  the energy flux of an upward or downward wave is always attenuated by a factor 
\begin{equation}
T_\mathrm{F}=\exp\{-2\pi|\mu|\}.
\end{equation} 
This attenuation factor is shown in Fig. \ref{atnuO2} versus $\Ro$ and $\ak$. We observe that the wave is largely absorbed at the critical level and thus deposits most, if not all its energy for most couples $(\alpha_k,\Ro)$.
\subsubsection{Possible unstable regime ($R<1/4$)}
\label{inst}
We now deal with the case where $\Od<1/4$, i.e. $\mu$ is real.
Contrary to the situation where $\Od>1/4$, we can no longer assimilate solutions to wave-like  functions. The exponential form of solutions for $\Od<1/4$ near the critical level reads
\begin{equation}
v(y)=\left\{\begin{aligned}
&a_0(y-\yo)^{\frac{1}{2}+|\mu|}+b_0(y-\yo)^{\frac{1}{2}-|\mu|}, &y>\yo\\
&-i\varsigma\left[\e^{-i\varsigma\pi|\mu|}a_0(y-\yo)^{\frac{1}{2}+|\mu|}\right.\\
&\left.~~~~~~~+\e^{i\varsigma\pi|\mu|}b_0(y-\yo)^{\frac{1}{2}-|\mu|}\right], &y<\yo
\end{aligned}\right.
\label{solpol_inst}
\end{equation}
and makes this region fully evanescent. Furthermore, the associated wave action flux is 
\begin{equation}
\mathcal{A}=-\frac{\rom}{k_\perp^2}\mu\left\{
\begin{aligned}
&\Im{a_0b_0^*},&y>\yo\\
&-\Im{a_0b_0^*\e^{-2i\varsigma\pi\mu}},&y<\yo
\end{aligned}\right..
\label{Apol_inst}
\end{equation}
\begin{table*}[ht!]
\centering
\begin{tabular}{c|c|| c|c}
 \rule[-1ex]{0pt}{5ex} Box & critical level & attenuation  & amplification \\\hline\hline
 \rule[-1ex]{0pt}{4ex} &  $\sigma=\ft$ 
 & yes for  waves $\downarrow$ & no \\
\rule[-2ex]{0pt}{6ex} \makecell{Inclined box\\$\left(\theta_0\in]0,\pi/2[\right)$} &  $ \sigma=-\ft$ & yes for waves $\uparrow$ & no \\
\rule[-3ex]{0pt}{6ex} & $\sigma=0$ & no & no \\ \hline
\rule[-1ex]{0pt}{7ex} \makecell{Pole \\ $(\theta_0=0)$}& $\sigma=0$ & 
 \begin{tabular}{l}
 yes if $R<1/4$\\
yes if $\left\{\begin{aligned}R>1/4 \\\varsigma\mA^+>0 \end{aligned}\right.$
 \end{tabular}
 & yes if $\left\{\begin{aligned}R>1/4 \\\varsigma\mA^+<0 \end{aligned}\right.$  \\\hline
 \rule[-1ex]{0pt}{4ex} \multirow{3}{*}{\makecell{Equator\\($\theta_0=\pi/2$)}} &   \aava{$\sigma=\pm1$} & \aava{\Large$\smallsetminus$} & \aava{\Large$\smallsetminus$} \\
\rule[-2ex]{0pt}{6ex}  &  $\sigma=0$  & no & no \\ \hline
\end{tabular}
\vspace{0.5cm}
\caption{Summary table of analytical results at each critical level when the box is inclined or not with respect to the rotation axis, in the $[0,\pi/2]$ quadrant and for positive wavenumbers. It indicates whether each critical level can cause attenuation or amplification of the (upward $\uparrow$ and downward $\downarrow$) travelling wave \aava{ in the y-direction, which depends notably, at the pole, on $R=\ak(1-\Ro)/\Ro^2$, and on} the wave action flux above the critical level $\mA^+$. \aava{The symbol $\smallsetminus$ means that no wave action flux is carried across the critical level.}
 Moreover, the results are analogous in the other quadrants of the \aava{spherical body (with the direction of the attenuated wave through $\sigma=\pm\ft$ varying according to $\sign(k_zf)$).}}
\label{recap_ana}
\end{table*}
Without knowing the direction of the wave or the energy flux since $\sign(\wg)=\varsigma\sign\left(\Im{a_0b_0^*}\right)$, it is difficult to assess the impact of the critical level on wave propagation, i.e. whether it will attenuate waves or on the contrary amplify them.  

\cite{LB1985} found a way to investigate the behaviour of internal gravity waves \aava{in the presence of a vertical shear,} passing through a critical level in a regime similar to ours ($\mathrm{Ri}<1/4$) where solutions are fully evanescent. 
Their work\aava{, which was carried out in local Cartesian geometry,} has been taken up by \cite{AM2013} \aava{in global} spherical geometry \aava{applied to the radiative zone of solar-like stars and evolved stars}. The method is to determine the reflection and transmission coefficients in a three-zone model. The evanescent region where the Richardson number \aava{satisfies} $\mathrm{Ri}<1/4$ and where the critical level is located (zone II), is sandwiched between two propagating wave layers (zones I and III). Using a linear mean flow profile so as to establish solutions inside zone II, \cite{LB1985} and \cite{AM2013} both used continuity relations of the perturbed vertical (or radial) velocity and its derivatives at the interfaces between zones in order to get the transmission and reflection coefficients. The critical level is located in the middle of zone II of width $\delta$. By consequence, the reflection and transmission coefficients depend, in their works, on the shear and more precisely the Richardson number, and on the width $\delta$. 
They both found that, depending on the Richardson number and $\delta$, the reflection and transmission coefficients can be greater than one, meaning that the wave can be over-reflected and/or over-transmitted\aava{, and thus extract energy and angular momentum fluxes from the mean flow, which can lead to potential shear instabilities after successive encounters of the wave with the critical layer}. However, this result is conditioned by the geometry of the model. As shown by \cite{L1988} in his review and references therein, models with one or even two layers with evanescent and eventually a wave-like region, do not allow such phenomena. A first region that allows the wave propagation is mandatory and is combined with a ``sink'' that pulls the wave to cross the critical level. According to \cite{LB1985} and \cite{L1988}, the nature of the sink for wave flux can be either another propagative region or an evanescent region, as in zone II, subject to friction processes. Given this peculiar geometry, instabilities can occur under boundary conditions that allow the wave to return successively to the critical level. Many studies have tried to relate over-reflection and shear instability for a specific wave geometry \cite[see in particular the reviews of][for internal gravity waves and Rossby waves]{L1988,HH2007}. 

In the present study, we do not investigate further shear instability by doing, for instance, a temporal analysis to estimate the waves' growth rate \citep[as in][who considered an initial value problem]{LB1985,WA2004}. On the contrary, we give arguments, such as $R<1/4$, of necessary but not sufficient condition to find instabilities. It is important to note that $R$ is constant in the whole domain for a linear mean flow profile, and thus one is stuck with either a propagative (stable) or an evanescent regime. Therefore, finding an adequate geometry to allow over-reflection and over-transmission requires at least that the Rossby number is not the same in the whole domain, by using for instance a non-linear mean flow profile. 
Furthermore, in the particular case where $\Od=0$, (i.e. $\Ro=1$ or $\alpha_{k}=0$ in Eq. (\ref{Pcpole}) when the wave with $k_{z}=0$ propagates in the $(x,y)$-plane), a necessary condition for instability is given by the Inflection Point Theorem \citep{SH2002}. This theorem is particularly used to study barotropic instabilities for Rossby waves \citep[see e.g.][]{LT1978}. In other words, a necessary condition to have unstable modes for $\Ro=1$ is that $\Um''$ cancels out in the domain of wave propagation. \\

We summarise in Table \ref{recap_ana} the main analytical results of Sects. \ref{Fro} and \ref{polar_case} about wave and wave action flux transmission, either when the box is \aava{tilted} relative to the rotation axis, at the North pole\aava{, or at the equator, in the inviscid limit}. \aava{Note that when ``no'' is given in both attenuation and amplification columns, the wave is fully transmitted across the critical level, regardless of the wavenumbers and of the mean flow profile.}

\section{A three-zone numerical model}
\label{secnum}
In order to test the analytical predictions of the previous section, we have built up a three-zone numerical model to simulate waves passing through critical levels. A similar model has been used, for instance, by \citet{J1967} to explore the behaviour of internal gravity waves passing through critical levels in a fluid with rotation and vertical shear. In our model, we solve the two first-order ODEs satisfied by $v$ and $\Pi$, the combination of which led to the \aava{wave propagation} equation~(\ref{Pc}).
By imposing boundary conditions such that waves satisfy the dispersion relations (see also Appendix \ref{solI_III}), we examine the dynamics of inertial waves propagating in the shear region. Also, whenever possible, we calculate analytically the wave transmission and reflection coefficients as the wave-like solution crosses the shear region. 
\begin{figure}[t!]
\centering
\includegraphics[width=0.4\textwidth]{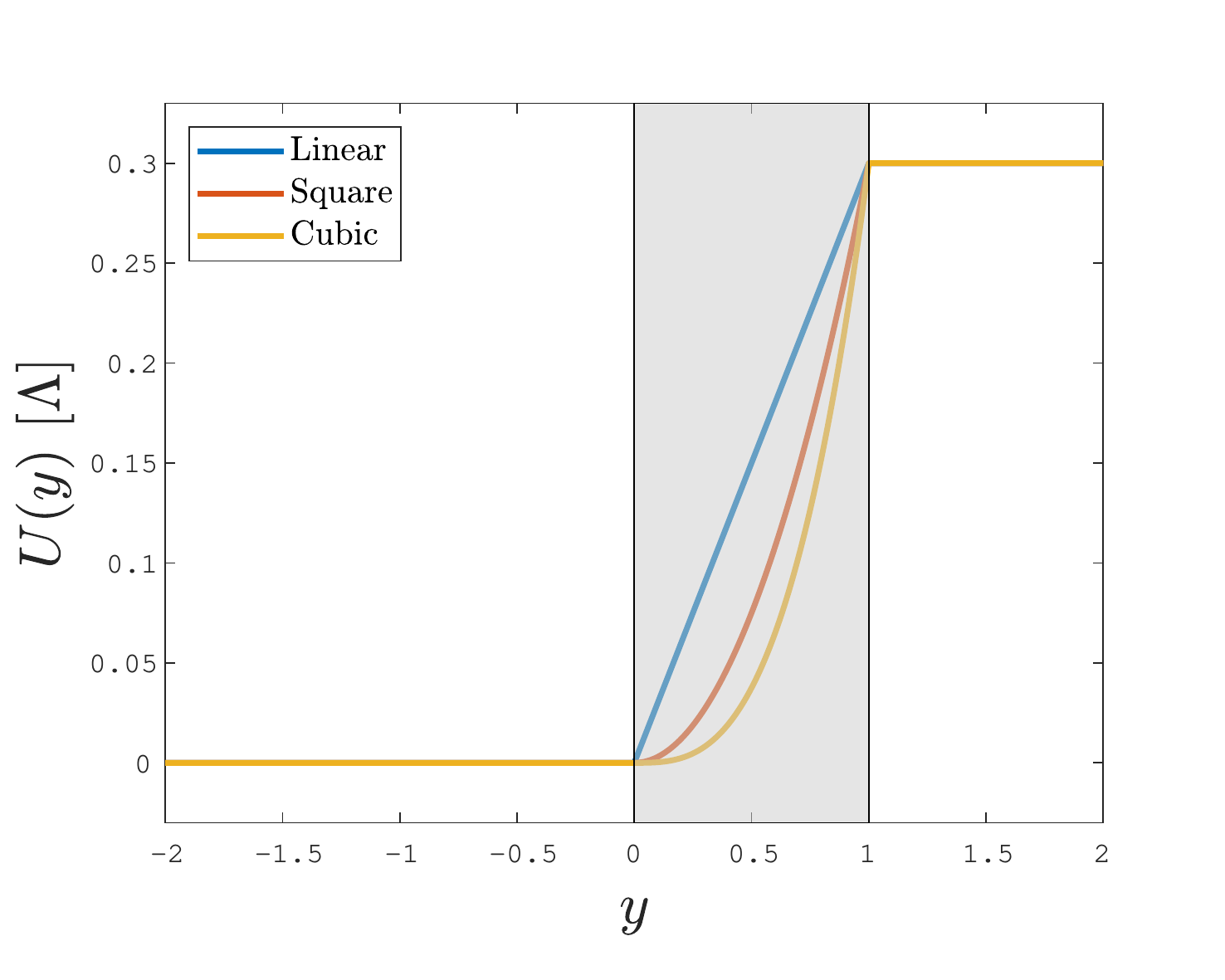}
\caption{Mean flow profiles used in the three-zone numerical model against $y$. In the no-shear regions I ($y<0$) and III ($y > 1$), the mean flow is uniform and set respectively to $\Um=0$ and $\Um=\Lambda$. The sheared region II (\textit{gray-shaded}) can have a linear, square, or cubic mean flow profile.}
\label{mean_flow}
\end{figure}
\subsection{Description of the model}
\label{numMod}
The mean flow profile that is used in the three-zone model is illustrated in Fig. \ref{mean_flow}.
The zone with shear (zone II) is surrounded by two no-shear regions, one with no mean flow (zone I), and one with a uniform mean flow (zone III). In the whole domain, the mean flow profile that we adopt is expressed as
\begin{equation}
U(y)=\left\{\begin{aligned}
&0~~~&\mathrm{for}&~y<0: & \textit{zone I},&\\
&\Lambda y^n~~~&\mathrm{for}&~0\leq y\leq1:&\textit{zone II},&\\
&\Lambda~~~&\mathrm{for}&~y>1:& \textit{zone III}.&\\
\end{aligned}\right.
\label{U_pro}
\end{equation}
where $U(y)$ is continuous at each interface, and $n$ is an integer: $n=1$ for a linear shear flow, $n=2$ for a square shear flow or $n=3$ for a cubic shear flow (see also Fig.~\ref{mean_flow}). In zone I, we assume that there is an incident wave that enters the shear zone as well as a wave that is reflected at the interface between zones I and II or in zone II, i.e.:
\begin{equation}
    v(y)=A_{I}\exp(ik_{I}y)+A_{R}\exp(ik_{R}y),
    \label{IR_waves}
\end{equation}
where $A_{I}$ and $A_{R}$ are the amplitudes, $k_{I}$ and $k_{R}$ are the wavenumbers of the incident and reflected waves, respectively. We further impose as boundary condition in zone III a transmitted wave that propagates towards positive $y$-values:
\begin{equation}
    v(y)=A_{T}\exp(ik_{T}y),
    \label{T_wave}
\end{equation}
where $A_{T}$ and $k_{T}$ are the amplitude and wavenumber of the transmitted wave, respectively.
We impose $A_{T}=1$ without loss of generality and compute the remaining amplitudes $A_{I}$ and $A_{R}$.
More details on the solutions and the dispersion relations of the waves in zones I and III can be found in Appendix \ref{solI_III}. We have ensured that the transmitted wave carries energy upwards, by deriving the wave action flux in zones I and III (see Appendix \ref{wAI_III}). 

We impose the continuity of the latitudinal velocity and reduced pressure at the interfaces (at $y=0$ and $y=1$). By doing so, the wave action flux is continuous at both interfaces. Thus, in the absence of critical points, the wave action flux is conserved in the whole domain, namely $\mA_\mathrm{T}=\mA_\mathrm{I-R}$ where $\mA_{\mathrm{T}}$ is the wave action flux of the transmitted wave and $\mA_{\mathrm{I-R}}=\mA_{\mathrm{I}}-\mA_{\mathrm{R}}$ with $\mA_{\mathrm{I}}$ and $\mA_{\mathrm{R}}$ for the incident and reflected wave action fluxes, respectively.

To solve the ODE in the three zones and in particular near singularities, we have used MATLAB's solver ode15s, which is suitable for solving stiff differential equations \citep[][]{SR1997}. 
To avoid strict singularities at $\sigma^{2}=\tilde{f}^{2}$, we have added a small friction $\sigma_f=\tt{-8}$ in our set of units. Given the boundary conditions, the numerical solver deals with two first-order ODEs for $v$ and $\Pi$, which take the form
\begin{equation}
\left\{
\begin{aligned}
F_{1,v}v+F_{1,p}\Pi=Av',\\
F_{2,v}v+F_{2,p}\Pi=A\Pi',
\end{aligned}\right.
\label{ODEcode}
\end{equation}
where
\begin{equation}
\begin{aligned}
F_{1,v}&=\ft\frac{\Up}{\sigma}\left(k_x\ft+ik_zs\right)-(\Up-f)\left(k_xs+ik_z\ft\right),\\
F_{1,p}&=-ik_\perp^2s,\\
F_{2,v}&=is\left[s^2+f(\Up-f)\right]-i\ft^2\left(s+f\frac{\Up}{\sigma}\right),\\
F_{2,p}&=ik_zf\ft-k_xfs,\\
\end{aligned}
\end{equation}
and where we recall that $s=\sigma+i\sigma_{f}$ is the modified Doppler-shifted frequency due to  Rayleigh friction. While $A_{T}=1$ is imposed by the boundary condition, we compute $A_{I}$ and $A_{R}$ by comparing numerical solutions of the system Eq. (\ref{ODEcode}) at $y=0$ with the definition of velocity in zone I (Eq. (\ref{IR_waves})) and its  associated reduced pressure (see Eq. (\ref{Pivis})).
\subsection{Numerical exploration at the pole for a constant shear}
\subsubsection{Reflection and transmission coefficients}
\begin{figure*}[ht]
\resizebox{\hsize/2}{!}{\includegraphics[trim=0cm 0cm 0cm 1.4cm,clip]{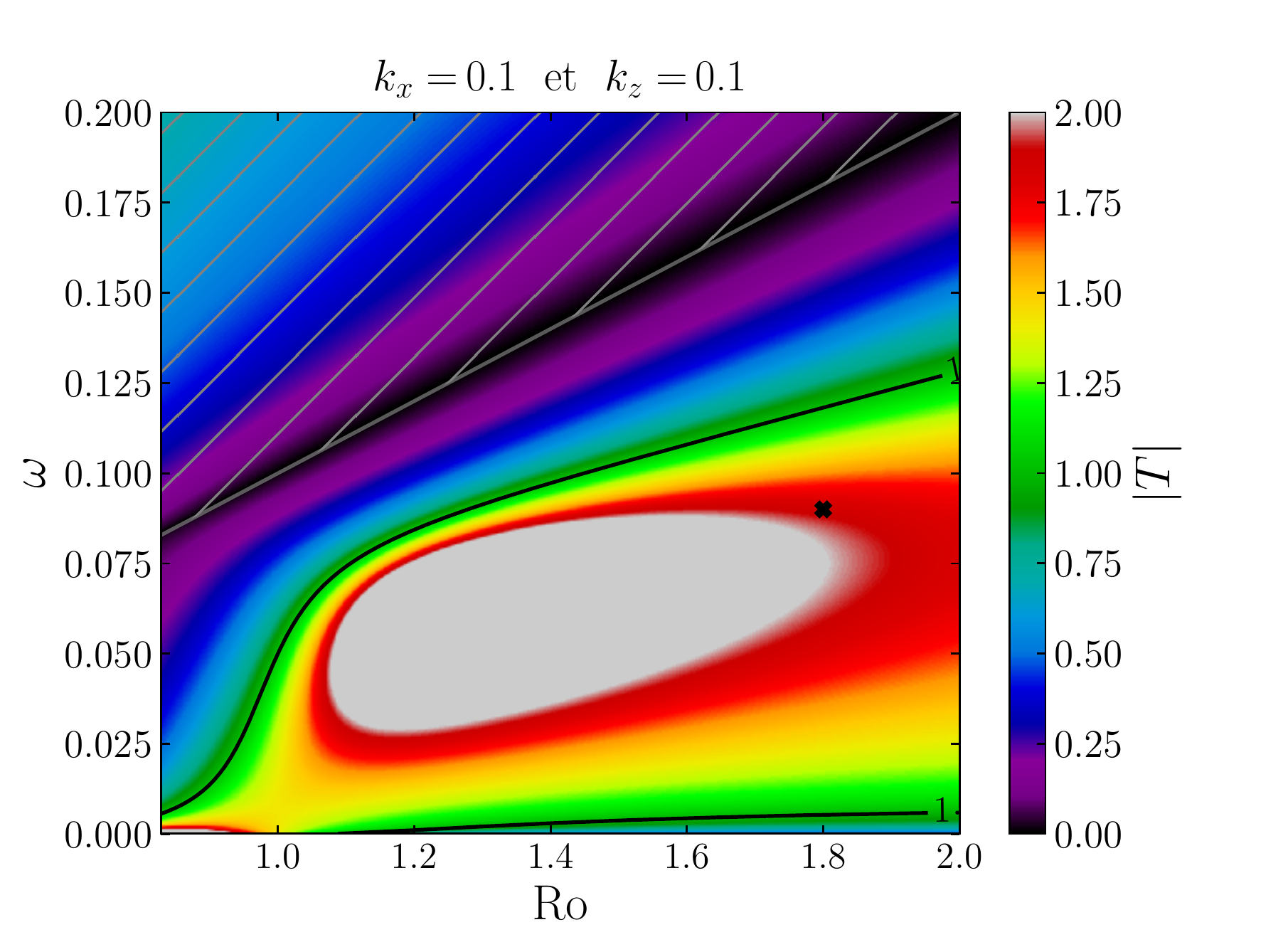}}
\resizebox{\hsize/2}{!}{\includegraphics[trim=0cm 0cm 0cm 1.4cm,clip]{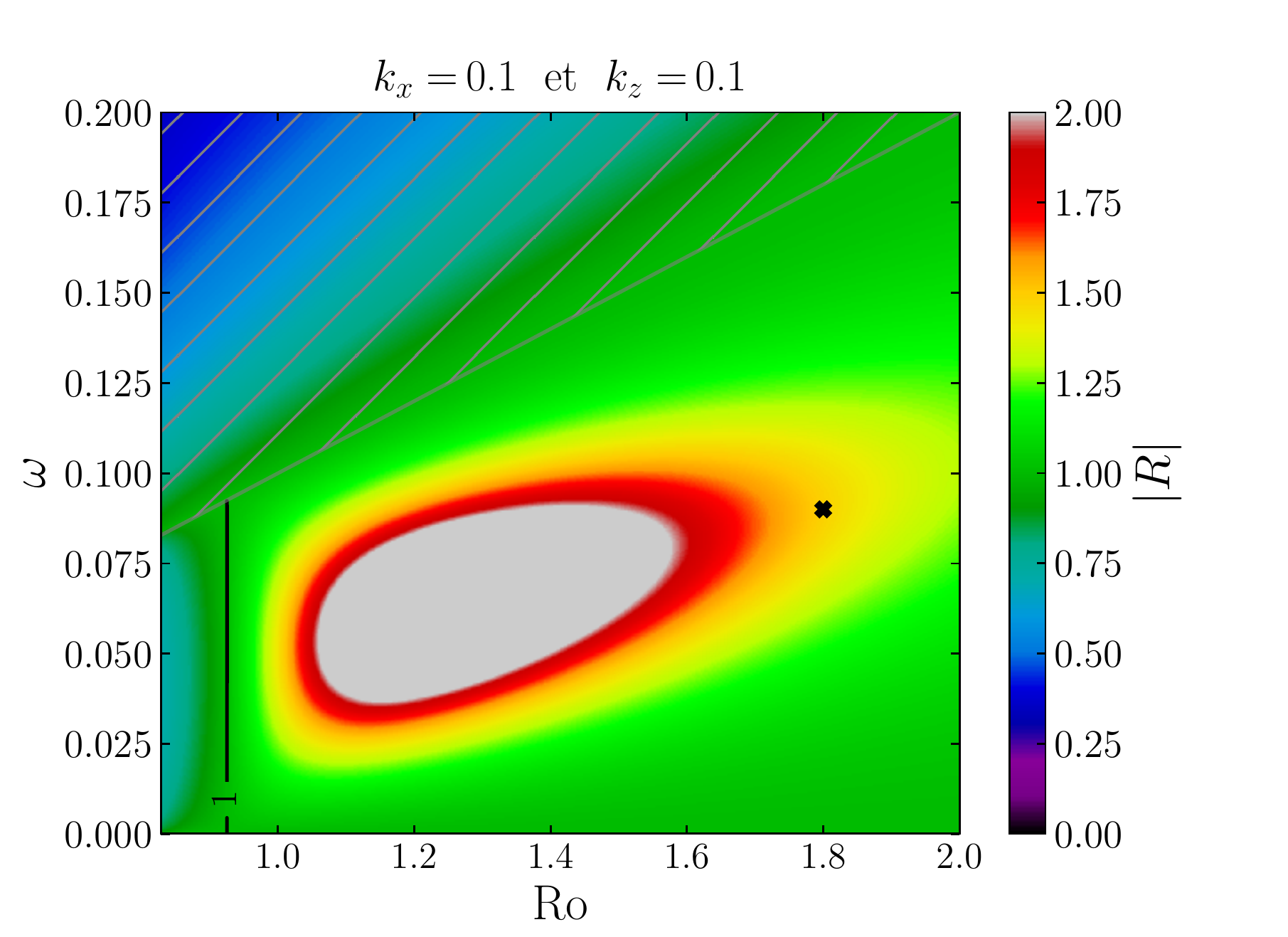}}
\caption{Transmission coefficient ($|T|$, left panel) and reflection coefficient ($|R|$, right panel) when the box is at the pole. They are plotted in the regime $\Od<1/4$ (possibly unstable case, see Sect. \ref{inst}) as a function of the Rossby number $\Ro$ and the inertial frequency $\omega$. The hatched areas do not feature critical points, which correspond to regions where $\omega>k_x\Ro$ in our peculiar geometry (see Appendix \ref{wAI_III} for this particular matter). Vertical and longitudinal wave numbers are fixed:  $k_x=0.1$ and $k_z=0.1$. Moreover, the contours that correspond to the coefficients $|R|$ and $|T|$ equal to one are indicated by solid black lines. 
Crosses mark the set of parameters used in Fig.~\ref{vA_pole} for the analysis of the behaviour of the velocity in the three-layer model.}
\label{RT_inst}
\end{figure*}
\begin{figure*}[ht]
\resizebox{\hsize/2}{!}{\includegraphics[trim=0cm 0cm 0cm 1.4cm,clip]{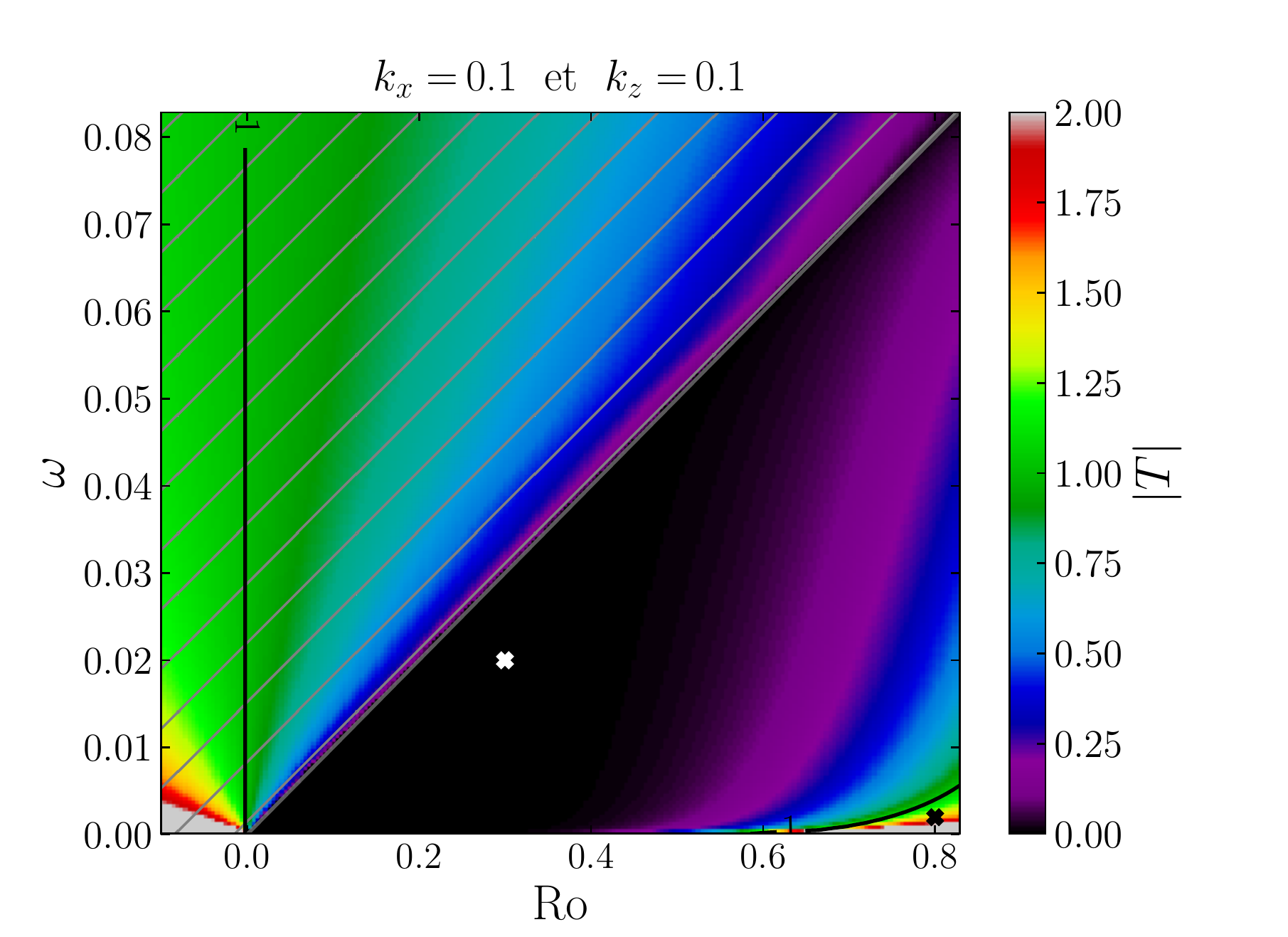}}
\resizebox{\hsize/2}{!}{\includegraphics[trim=0cm 0cm 0cm 1.4cm,clip]{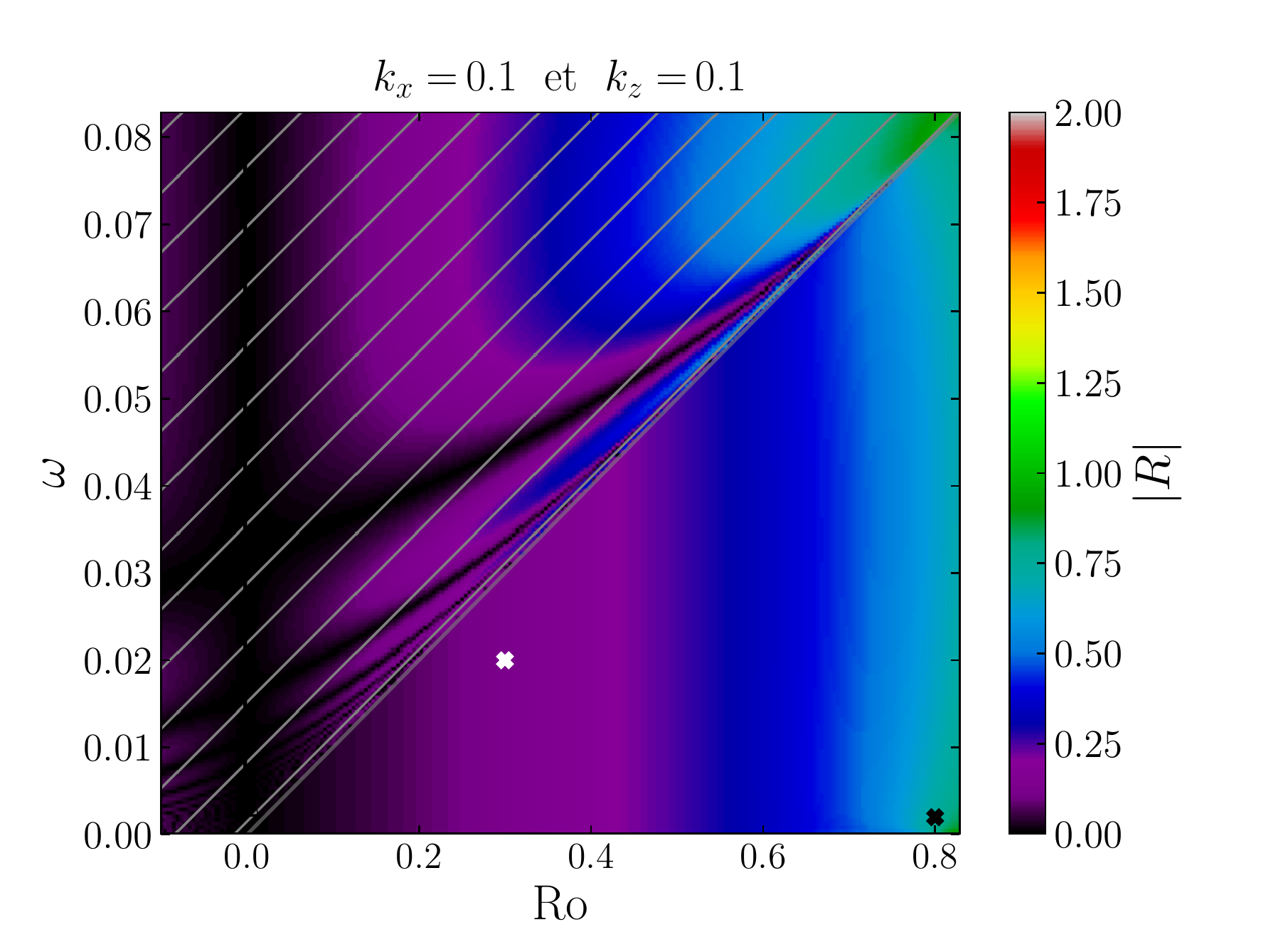}}
\caption{Same as Fig. \ref{RT_inst}, but for $\Od>1/4$. When $\Ro<0$ in these panels, hatched areas are also displayed for all $\omega>0$ (see Appendix \ref{wAI_III}).}
\label{RT_st}
\end{figure*}
In most cases, for any inclination of the box and any mean flow profile, there is no analytical solutions in zone II. 
Nevertheless, we have shown in Sect. \ref{ana_pole} that, when the box is at the pole and for a linear mean shear flow, solutions can be found in terms of Whittaker functions. 
In this section, we will find reflection and transmission coefficients similarly as in \cite{LB1985} and \cite{AM2013}, though there are a few differences. \aava{In particular, our present study differs from the latter by the treatment of inertial waves in convective regions (instead of gravity waves in stably stratified radiative regions in their case) with a latitudinal shear (instead of a vertical/radial shear). Our study, however, uses a local Cartesian model as in \cite{LB1985}}.
Moreover, our boundary conditions are different, as detailed in Sect. \ref{inst}, and the thickness of our shear region is fixed to one in scaled units while \cite{LB1985} and \cite{AM2013} leave the thickness $\delta$ as a control parameter.
We also check the existence of a critical level in the shear zone and the frequency range that delineates the regimes with and without the critical level. 

We consider that the perturbed reduced pressure $\Pi$ and velocity $v$ are continuous at the interfaces $y=0$ and $y=1$. In the presence of the critical level $\yo$ in  zone II, we have a set of four analytical solutions whose values at $y=0$ and $y=1$ allow us to determine the reflection and transmission coefficients. The solutions to \aava{the wave propagation} equation in zones I, II (below and after the critical level) and III are:
\begin{equation}
v=\left\{\begin{aligned}
&A_\mathrm{I}\e^{ik_\mathrm{I}y}+A_\mathrm{R}\e^{ik_\mathrm{R}y},\ &\text{for }y<0\\
&-i\left[A\e^{-i\pi\mu}M_{0,\,\mu\,}(-\tilde y)+B\e^{i\pi\mu}M_{0,\,-\mu\,}(-\tilde y)\right],\ &\text{for }0<y<\yo\\
&AM_{0,\,\mu\,}(\tilde y)+BM_{0,\,-\mu\,}(\tilde y),\ &\text{for }\yo<y<1\\
&A_\mathrm{T}\e^{ik_\mathrm{T}y},\ &\text{ for }y>1
\end{aligned}\right.
\end{equation}
where $A$ and $B$ are complex coefficients which we will express below. We remind $A_T=1$ in our numerical model. In the shear region (zone II), the reduced pressure perturbation is given by Eqs. (\ref{pol}), which, at the pole, can be recast as
\begin{equation}
\Pi=\frac{i}{k_\perp^2}\left[\sigma v_y+k_x(\Ro-1)v \right]
\label{Pipole}, 
\end{equation}
with $\Ro=\Lambda$.
In the regions with no shear (zones I and III), $\Pi$ takes a simpler expression with $\Ro=0$, and $\sigma=\omega$ in zone I and $\sigma=\omega-k_x\Um(1)=\omega-k_x\Ro$ in zone III (noted $\sn$ in the following). Note that while the reduced pressure is kept continuous to conserve the wave action flux across the interfaces, the first derivative of the latitudinal velocity $v'$ is not necessarily continuous at the interfaces.
To find the transmission and reflection coefficients, we solve the system of equations that consist of matching conditions at interfaces as follows:
\begin{my_enumerate}
\item $v~\mathrm{is~continuous~at}~y=0:$\\
$A_\mathrm{I}+A_\mathrm{R}=v_\mathrm{W}(0)$;
\item $\Pi~\mathrm{is~continuous~at}~y=0:$\\
$\left(ik_\mathrm{I}\omega-k_x\right)A_\mathrm{I}+\left(ik_\mathrm{R}\omega-k_x\right)A_\mathrm{R}=\omega v'_\mathrm{W}(0)+k_x(\Ro-1)v_\mathrm{W}(0),$
\item $v~\mathrm{is~continuous~at}~y=1:$\\
$\e^{ik_\mathrm{T}}=v_\mathrm{W}(1),$
\item $\Pi~\mathrm{is~continuous~at}~y=1:$\\
$\left(ik_\mathrm{T}\sn-k_x\right)\e^{ik_\mathrm{T}}=\sn$ $v'_\mathrm{W}(1)+k_x(\Ro-1)v_\mathrm{W}(1),$
\end{my_enumerate}
with the Whittaker functions $v_\mathrm{W}$. At the interfaces below and above the critical level $\yo$ (i.e. $y=0$ and $y=1$), we have
\begin{equation}
\begin{aligned}
v_\mathrm{W}(0)&=-i\left[A\e^{-i\pi\mu}M_{0,\,\mu\,}(2k_\perp\yo)+B\e^{i\pi\mu}M_{0,\,-\mu\,}(2k_\perp\yo)\right],\\
v_\mathrm{W}(1)&=AM_{0,\,\mu\,}\left(2k_\perp(1-\yo)\right)+BM_{0,\,-\mu\,}\left(2k_\perp(1-\yo)\right).
\end{aligned}
\label{whit_01}
\end{equation}
Please note that the first derivative of the Whittaker functions can be computed either numerically or analytically via the relationships in \cite{AS1972}.

The equations in the above continuity relationships 1. to 4. are independent two by two (1. and 2., 3. and 4.), and $A$ and $B$ can be found first:
\begin{equation}
\begin{aligned}
 A = &\e^{ik_\mathrm{T}}\frac{\sn(M'_{-\mu,1}-ik_\mathrm{T}M_{-\mu,1})+M_{-\mu,1}k_x\Ro}{\sn(M_{\mu,1}M'_{-\mu,1}-M'_{\mu,1}M_{-\mu,1})},\\
 B = &\e^{ik_\mathrm{T}}\frac{\sn(M'_{\mu,1}-ik_\mathrm{T}M_{\mu,1})+M_{\mu,1}k_x\Ro}{\sn(M'_{\mu,1}M_{\mu,1}-M_{\mu,1}M'_{-\mu,1})}.
\end{aligned}
\end{equation}
The amplitude of the incident and reflected waves can be written, in terms of $A$ and $B$, as:
\begin{equation}
\begin{aligned}
 A_\mathrm{I}=& \frac{A\e^{-i\mu\pi}X_\mu(k_\mathrm{R})+B\e^{i\mu\pi}X_{-\mu}(k_\mathrm{R})}{\omega(k_\mathrm{R}-k_\mathrm{I})},\\
A_\mathrm{R}=& \frac{A\e^{-i\mu\pi}X_\mu(k_\mathrm{I})+B\e^{i\mu\pi}X_{-\mu}(k_\mathrm{I})}{\omega(k_\mathrm{I}-k_\mathrm{R})},
\end{aligned}
\label{AIR}
\end{equation}
where
\begin{equation}
\begin{aligned}
X_{\mu}(k)=\omega(M'_{\mu,0}-ikM_{\mu,0})+k_x\Ro M_{\mu,0}.\\
\end{aligned}
\end{equation}
The transmission and reflection coefficients are then:
\begin{equation}
|T|=\frac{|A_{\mathrm{T}}|}{|A_{\mathrm{I}}|}=\frac{1}{|A_\mathrm{I}|}\ \text{ and }\ |R|=\frac{|A_\mathrm{R}|}{|A_\mathrm{I}|}.
\end{equation}
We emphasise that these factors depend notably on the location of the critical level and on the inertial frequency, which was not the case in \cite{LB1985} and \cite{AM2013}.

\begin{figure*}[ht!]
\resizebox{\hsize/2}{!}{\includegraphics[trim=0cm 0cm 0cm 1.4cm,clip]{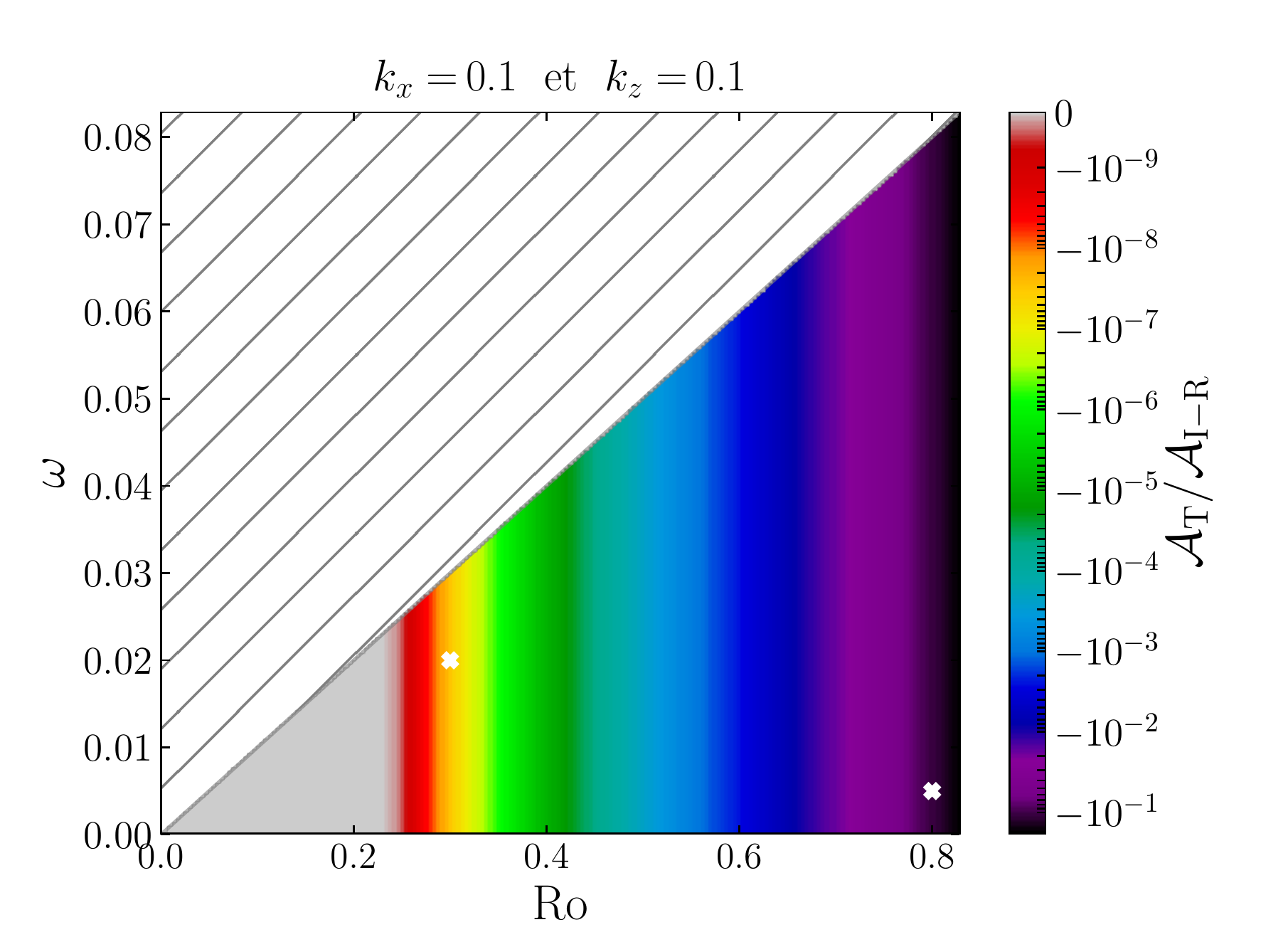}}
\resizebox{\hsize/2}{!}{\includegraphics[trim=0cm 0cm 0cm 1.4cm,clip]{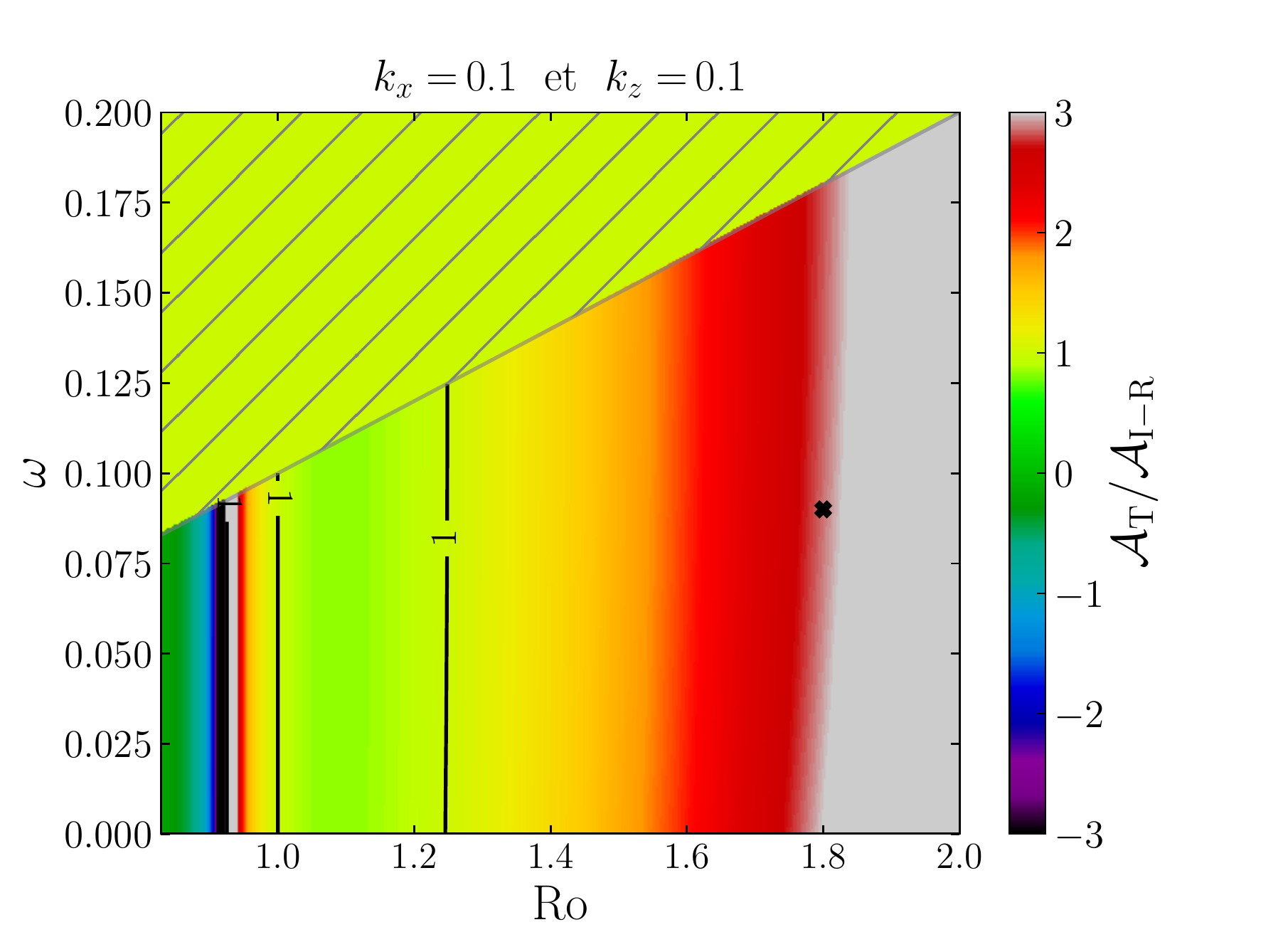}}
\caption{Ratio of the wave action flux above and below the critical level against the Rossby number and the inertial frequency, for $R>1/4$ (\textit{left panel}) and $R<1/4$ (\textit{right panel}), when the box is at the pole. Like in Fig. \ref{RT_st},  $k_x=k_z=0.1$, and hatched zones represent areas without critical levels. Again, crosses indicate the set of parameters chosen to analyse the behaviour of the velocity in the three-layer model.}
\label{wA_IRT}
\end{figure*}
 We display in Figs. \ref{RT_inst} and \ref{RT_st} the transmission and reflection coefficients as a function of the Rossby number $\Ro$ and the normalised inertial frequency $\omega$ in the regimes $R<1/4$ and $R>1/4$, respectively. We choose $k_{x}=k_{z}=0.1$ in these plots and $R=1/4$ gives $\Ro=-2\pm\sqrt{8}\approx \{-4.8,0.8\}$, which delineates the two regimes as we can see in Fig. \ref{O2}. 
 Areas that are hatched do not possess the corotation point $\sigma=0$.
 In addition, the wavenumber $k_\mathrm{T}$ was chosen with a positive sign in these regions to maintain an upward propagating wave (see Appendices \ref{solI_III} and \ref{wAI_III}).
 Areas that are not hatched feature a critical level, according to the table of Appendix \ref{wAI_III}. 
 In the case where $R<1/4$, over-reflection and over-transmission are both possible (see Fig.~\ref{RT_inst}).
 One should notice that for $\Ro>1$, we always have by definition of $R$
\begin{equation}
\frac{R}{(y-\yo)^2}-k_\perp^2<0,
\end{equation}
regardless of $y$, which makes the solutions of Eq. (\ref{pole_cstUp}) tend towards pure exponential functions, i.e. without any imaginary part. Also, we do not see any over-reflection, nor over-transmission in the hatched areas where there is no corotation point. 
This highlights the essential role of the critical level in inducing over-reflection or over-transmission of inertial waves crossing the shear region in this regime.

The regime where $R>1/4$ (in Fig. \ref{RT_st}) is more delicate to analyse. 
According to our discussion in Sect. \ref{stab_pole}, we expect a strong attenuation of the wave and of the wave action flux as shown in Fig. \ref{atnuO2}. 
From this figure and for $\ak=0.1$, the damping is very strong for low positive Rossby numbers. 
This tendency is also found for both transmission and reflection coefficients.
Nevertheless, one can also observe an unexpected regime 
of over-transmission near $\Ro=0.8$ and low frequency $\omega$. 
Still, we must not forget that solutions in this regime, even near the critical level (see Eq.~(\ref{solpol})), are not rigorously equivalent to wave-like functions. 
In particular, the amplification term $(y-\yo)^{1/2}$ that can be found at the first-order in the Frobenius solutions becomes more prominent as the thickness of the shear zone is larger. 
This is especially true for the transmission coefficient. Assuming that Eq.~(\ref{solpol}) holds throughout zone II and corresponds to upward and downward waves, the transmission coefficient is modulated by $|1-\yo|^{1/2}/|0-\yo|^{1/2}$, the amplitude ratio between the transmitted and incident waves.  
This term can be greater than one in the shear region. 
In particular, it is always greater than one when $\yo<0$, i.e. no critical level in the regime $R>1/4$ (hatched areas in Fig. \ref{RT_st}).
On the contrary, this ratio is not present for the reflection coefficient since $|R|$ is function of the incident and reflected waves evaluated at $y=0$.
Though this hand-waving explanation does not formally demonstrate the origin of this amplification, it stresses the important role of the shear-region thickness and more generally of the geometry of the model.

In order to clarify whether the amplification is due to the geometry or the critical level, we need to investigate how the wave action flux changes before and after the critical level. The wave action flux is indeed the relevant quantity to investigate energy flux exchanges at a critical level.
\subsubsection{Wave action fluxes below and above the shear region}
Since $v$ and $\Pi$ are continuous at the interfaces between the shear and no-shear regions, the wave action flux is preserved and continuous in all three zones in the absence of \aava{friction} and critical levels. However, it is discontinuous at the corotation point as demonstrated in Sects. \ref{stab_pole} and \ref{inst}. Given the amplitude of the incident and reflected waves (Eq. (\ref{AIR})\aava{)}, we can calculate the ratio of the wave action flux below and after the corotation (see Appendix D.2 for the detailed calculation):
\begin{equation}
\frac{\mA_\mathrm{T}}{\mA_\mathrm{I-R}}=\pm\frac{\omega^2\sqrt{\sn^2\left[k_z^2f^2+k_\perp^2\left(\ft^2-\sn^{2}\right)\right]}}{\sn^2\sqrt{\omega^2\left[k_z^2f^2+k_\perp^2\left(\ft^2-\omega^{2}\right)\right]}\left(\left|A_\mathrm{I}\right|^{2}-\left|A_\mathrm{R}\right|^{2}\right)}.
\label{rwA}
\end{equation}
The signs $+$ or $-$ can be chosen in regards to the wave action flux of the transmitted wave that can be positive or negative depending on the presence of the critical level, while the energy flux is always positive in order to have an upward propagating wave in zone III (see Appendix \ref{wAI_III} for a more detailed discussion).
This wave action flux ratio is displayed in Fig. \ref{wA_IRT} in the two regimes $R\lessgtr1/4$. As expected, this ratio is equal to one when no critical level is present (hatched areas).
Unlike in the previous section, the regime where $R>1/4$ has no longer amplification areas, $|\mA_\mathrm{T}/\mA_\mathrm{I-R}|<1$ everywhere. This supports the idea that the critical level has nothing to do with the amplification phenomenon observed in the left panel of Fig. \ref{RT_st}. As already observed in Fig. \ref{atnuO2}, the damping due to the critical level is strong except for Rossby numbers close to the threshold between the two regimes.
Moreover, $\mA_\mathrm{T}/\mA_\mathrm{I-R}<0$ means that $\left|A_\mathrm{I}\right|^{2}>\left|A_\mathrm{R}\right|^{2}$ since the minus sign is taken in Eq. (\ref{rwA}). Therefore, no over-reflection due to the critical level is expected in this regime.
The other regime ($R<1/4$, right panel) features areas where the wave is over-reflected for $\left|A_\mathrm{I}\right|^{2}<\left|A_\mathrm{R}\right|^{2}$ (i.e. when $\mA_\mathrm{T}/\mA_\mathrm{I-R}>0$) and areas where the wave is over-transmitted for  $|\mA_\mathrm{T}/\mA_\mathrm{I-R}|>1$. For the first inequality ($\left|A_\mathrm{I}\right|^{2}<\left|A_\mathrm{R}\right|^{2}$), the threshold between under and over reflection (around $\Ro\approx0.9$) is the same than for the reflection factor (in the right panel of Fig. \ref{RT_inst}). For the second one ($|\mA_\mathrm{T}/\mA_\mathrm{I-R}|>1$), the comparison with the transmission factor (in the left panel of Fig. \ref{RT_inst}) is more questionable. 
Still, these two points suggest that the critical level can induce the over-reflection and over-transmission phenomena in the regime where $R<1/4$. 
\subsubsection{Numerical solutions}
\begin{figure*}[ht]
\resizebox{\hsize/3}{!}{\includegraphics[trim=0cm 0cm 0cm 1cm,clip]{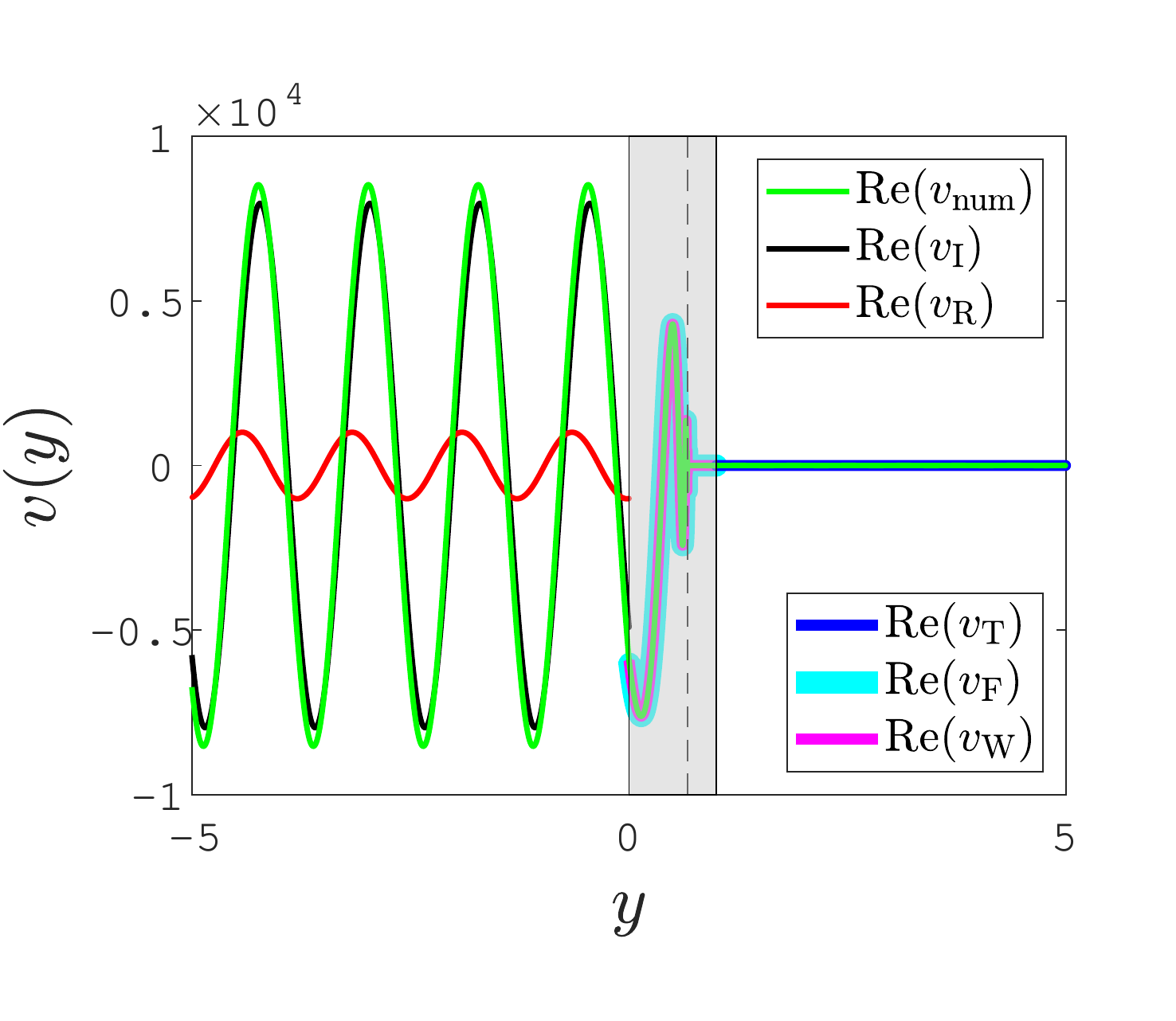}}
\resizebox{\hsize/3}{!}{\includegraphics[trim=0cm 0cm 0cm 1cm,clip]{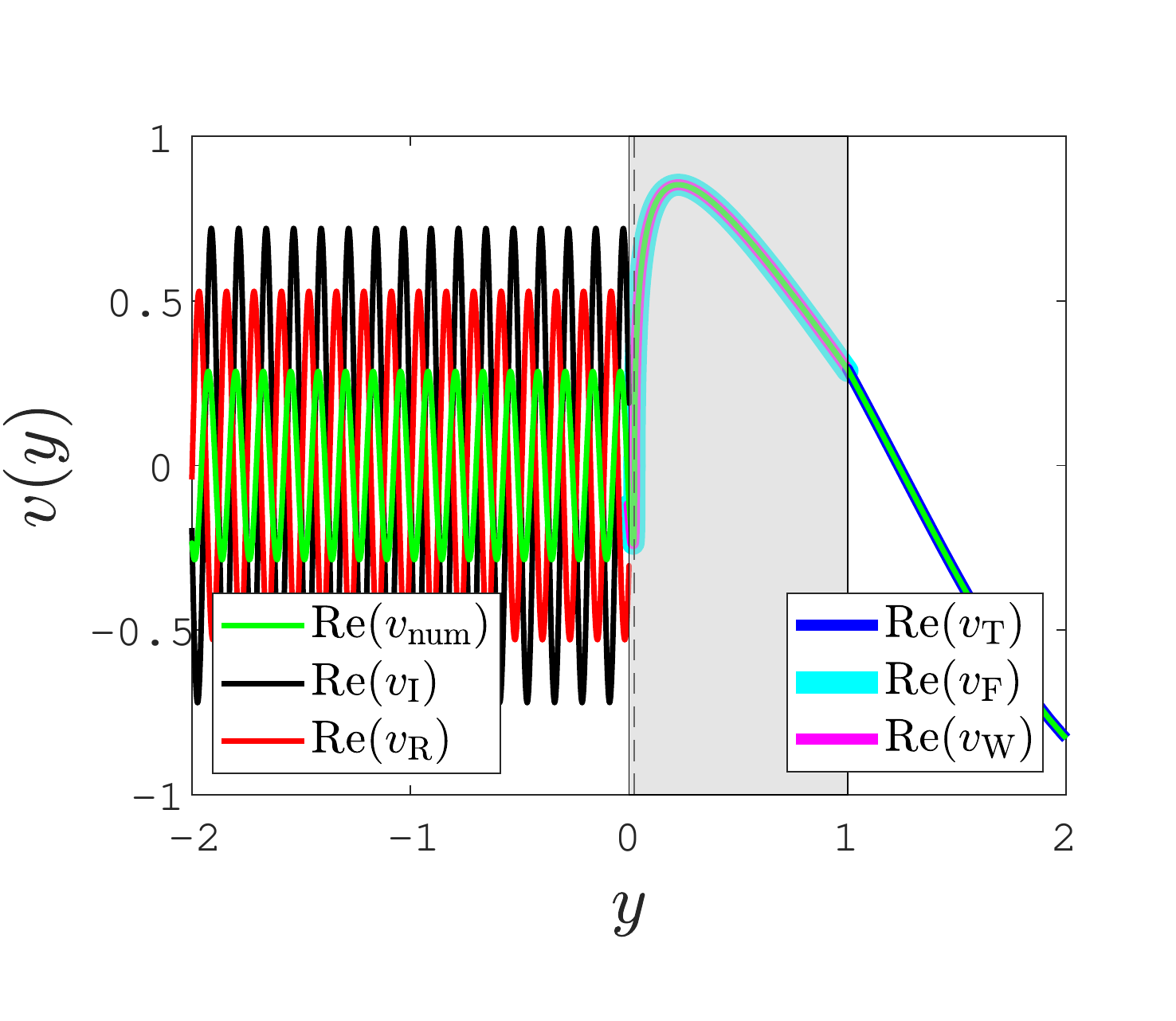}}
\resizebox{\hsize/3}{!}{\includegraphics[trim=0cm 0cm 0cm 1cm,clip]{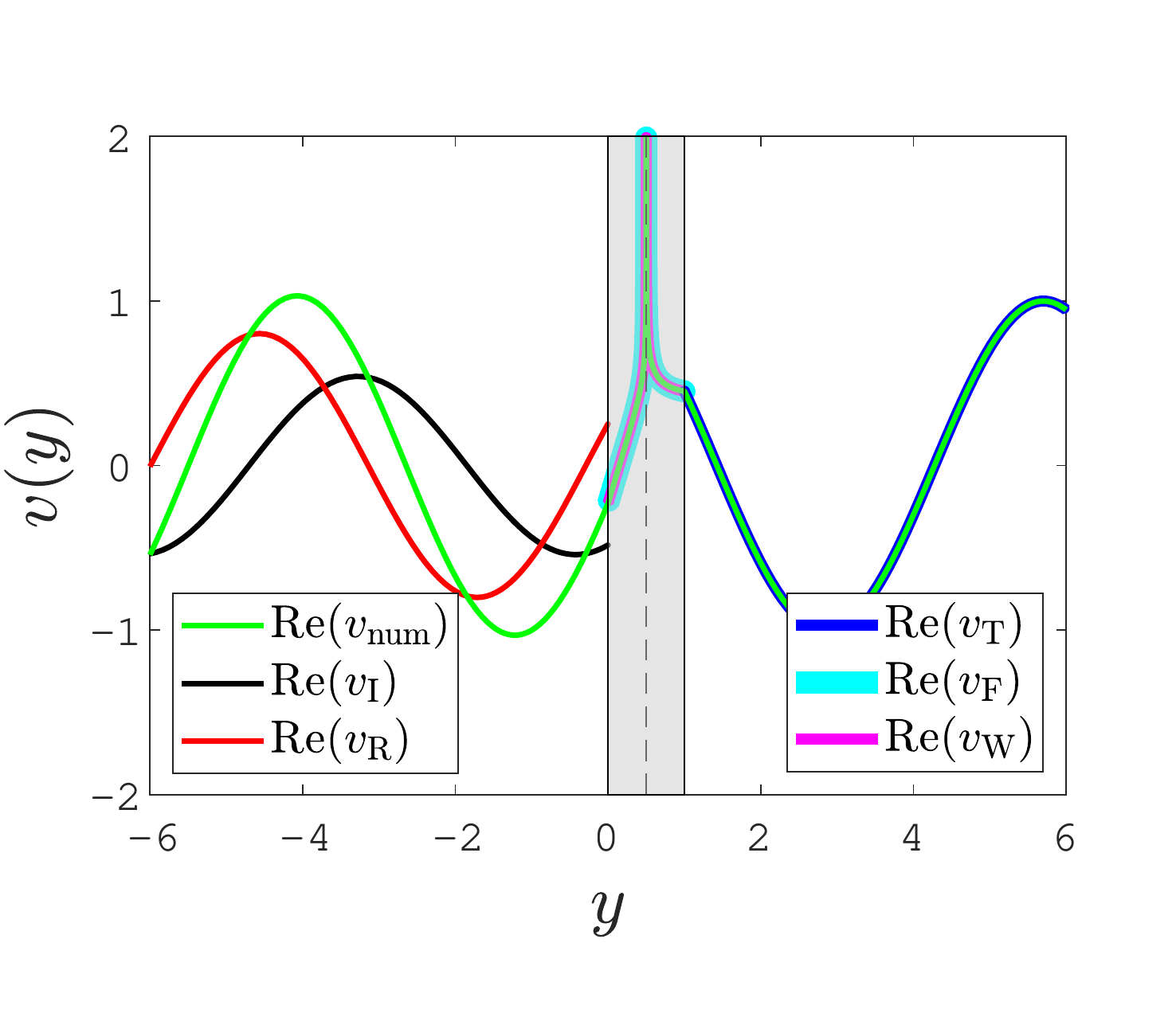}}
\resizebox{\hsize/3}{!}{\includegraphics[trim=0cm 0cm 0cm 1cm,clip]{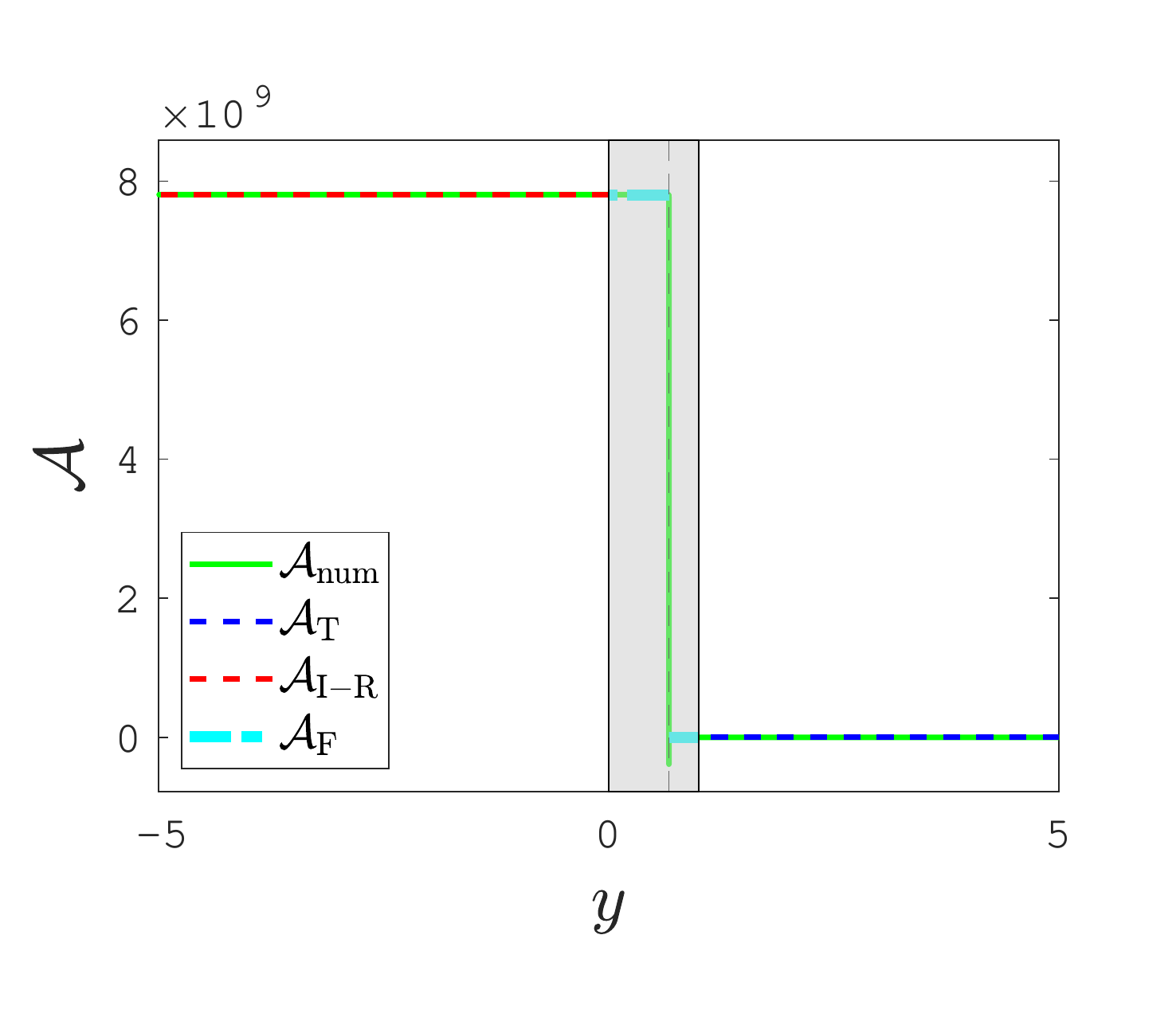}}
\resizebox{\hsize/3}{!}{\includegraphics[trim=0cm 0cm 0cm 1cm,clip]{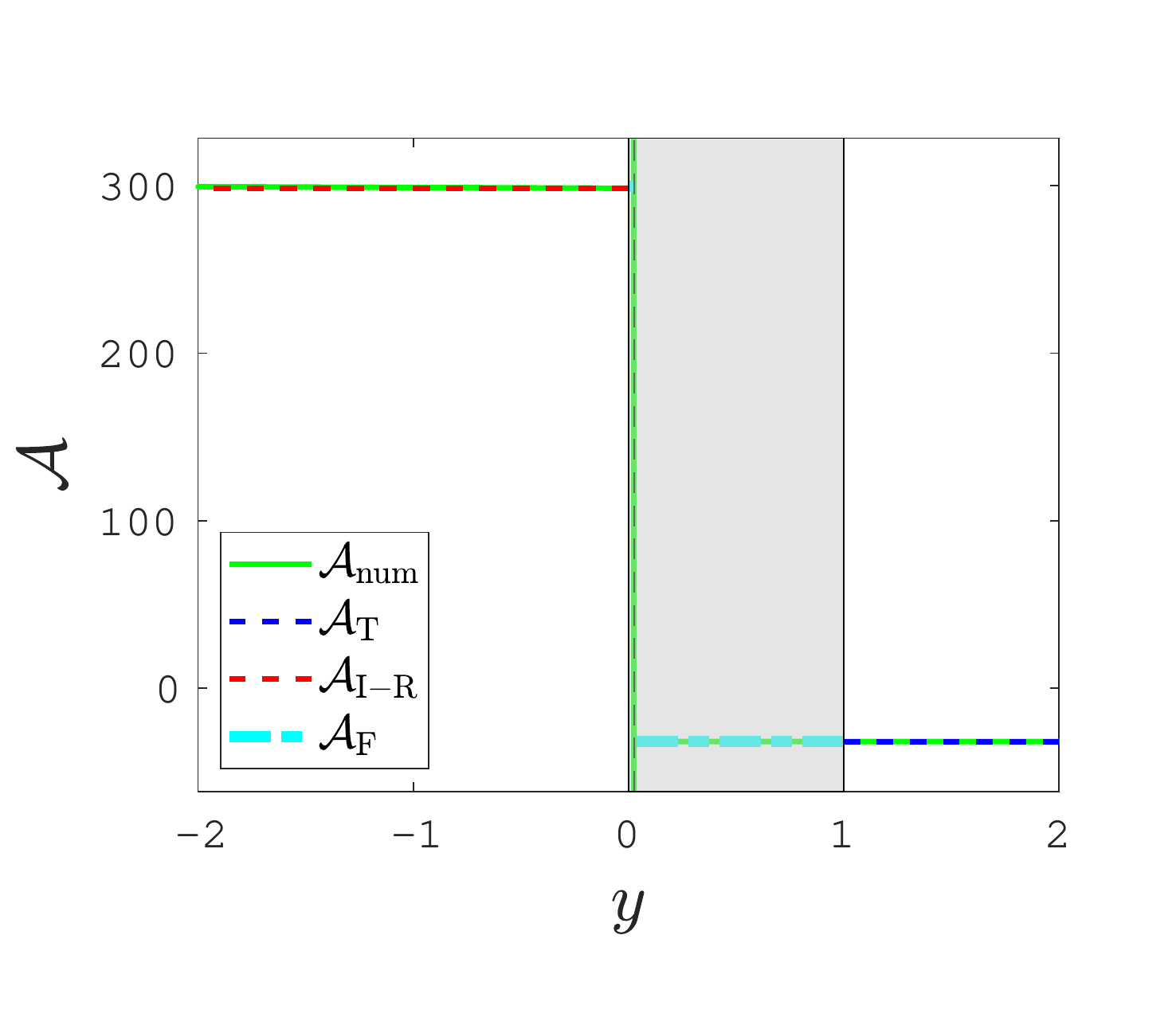}}
\resizebox{\hsize/3}{!}{\includegraphics[trim=0cm 0cm 0cm 1cm,clip]{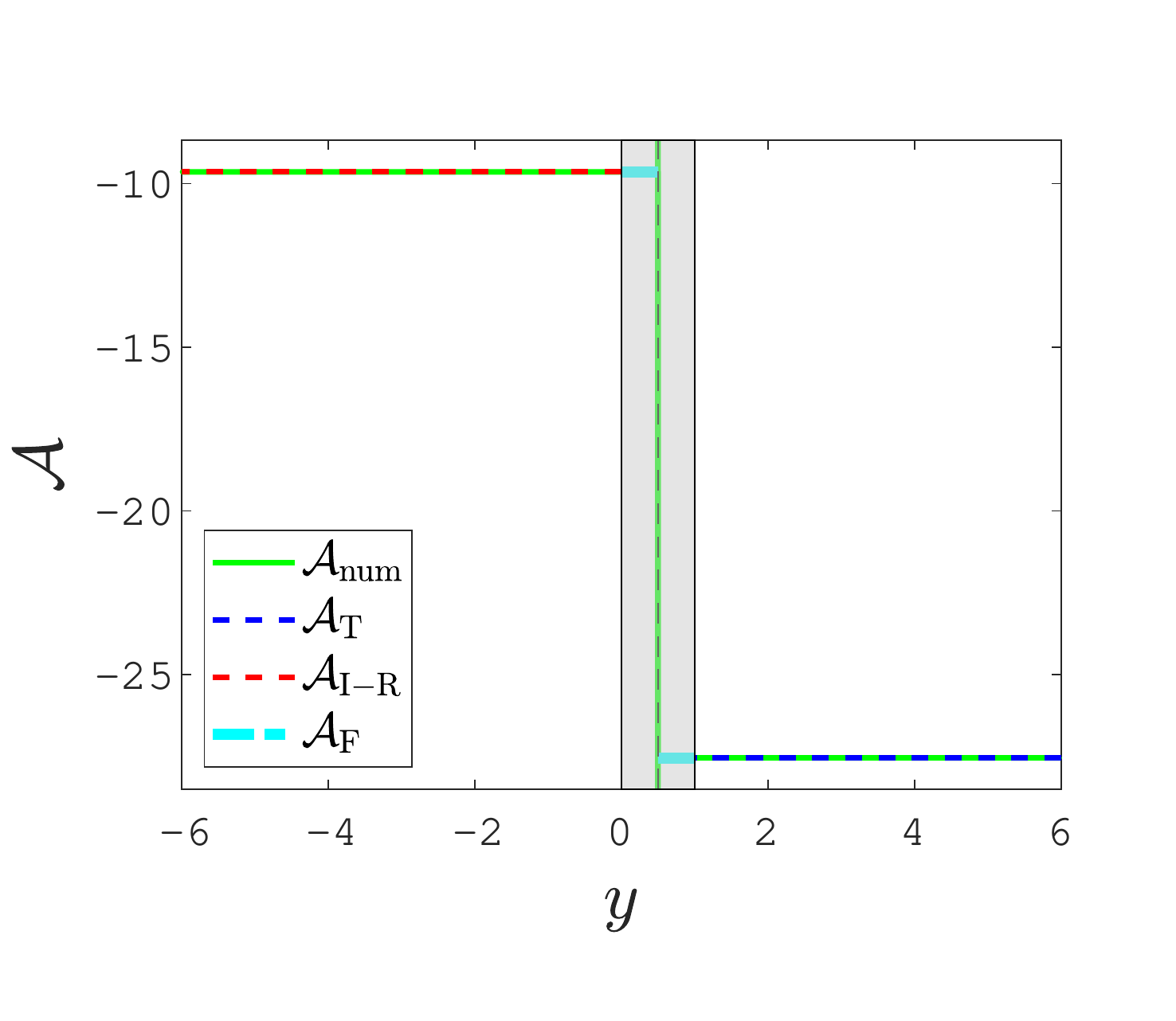}}
\caption{\aava{\textit{Top:}} Real part of the latitudinal velocity $v$ against $y$. 
The quantities $v_\mathrm{num}$, $v_\mathrm{I}$, $v_\mathrm{R}$, $v_\mathrm{T}$, $v_\mathrm{F}$ and $v_\mathrm{W}$ are the numerical, incident, reflected, transmitted,  first-order Frobenius and Whittaker velocities, respectively.
\aava{\textit{Bottom:}} Wave action flux against $y$. The quantities $\mA_\mathrm{num}$,  $\mA_\mathrm{T}$, $\mA_\mathrm{I-R}$, $\mA_\mathrm{F}$ are the numerical, transmitted, incident and reflected, Frobenius wave action fluxes, respectively. 
For all panels, $\theta_0=0$, the mean flow is linear in the grey-shaded shear regions, and the critical level is marked by dashed lines. The horizontal wave numbers are set to $k_x=k_z=0.1$. 
\textit{From left to right:} (i) $\omega=0.02$ and $\aava{\Ro}=0.3$ ($R>1/4$), (ii) $\omega=0.002$ and $\aava{\Ro}=0.8$ ($R>1/4$), and (iii) $\omega=0.09$ and $\aava{\Ro}=1.8$ ($R<1/4$).
}
\label{vA_pole}
\end{figure*}
In the previous sections, we have examined how the shear parameter and the inertial wave frequency impact the reflection and transmission coefficients as well as the wave action flux. We now study particular cases of wave propagation through the critical level for fixed sets of parameters in both regimes $R\lessgtr1/4$. 
To do that, we have numerically solved Eqs. (\ref{ODEcode}) for the three-layer model described in Sect. \ref{numMod}, for $\theta_0=0$ and a linear shear flow ($n=1$ in zone II). We have selected  three pairs of values for the inertial frequency and the shear, two in the regime $R>1/4$ and one in the regime $R<1/4$. These values are marked by crosses in Figs. \ref{RT_inst}, \ref{RT_st} and \ref{wA_IRT}. In each case, the latitudinal velocity and the wave action flux have been calculated through the three zones successively and plotted in Fig. \ref{vA_pole}. The numerical solution, which has been computed by imposing the boundary condition $A_{\mathrm{T}}=1$ and the continuous interfacial conditions for $v$ and $\Pi$ at $y=\{0,1\}$, is the sum of incident and reflected waves in zone I, and equal to a transmitted wave in zone III. The expressions for the incident, reflected and transmitted waves are given by Eqs. (\ref{IR_waves}) and (\ref{T_wave}) (see also Appendix \ref{solI_III}). In the shear region (zone II) of the upper panel of Fig. \ref{vA_pole}, the Whittaker solution has been added and it matches perfectly with the numerical solution below and above the critical level in each case. 
Moreover, Frobenius approximations for the latitudinal velocity (Eqs. (\ref{solpol}) when $R>1/4$, and  (\ref{solpol_inst}) when $R<1/4$) and for the wave action flux (Eqs. (\ref{Apol_st}) when $R>1/4$, (\ref{Apol_inst}) otherwise) have also been included. The coefficients $a_0$ and $b_0$ have been determined by matching the numerical solution and its derivative to the Frobenius approximation for the velocity close to the critical level. For both the latitudinal velocity and the wave action flux, this first-order approximation  gives satisfactory agreement with numerical solutions, although a slight deviation (regarding velocity) from the numerical solution can be observed as one moves away from the critical level. In particular, it should be mentioned that for the far left panels, $a_0(y-yc)^{1/2+i|\mu|}$ corresponding to an upward wave is sufficient to fit correctly the numerical solution, meaning that the counter-propagating wave in zone I is reflected at $y=0$.
However, for the middle and right-hand side panels, it is not clear whether the first-order Frobenius solutions can be reconnected to the incident and reflected waves at $y=0$. 

We now examine attenuation or amplification phenomena in each column of panels in Fig. \ref{vA_pole}, from left to right. In the left panels, for which $R>1/4$, the latitudinal velocity is strongly attenuated at the critical levels, and so is the wave flux action as we can expect from the left panel of Fig. \ref{wA_IRT}. While the transmitted wave is totally absorbed, the reflected wave remains, which is consistent with analytical values of the transmission and reflection coefficients in Fig. \ref{RT_st} (see white crosses). In the middle panels, where we also have $R>1/4$, the wave is over-transmitted but not over-reflected, which is also consistent with Fig. \ref{RT_st} (see black crosses). 
In view of the wave action flux, the amplification of the transmitted wave does not seem to be related to the critical level because this quantity is greatly reduced after the critical level (see also the white cross in the left panel of Fig. \ref{wA_IRT}).
The third column of panels now refers to the regime where $R<1/4$. The wave is over-reflected by a factor of $\sim1.5$ and over-transmitted by a factor of $\sim2$ in concordance with the reflection and transmission coefficients plotted in Fig. \ref{RT_inst}. The wave action flux is negative and $|\mA_\mathrm{I-R}|<|\mA_\mathrm{T}|$ by a factor of $3$ as observed in Fig. \ref{wA_IRT}.
These three case studies reinforce the idea of \cite{BB1967} in the case of stratified z-sheared flows that the wave energy can be lost to the mean flow, or on the contrary that the wave can take energy from the mean flow.

\begin{figure*}[ht]
\resizebox{\hsize/3}{!}{\includegraphics[trim=0cm 0cm 0cm 1cm,clip]{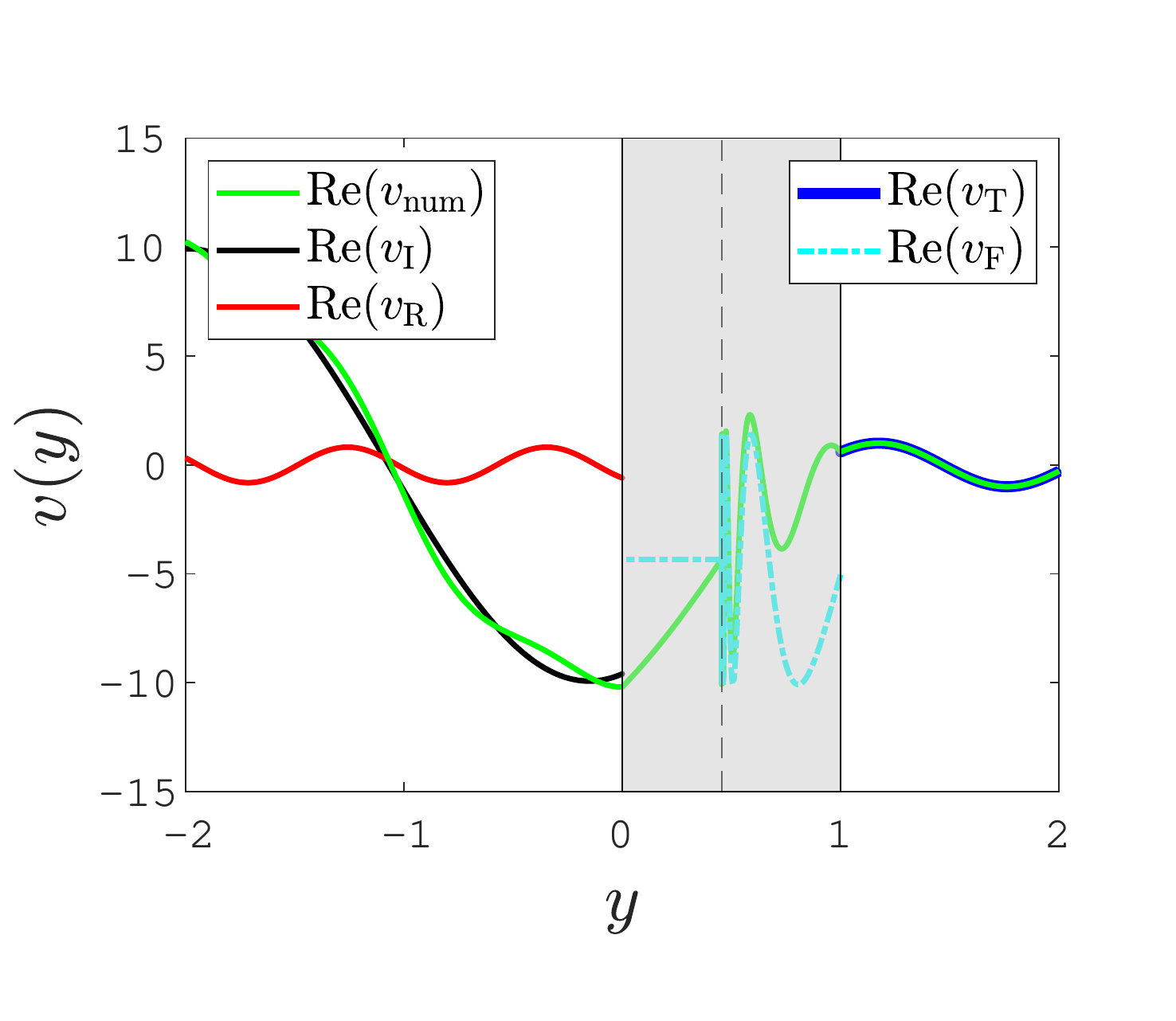}}
\resizebox{\hsize/3}{!}{\includegraphics[trim=0cm 0cm 0cm 1cm,clip]{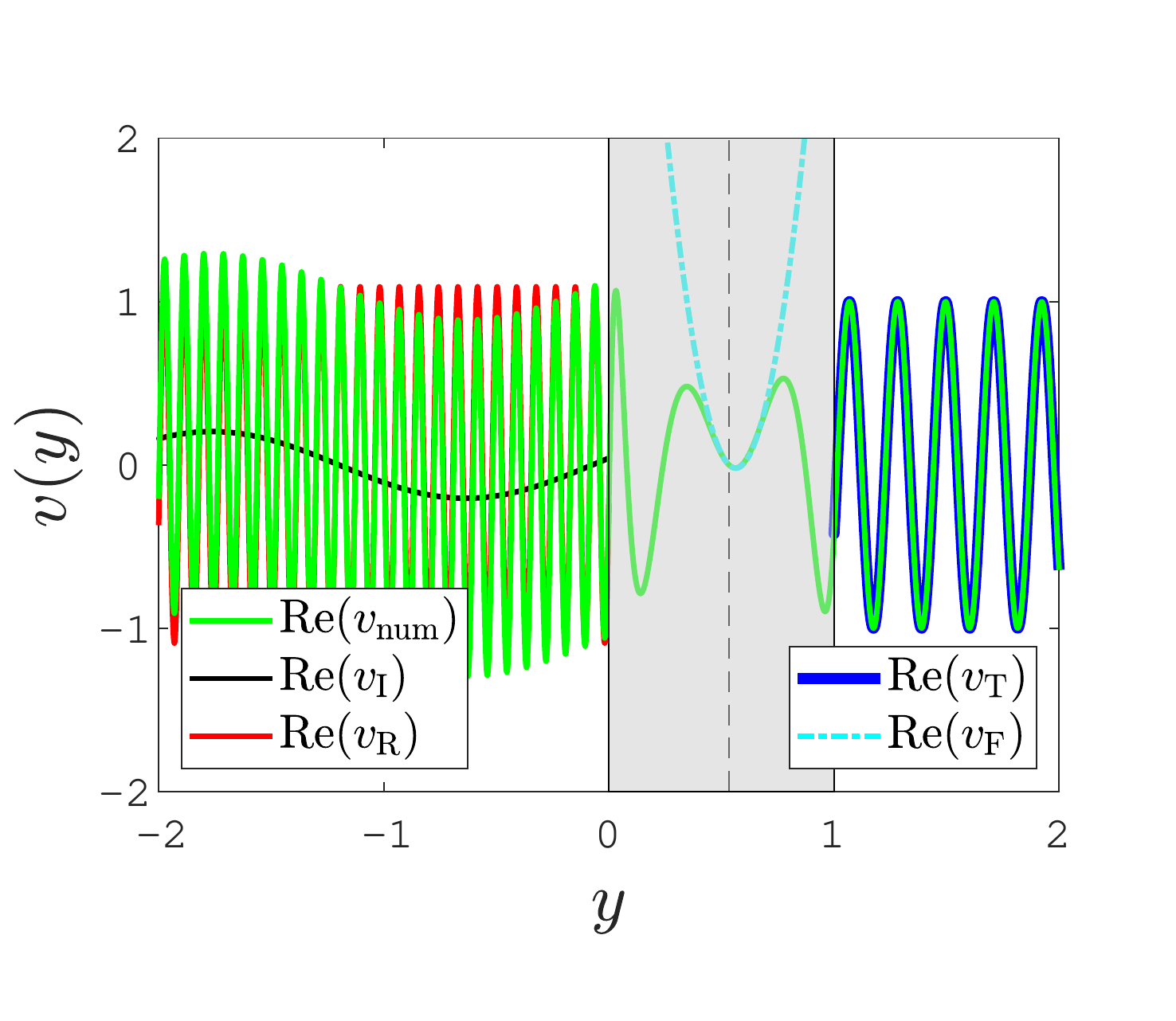}}
\resizebox{\hsize/3}{!}{\includegraphics[trim=0cm 0cm 0cm 1cm,clip]{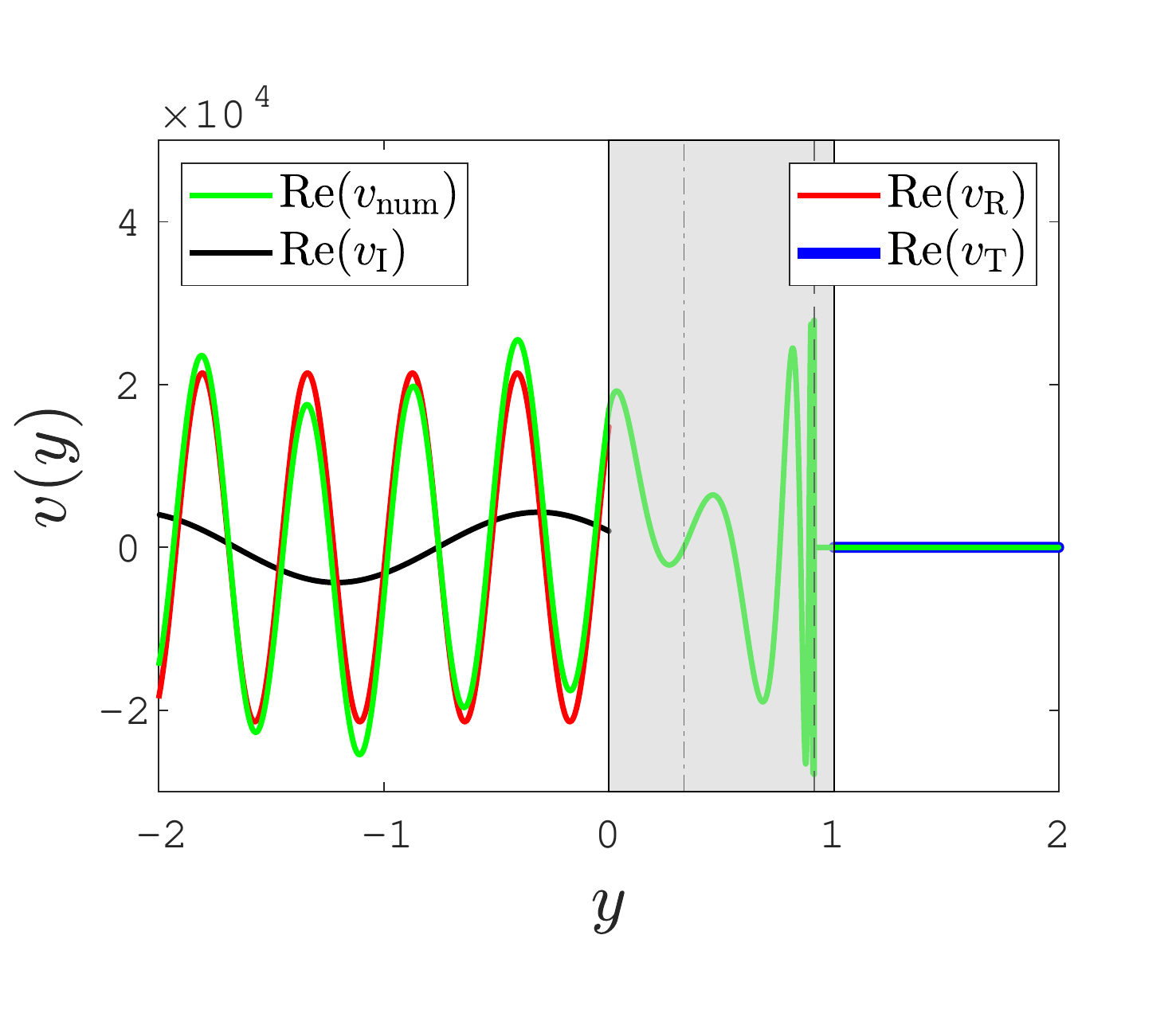}}
\resizebox{\hsize/3}{!}{\includegraphics[trim=0cm 0cm 0cm 1cm,clip]{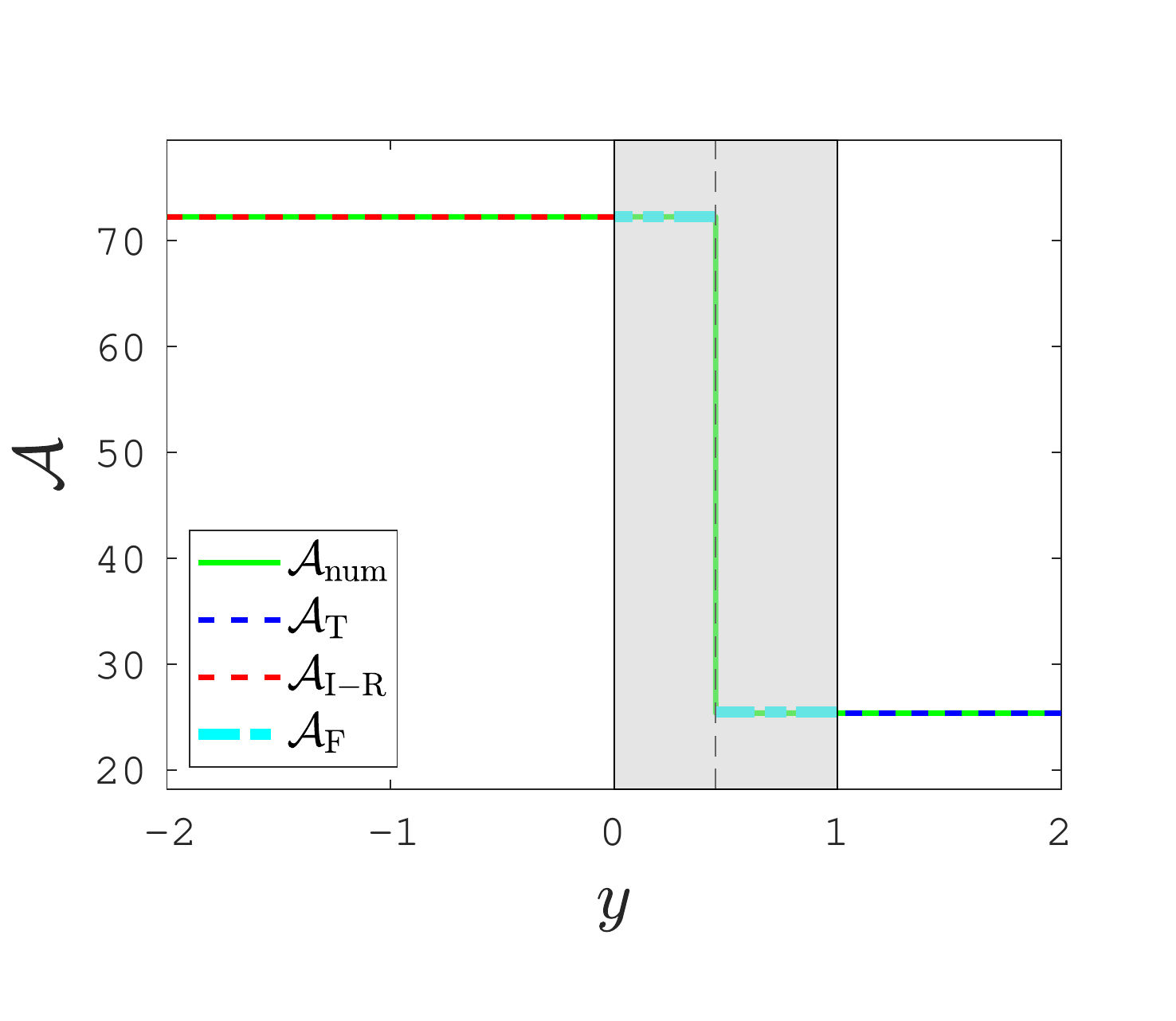}}
\resizebox{\hsize/3}{!}{\includegraphics[trim=0cm 0cm 0cm 1cm,clip]{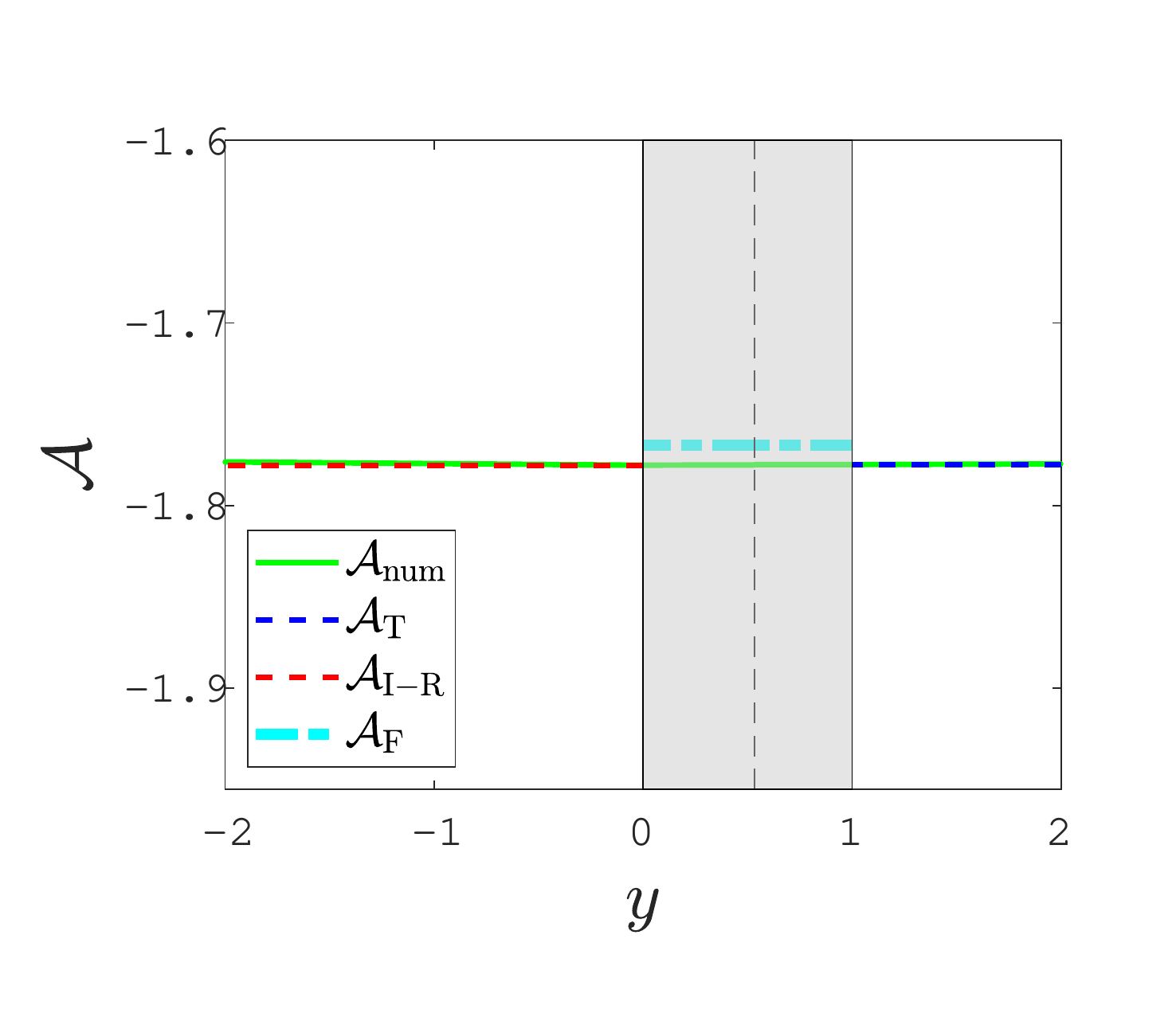}}
\resizebox{\hsize/3}{!}{\includegraphics[trim=0cm 0cm 0cm 1cm,clip]{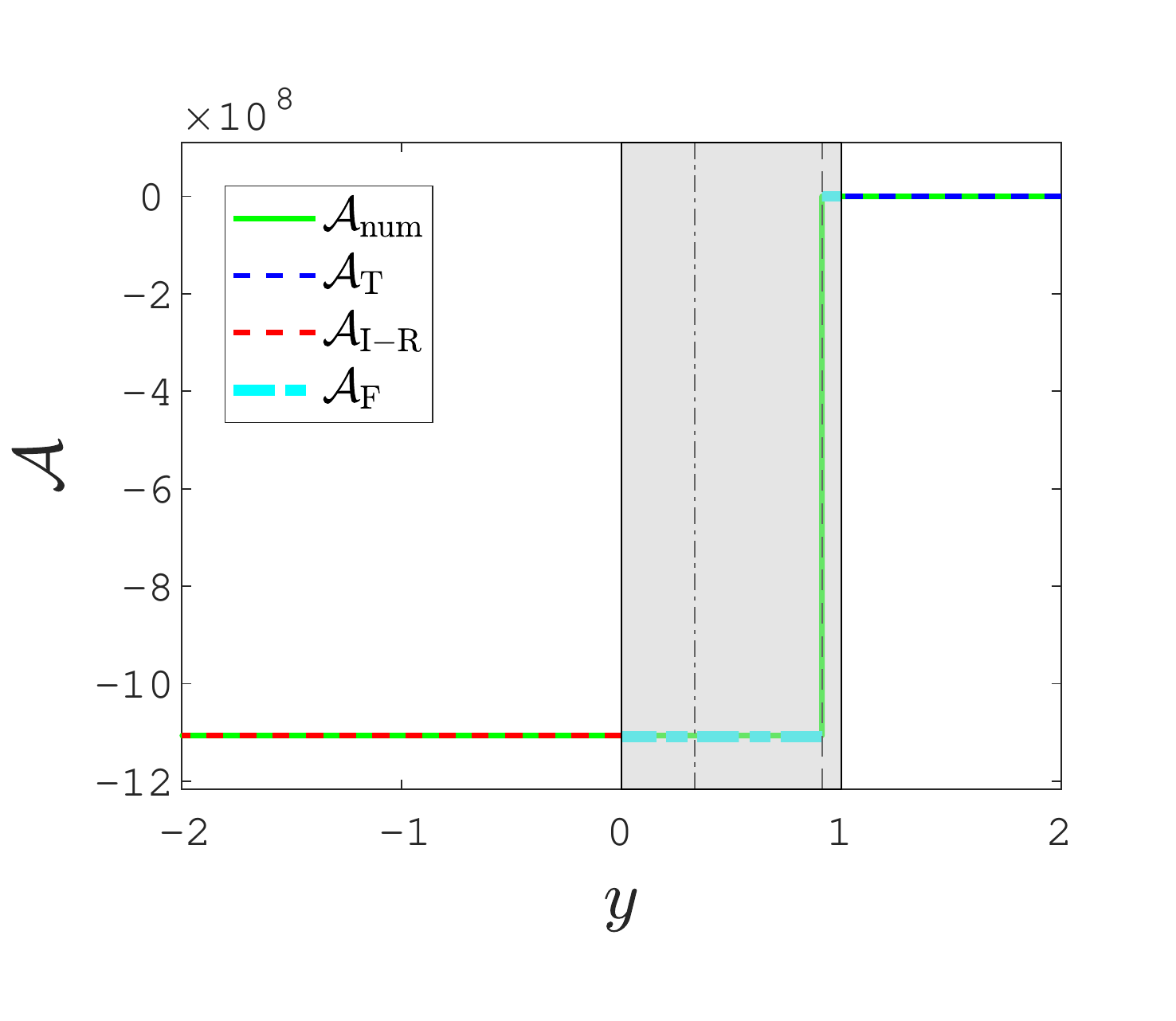}}
\caption{\aava{Same quantities as in Fig. \ref{vA_pole} (when the box is at the pole), but for a box tilted by 10 degrees relative to the pole and for different values of the shear, wavenumbers, and inertial frequencies.} Note that unlike Fig. \ref{vA_pole}, we do not have analytical solutions in the shear region. In all panels, the horizontal wave numbers are set to $k_x=k_z=1$, and the shear is $\aava{\Ro}=0.3$.
\textit{From left to right:} The inertial frequency is set to $\omega=0.31$, $\omega=0.16$,  and $\omega=0.1$. 
In the third panels, we indicate $\yo$ and $y_-$ by dashed dotted and dashed vertical lines, respectively.
}
\label{vA_10deg}
\end{figure*}
\subsection{Numerical exploration at constant shear when the box is inclined}
We now investigate wave propagation through the different critical levels when the box is inclined with respect to the rotation axis. We still assume that the shear region (zone II) has a linear shear flow profile ($n=1$). In contrast to the polar configuration, we do not have analytical solutions to the ordinary differential equation. Instead of going for an extensive numerical investigation of the parameter space, we will rather focus on the dynamics of inertial waves in our three-layer model as they cross the critical levels $\sigma=\pm\ft$ and $\sigma=0$.
Our results are presented in Fig. \ref{vA_10deg} for a box inclination of $10\degree$ relative to the rotation axis, a shear fixed to $\Ro=0.3$ and wave numbers set to $k_x=k_z=1$. 
The value of the frequency $\omega$ determines the existence and the nature of the critical level as detailed in the table of Appendix \ref{wAI_III}.  
Like in Fig. \ref{vA_pole}, we plot in each column of Fig. \ref{vA_10deg} the latitudinal velocity and the wave action flux. 
From left to right, we illustrate our results for the critical levels $\sigma=+\ft$, $\sigma=0$, and $\sigma=\{0,-\ft\}$ (i.e., there are two critical levels in the rightmost panels). 
One can notice that the first-order Frobenius approximation is not in good agreement anymore with the numerical solution in the entire shear region, though it remains a reasonable approximation close enough to a critical level. 
Unlike the polar case, the discrepancy far outside the critical levels is due to the linear approximation that the governing ODE takes around critical levels (see Eq. (\ref{PcDL_ft}) and (\ref{Pcorot})).

In the left panels of Fig.~\ref{vA_10deg}, the reflected and transmitted waves are strongly attenuated at the critical level $y=y_+$. 
Part of the wave energy is laid down to the mean flow, as corroborated by the drop in the wave action flux. 
In the middle panels, we do not see any discontinuity at the corotation $y=\yo$, which is in line with the theoretical analysis in Sect. \ref{corot_inclined} for a constant shear. 
However, the wave is over-reflected and over-transmitted, possibly due to the polynomial form of the solutions in the Frobenius series around $y=\yo$. 
In the right panels, the wave encounters successively critical levels at $y=\yo$ and at $y=y_-$. 
Although the wave going through the shear region is not attenuated at the corotation $y=\yo$, it is completely absorbed at the second critical level $y=y_-$ where the wave action flux drops to zero. This is consistent with the transmission coefficient in the left panel of Fig. \ref{atnu_deg}. 
The latitudinal velocities displayed in the top-left and top-right panels of Fig. \ref{vA_10deg} support the concept of a valve effect. 
Indeed, according to our analysis in Sect. \ref{cc_ft} with the Frobenius method and given the shear and wave numbers of Fig. \ref{vA_10deg}, the attenuation is strong for a downward wave meeting the critical level $y_+$ (first panel) while the attenuation is strong for an upward wave that meets the critical level $y_-$ (third panel). 
Before and after the critical levels $y_-$ and $y_+$, respectively, we observe fast oscillations of shorter and shorter period close to the critical level as already evidenced by \cite{BB1967}.
The analysis to determine how the wave is reflected in the shear zone can hardly be taken any further, because Frobenius solutions are not fully separable into upward and downward waves.

We emphasise that the behaviour of the wave at corotation $y=\yo$ when the box is at the pole stands out clearly different from the case when the box is inclined for a linear mean flow profile. 
This is particularly true in terms of the absolute value of the wave action flux that is subject to rise and drop in the polar configuration, whereas it remains conserved when the box is inclined. 
In this inclined case, the only way for a wave to be attenuated without \aava{friction} is that it meets critical levels $\sigma=\pm\ft$. 
Depending on the critical level encountered, an upward or downward wave will not be attenuated in the same way as known by the valve effect.
In addition, we no longer observe amplification due to the critical level but still ``geometric'' amplification (for instance in the middle panels where we can observe over-transmission and over-reflection), that can be explained by the exponential form of the Frobenius series. 
\subsection{Numerical results with a non-constant shear}
\label{nonlin}
\begin{table*}[ht!]
\begin{tabular}{ c|c ||c c c | c c c | c c c}
 &&\multicolumn{3}{c|}{$R$} & \multicolumn{3}{c|}{$T$} & \multicolumn{3}{c}{$\mA_\mathrm{T}/\mA_\mathrm{I-R}$}\\
\rule[-1ex]{0pt}{3.5ex}
&\backslashbox{\makecell{case}}{$n=$} & \makecell{$1$\\$\ $} & \makecell{$2$\\$\ $} & \makecell{$3$\\$\ $} & \makecell{$1$\\$\ $} & \makecell{$2$\\$\ $} & \makecell{$3$\\$\ $} & \makecell{$1$\\$\ $} & \makecell{$2$\\$\ $} & \makecell{$3$\\$\ $}\\
 \hline\hline
\rule[-1ex]{0pt}{4ex} \multirow{3}{*}{\rotatebox[origin=r]{90}{$\theta_0=0\degree$}} &$\Ro=0.3$ & $1.3\tt{-4}$ & $9.1\tt{-3}$ & $1.5\tt{-2}$ & $1.3\tt{-1}$ & $ 2.1\tt{-2}$ & $7.4\tt{-2}$ & $-3.2\tt{-8}$ & $-1.6\tt{-4}$ & $-1.1\tt{-2}$\\
\rule[-1ex]{0pt}{4ex}& $\Ro=0.8$ & $7.4\tt{-1}$ & $5.7\tt{-3}$ & $1.7\tt{-4}$ & $1.4$ & $9.1\tt{-5}$ & $4.0\tt{-6}$ & $-1.1\tt{-1}$ & $-2.1\tt{-10}$ & $-4.1\tt{-13}$\\
\rule[-1ex]{0pt}{4ex} &$\Ro=1.8$ & $1.5$ & $6.4\tt{-1}$ & $7.0\tt{-1}$ & $1.8$ & $1.1$ & $9.7\tt{-1}$ & $2.9$ & $-1.9$ & $-1.8$ \\\hline
\rule[-1ex]{0pt}{4ex}  \multirow{3}{*}{\rotatebox[origin=r]{90}{$\theta_0=10\degree$}}  &$y_+$ & $8.2\tt{-2}$ & $2.2\tt{-2}$ & $2.4\tt{-2}$ & $1.0\tt{-1}$ & $8.9\tt{-2}$ & $8.3\tt{-2}$ & $3.5\tt{-1}$ & $2.7\tt{-1}$ & $2.4\tt{-1}$\\
\rule[-1ex]{0pt}{4ex}& $\yo$ & $5.3$ & $4.9$ & $4.7$ & $4.9$ & $4.4$ & $4.3$ & $1.0$ & $1.0$ & $1.0$\\
\rule[-1ex]{0pt}{4ex} &$\yo$ and $y_-$ & $5.0$ & $6.2$ & $6.1$ & $2.3\tt{-4}$ & $4.0\tt{-2}$ & $2.2\tt{-1}$ & $1.1\tt{-9}$ & $2.0\tt{-5}$ & $6.7\tt{-4}$\vspace{0.5cm}
\end{tabular}
\caption{Reflection $R$, transmission $T$ coefficients and ratio of the wave action flux above and below a critical level $\mA_\mathrm{T}/\mA_\mathrm{I-R}$, for a linear, square and cubic mean flow ($n=1,\ 2, \text{ and }3$, respectively). The six cases presented here are, in order, the three cases examined at the pole in Fig. \ref{vA_pole} and the three cases of a tilted configuration   examined in Fig. \ref{vA_10deg}.}
\label{tab_RTA}
\end{table*}
The choice of a linear mean flow profile allows the resolution of the ordinary differential equation at the pole and a simpler implementation of the Frobenius method. Nevertheless, the correspondence between a global and a local mean flows as presented in Sect. \ref{mean_flow_sect} involves higher-order terms than a simple linear dependence. 
Therefore, it is important to examine the effect of different mean flow profiles with non-zero $U''$. Now assuming $n > 1$ for the shear flow profile in zone II (see Eq.~\ref{U_pro}), the Frobenius method in the inclined and polar cases still holds provided that
\begin{equation}
\left\{\begin{aligned}
&\yo=\sqrt[n]{\frac{\omega}{k_x\Lambda}},\ \text{ and }\ y_\pm=\sqrt[n]{\frac{\omega\mp\ft}{k_x\Lambda}}\\
&\Ro=\Lambda n(y_{0,\pm})^{n-1}.
\end{aligned}\right.
\end{equation}
Moreover, we specify that the critical level in $y<0$ for even values of $n$ is not examined given our numerical set up where the shear region is located in the range $y\in[0,1]$. 
The conditions to have critical levels inside the shear region (zone II) are the same as those for a constant shear (see Appendix \ref{wAI_III}). 

We show in Table \ref{tab_RTA} numerical values of the reflection and transmission coefficients along with the ratio of the wave action flux below and after the critical level in the six parameter sets illustrated in Figs. \ref{vA_pole} and \ref{vA_10deg} for linear, square and cubic mean flow profiles.
For all cases, $R$, $T$ and $\mA_\mathrm{T}/\mA_\mathrm{I-R}$ change quite significantly between $n=1$, 2 and 3.
At the pole, the (``geometric'') over-transmission found for $\Ro=0.8$ disappears for $n=2$ and $n=3$, where
the reflected and transmitted waves are strongly attenuated. 
Similarly, in the case where $\Ro=1.8$, the over-transmission and over-reflection disappear when $n=3$, whereas $|\mA_\mathrm{T}/\mA_\mathrm{I-R}|>1$ which entails that the transmitted wave is taken wave action flux from the mean flow just after the critical level. 
We find that the Frobenius method does not give consistent results for the case where $\Ro=1.8$ and $n=\{2,3\}$. 
When the box is tilted and for the critical level $y_+$, the reflected wave is more attenuated than the transmitted one, which is consistent with stronger attenuation of the counter-propagating wave discussed in the previous section. 
At the corotation and when the box is inclined (fifth row of Table \ref{tab_RTA}), 
 \aava{the wave action flux remains the same in the whole domain for linear or non-linear mean flow profiles, in agreement with the wave action flux derived analytically in Eq. (\ref{wAconcor})}.
We speculate that the over-transmission and over-reflection are due to the polynomial form of the solutions in the shear region. \aava{These wave amplifications may be related to shear instabilities in this particular three-layer configuration, and are probably not linked with the presence of a critical level whose implications on the flow are well diagnosed by the wave action flux}. 
In the last case (sixth row), where two critical levels co-exist, the transmitted wave is less attenuated at $y=y_-$ from $n=1$ to $n=3$ but it remains more attenuated than the reflected wave as discussed in the previous section. Again, no jump is found in the wave action flux at the corotation despite analytical predictions.

The same analysis was carried out for different amplitudes of the Rayleigh friction force, up to $\sf=\tt{-2}$. 
Of course, the wave action flux is no longer constant, but we observe that all three parameters of Table \ref{tab_RTA} change very little compared to the case where $\sf=\tt{-8}$. 
This result is consistent with \cite{AM2013} in the context of gravity waves and  vertical shear. 
We comment that while a low \aava{friction} is mandatory in the numerical code to solve the ODE at singularities $y_\pm$, it is not the case at the corotation $y=\yo$. 
\section{Astrophysical discussion}
\label{astrodi}
\begin{figure}
\includegraphics[width=\hsize]{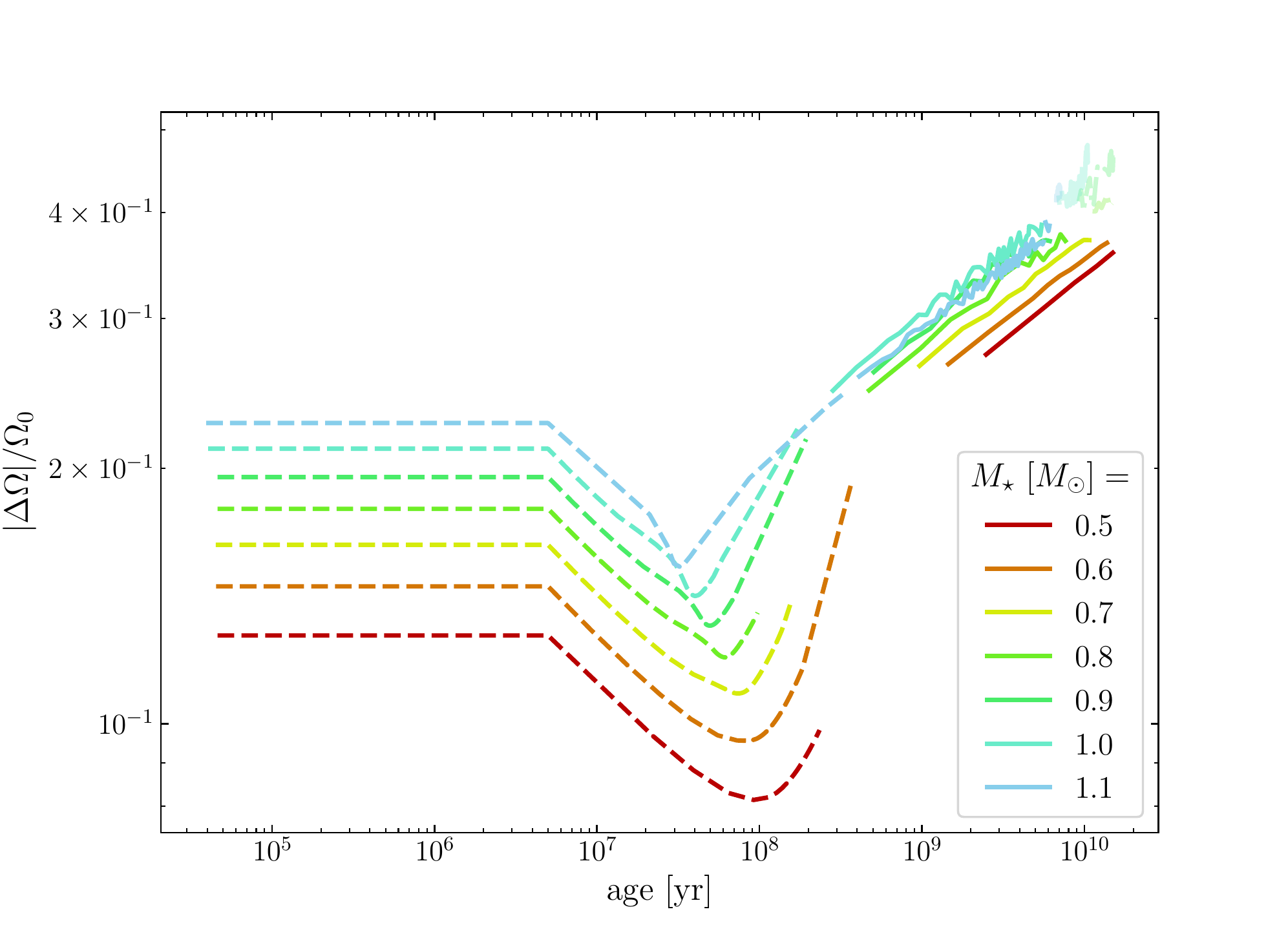}
\caption{Absolute values of the shear contrast normalised by the rotation at the pole against the age of stars of K and G spectral types. Solid lines feature $0.3<R_\mathrm{of}<0.9$, dashed lines $R_\mathrm{of}\lesssim0.3$ and transparent lines $R_\mathrm{of}>1$.}
\label{Dom}
\end{figure}
\subsection{Latitudinal differential rotation in stars}
\label{appli_lat}
In stars, latitudinal differential rotation is often characterised by the difference in rotation frequency between the equator and the pole, that is the quantity  $\Delta\Omega=\Omega_\mathrm{eq}-\Omega_0$, where $\Omega_\mathrm{eq}$ is the rotation frequency at the equator \cite[e.g.,][]{BJ2017}. We will now refer to $\Delta\Omega$ as the shear contrast. Different regimes are distinguished according to the value of $\Delta\Omega$: anti-solar-like rotation for $\Delta\Omega<0$, cylindrical rotation for $|\Delta\Omega|\ll1$, and solar-like rotation rotation for (not-too-low) positive $\Delta\Omega$ (e.g. $\Delta\Omega/\Omega_{0}\simeq0.3$ for the Sun).
Several works based on three-dimensional numerical simulations have explored the range of physical parameters leading to each aforementioned regime in stars and in giant planets \cite[e.g.,][]{GW2013,VS2016,BS2018}
In particular, \cite{BS2017} have derived a criterion based on mixing length theory and calibrated with 3D simulations that determines the rotation profile of a solar-like star. This criterion is based on the fluid Rossby number $R_\mathrm{of}$, defined as
\begin{equation}
R_\mathrm{of}=R_\mathrm{of,\odot}\,\Omega_*^{-0.82}M_*^{1.53},
\label{B1}
\end{equation}
 where $R_\mathrm{of,\odot}=0.89$ is the solar fluid Rossby number, and $\Omega_*$ and $M_*$ are the mean rotation and the mass of the star respectively, normalised with their solar value. \cite{BS2017} highlighted the following three regimes:
\begin{itemize}
\item $R_{\mathrm{of}}>1$ for anti-solar-like rotation, 
\item $0.3<R_{\mathrm{of}}<0.9$ for solar-like rotation,
\item $R_{\mathrm{of}}\lesssim0.3$ for cylindrical rotation.
\end{itemize}
Furthermore, they introduced the shear contrast $\Delta\Omega_{\mathrm{S}}$ at the colatitude $30\degree$ as
 \begin{equation}
 \Delta\Omega_\mathrm{S}=\Omega_\mathrm{eq}-\Omega(\theta=30\degree),
 \end{equation}
 as the rotation frequency is often ill-defined at low colatitudes in 3D numerical simulations in spherical geometry. From their 3D simulations, \citet{BS2017} obtained the following scaling:
 \begin{equation}
\Delta\Omega_\mathrm{S}=\Delta\Omega_{\mathrm{S}, \odot}\,M_*^{0.73}\Omega^{0.66},
\label{B2}
\end{equation}
where $\Delta\Omega_{\mathrm{S},\odot}\simeq565\times\tt{-9}\second^{-1}$ is the solar value of $\Delta\Omega_\mathrm{S}$ calculated from \citet{GT2007}. Using the mean flow profile (Eq. (\ref{law})), $\Delta\Omega_\mathrm{S}$ can be related to our shear contrast $\Delta\Omega$ via
\begin{equation}
 \frac{\Delta\Omega}{\Omega_0}=\chi=\frac{4}{3}\frac{\Delta\Omega_\mathrm{S}}{\Omega_0}.
 \label{deltaOmega}
\end{equation}
We show in Fig. \ref{Dom} the quantity $\Delta\Omega/\Omega_0$, expressed in Eq. (\ref{deltaOmega}), versus the age of solar-like stars for K to G spectral types. 
To compute this quantity, we have used grids of the 1D stellar evolution code STAREVOL \cite[see][for details of the code]{AP2019}.
In light of Fig. \ref{Dom}, the stars in the pre-main sequence ($\mathrm{age}\lesssim 100\,\mega\mathrm{yr}$) exhibit cylindrical rotation as they 
are fast rotating. During the main sequence, stars mostly feature solar-like rotation, while anti-solar-like rotation is observed at the end of the main sequence from $0.8$ to $1.1\,M_\odot$. According to Fig. \ref{Dom}, a limit on the absolute value of the normalised shear contrast can be set to $|\Delta\Omega|/\Omega_0<0.5$. 
However, as already stressed by \cite{BB2018}, the latitudinal shear inferred by asteroseismology can be much larger than predicted by numerical simulations. This can actually be inferred by comparing the shear factors in their work \cite[see Table S3 from supplementary materials in][]{BB2018} with ours given in Fig. \ref{Dom}. Moreover, according to their study, cylindrical and anti-solar differential rotation are hardly unambiguously detectable.
Finally, one should recall that \cite{BS2017}'s scaling laws given in Eqs. (\ref{B1}) and (\ref{B2}) are derived for K and G spectral type stars only.\\ 

Since we now know values of the shear contrast, we can calculate the ``shear'' Rossby number $\Ro=\Up/(2\Omega)$, given by the following relationship:
\begin{equation}
\Ro=-\frac{3}{2}\cos\theta_0\sin^2\theta_0\,\chi,
\end{equation}
which has been derived from Eq. (\ref{U_conic}) by keeping only zero-order terms in $y$.
Taking $\chi\simeq0.3$ as a representative value of the shear contrast for main sequence G and K stars, we find that the Rossby number is maximal when $\theta_0\simeq55\degree$, its maximum value being $\Ro\simeq-0.17$. In particular, $\Ro\simeq-0.013$ for $\theta_0=10\degree$ and $\Ro\simeq-0.076$ for $\theta_0=80\degree$.
These values are useful to interpret wave flux action transmission at critical levels $\sigma=\pm\ft$, considering Fig. \ref{atnu_deg} \aava{(we recall that at the corotation the wave action flux is fully transmitted)}. A downward (upward) propagating wave through $\sigma=\ft$ ($\sigma=-\ft$) is
\begin{itemize}
\item totally absorbed provided that $k_z\gtrsim0.1 k_x$ ($\alpha_k\gtrsim0.1$) at $\theta_0=10\degree$, and that $k_z\gtrsim k_x$ ($\alpha_k\gtrsim1$) at $\theta_0=80\degree$,
\item strongly attenuated for $k_z\sim0.1 k_x$ ($\alpha_k\sim0.1$) at $\theta_0=80\degree$,
\item fully transmitted given that $k_z\ll \tt{-1}k_x$ ($\alpha_k\ll\tt{-1}$) for both inclinations.
\end{itemize}
 These results also hold for anti-solar-like differential rotation, since the transmission factor $T_{\theta_0}$ is function of $|\Ro|$. For larger values of $|\Ro|$, i.e. for larger values of the shear contrast, waves are less damped at critical levels $\sigma=\pm\ft$ at a given $\alpha_k=k_z/k_x$.
 The connection between this ratio of the vertical and azimuthal wavenumbers in the local model and an equivalent ratio of global wavenumbers in the spherical geometry is not straightforward.
  A first hint can be to state that $k_z\sim k_r$, where $k_r$ is the wavenumber in the global radial direction, while $k_x\sim m/(r_0\sin\theta_0)$, where $m$ is the azimuthal order of the  considered mode of the tidal potential \citep[when $m\neq 0$;][]{ZT1997}. Then, we have $\alpha_k\equiv k_r  r_0\sin\theta_0/m \equiv 2\pi\,r_0/\lambda_r\times \sin\theta_0/m$, by introducing $\lambda_r$ the radial wavelength of the tidal wave.
In the case where $r_0>\lambda_r m$, we should thus be in the regime where the tidal wave is attenuated. 
\subsection{Cylindrical differential rotation in Jupiter and Saturn}
\begin{figure}
\includegraphics[width=\hsize]{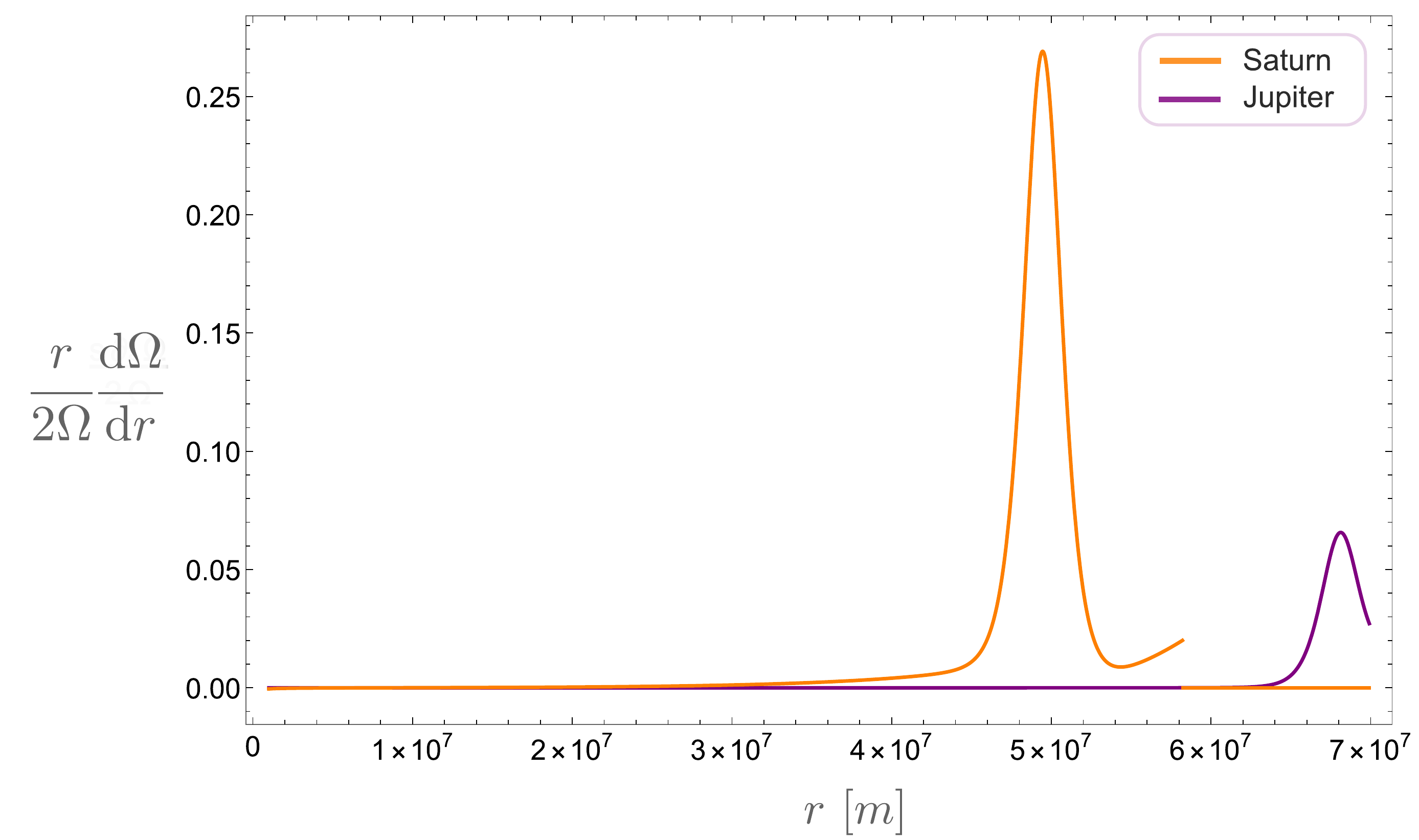}
\caption{Rossby numbers $\Ro=\frac{r}{2\Omega}\df\Omega r$ for Jupiter and Saturn as a function of the axial distance $r$.}
\label{JuSa}
\end{figure}
\begin{table*}[t!]
\centering
\begin{tabular}{c|cc|cc|c| c}
 \rule[-1ex]{0pt}{0ex} & \multicolumn{2}{c|}{ spherical } &\multicolumn{2}{c|}{ equatorial} & \makecell{\\$\alpha_k$} & \makecell{\\Regime}\\
 \rule[-1ex]{0pt}{0ex}& $l$ & $m$ &  $m$ & $l_\Theta$ & & \\\hline\hline
\rule[-1ex]{0pt}{4ex} asynchronous & 2 & 2 & 2 & \aava{2} & \aava{1} & \aava{total} absorption \\
\rule[-1ex]{0pt}{4ex} inclined &2 & 1 & 1 & 2 & $2$ & total absorption\\
\rule[-1ex]{0pt}{4ex} eccentric & 2 & 0 & 0 & 2 & $\rightarrow\infty$ & total absorption\vspace{0.5cm} 
\end{tabular}
\caption{Regime of wave transmission at corotation deduced from Figs. \ref{O2} and \ref{atnuO2} for three orbital states of a satellite around a  giant gaseous planet.}
\label{lm}
\end{table*}
Mathis et al. (in prep.) have developed an equatorial model to examine inertial wave properties in the outer convective layers of giant gaseous planets such as Saturn and Jupiter, which are subject to cylindrical differential rotation. 
In their model (built in cylindrical coordinates), they derived a Schrödinger-like differential equation for $\Psi=\sqrt{\rho}r^2v_r$ under the anelastic approximation, where $\rho$ is the density, $r$ the axial distance coordinate, and $v_r$ the axial velocity. For free inertial waves, their second-order differential equation is:
\begin{equation}
\dfd{\Psi}{r}+\left(\frac{\kappa_r^2l_\Theta^2}{\hat\sigma^2r^2}-\frac{l_\Theta^2+m^2}{r^2}\right)\Psi=0,
\label{equa}
\end{equation}
where $l_\Theta$ and $m$ denote the equatorial and azimuthal wavenumbers, respectively, 
$\hat\sigma=\omega+m\Omega(r)$ is the (linear) Doppler-shifted frequency, and $\kappa_r$ the ``axial'' epicyclic frequency defined as $\kappa_r^2=4\Omega^2+2\Omega r\df{\Omega}{r}$.
For cylindrical differential equation, the corotation resonance $\hat\sigma=0$ results in critical cylinders \citep[see][]{BR2013} characterised by a critical axial distance $r=\rc$. The Taylor expansion of Eq. (\ref{equa}) at first order around $\rc$ gives:
\begin{equation}
\dfd{\Psi}{r}+\left(\frac{ l_\Theta^2(1+\Ro)}{m^2\Ro^2(r-\rc)^2}-\frac{l_\Theta^2+m^2}{\rc^2}\right)\Psi=0,
\label{equa_exp}
\end{equation}
by setting $\Ro=\rc\left.\df{\Omega}{r}\right|_{r=\rc}/(2\Omega_\mathrm{c})$ the local Rossby number in cylindrical coordinates. By writing $\alpha_k^2=l_\Theta^2/m^2$, Eq. (\ref{equa_exp}) becomes very similar to our \aava{ODE} Eq. (\ref{Pc}) when the box is located at the South pole and for a constant shear (i.e. with $\ft=0$, $f=-1$ and $\Up'=0$). 
Note that, when the local shear box model is located at the equator, the latitudinal coordinate $y$ is directed along the (vertical) rotation axis, whereas when the box is at the poles, $y$ is directly the axial distance. That is why a polar configuration of the box best reproduces the ``equatorial'' model of Mathis et al. (in prep.).
Moreover, the convention of a plus sign in the Doppler-shifted frequency explains why Eq. (\ref{equa_exp}) is analogous to our \aava{wave propagation} equation when the box is at the South pole rather than at the North pole. 

In Fig. \ref{JuSa}, we show Jupiter's and Saturn's local Rossby number from Mathis et al. (in prep.). Cylindrical differential rotation extends in the outer layer of the convective envelope of both planets, in agreement with Juno and Cassini Grand Finale observations \citep[respectively]{KG2017,GK2019}. According to Fig. \ref{O2}, where we notably plotted $R=\alpha_k^2(1+\Ro)/\Ro^2$ at the South pole (dark red and purple areas), we can assess the role of the critical level for wave transmission across the corotation, in terms of the wavenumber ratio $\alpha_k$. In Fig. \ref{JuSa}, the Rossby number \aava{satisfies} $\Ro\lesssim0.27$ for Saturn and $\Ro\lesssim0.07$ for Jupiter. Given this range of values, two regimes can be evidenced for waves and wave action fluxes through the corotation:
\begin{itemize}
\item waves are strongly attenuated for $l_\Theta\gtrsim m$ (see also Fig. \ref{atnuO2} for $\Ro=-|\Ro|$ as the transmission factor in this figure is plotted for the North pole).
\item waves can be over-reflected and over-transmitted for $l_\Theta\ll m$, and can potentially lead to instabilities given specific boundary conditions.
\end{itemize}
To give an idea of the values these wavenumbers can take, we have listed in Table \ref{lm} three typical orbital states, where, in order, the asynchronous, eccentricity and obliquity tides are supposed to be dominant \citep{O2014}.
These states are described by  the ``spherical'' quadrupolar components of the dominant terms in the tidal potential, i.e. the degree $l$ and the order $m$ of the spherical harmonics. The analogy with the equatorial model is then made to get $m$, \aava{and $l_\Theta$ is chosen to approximate as best as possible the behaviour of the Legendre polynomial $P_l^m(\cos\Theta)$ around the equator with a simple trigonometric function $\Re\{\exp[i(l_\Theta \Theta+\phi)]\}$, where $\phi$ is the appropriate phase} (Mathis et al. in prep.). \aava{To find the associated wave attenuation at corotation for the three main tides, one can use Figs. \ref{O2} and \ref{atnuO2} for $\ak\geq1$ and $\Ro\lesssim0.27$, and look at the South pole (as Fig. \ref{atnuO2} is plotted at the North pole, one has to take the opposite Rossby number). From Fig. \ref{O2}, we can assess that waves excited by these tides are always in the so-called ``stable'' regime  for these ranges of $\Ro$ and $\ak$, which excludes an amplification of these waves. Moreover, from Fig. \ref{atnuO2}, we also observe that waves are completely absorbed at corotation for our given ranges of parameters. By consequence, waves excited by the asynchronous, inclined, or eccentric tides in Jupiter and Saturn are expected to transfer all their wave action flux to the mean flow at corotation}. 
\section{Conclusion and perspectives}
\label{conclu}
The present study was motivated by the works of \cite{BR2013} and \cite{GB2016,GM2016}, who showed 
that differential rotation can strongly affect the propagation and dissipation properties of (tidal) linear inertial waves. They considered 
different rotation profiles typical of stellar and planetary interiors and they pointed out that tidal waves can deeply interact with zonal flows at corotation resonances, leading to intense wave energy dissipation, alongside possible instabilities. 
In this paper, we have investigated the transmission of free inertial waves with latitudinal stratification and differential rotation, at the corotation resonance (characterised by a zero Doppler-shifted wave frequency) and more broadly at critical levels (any singularities of the governing second-order wave propagation equation in the inviscid limit). 
For this purpose, we have built a new local Cartesian box model with horizontal shear, modelling a small patch of the convective zone of a low-mass star or a giant planet. 
By considering the inclination of the local reference frame relative to the rotation axis, we have examined the effect on wave propagation through a critical level of a conical rotation profile at a general colatitude when the box is tilted, or of a cylindrical rotation profile when the box is at the North or South poles. 
These rotation profiles are inspired by those observed or expected in the Sun, low-mass stars, and giant gaseous planet in our solar system. Three critical levels can be identified when the box is inclined relative to the rotation axis: the corotation resonance and two others critical levels that arise from the inclination between the gravity and the rotation vectors, and which are defined by a Doppler-shifted frequency equal to plus or minus the latitudinal component of the rotation frequency. When the box is at the poles, critical levels are restricted to the corotation.

In order to diagnose the behaviour of a wave passing through a critical level for both aforementioned rotation profiles, we made use of an invariant called the wave action flux that is independent of the latitudinal coordinate in a non-dissipative fluid flow. 
This invariant was used when the ``directional'' flux of angular momentum (here latitudinal) can not be constructed easily from the mean perturbed velocity, as it is the case for example in \cite{LT1978} for Rossby waves in plane-parallel shear flows. 
The wave action flux has already been used in vertically-stratified shear flows in the presence of rotation or magnetic fields to interpret the role of critical levels \citep{G1975,AM1978,G1979,M2009,MB2012}. 
Using the condition that this invariant is discontinuous at critical levels, we have demonstrated in Sect. \ref{sec3} that waves can be either fully transmitted, damped or even amplified after passing through critical levels as a result of wave action flux exchanges. 
These different regimes of wave transmission are found both with conical and cylindrical rotation profiles; they depend on the critical level encountered, on the wave properties (e.g. the propagation direction, wavenumbers) and on the profile of the mean flow. Table \ref{recap_ana} summarises the main analytical results.

We have then confronted our analytical results with a three-layer numerical model that comprises a shear zone in which the critical level is located and two surrounding shear-free zones that allow incident, reflected and transmitted waves. 
A difference with the analytical model is the introduction of small dissipative force under the form of a Rayleigh friction \citep[also called frictional force by][]{O2009} to avoid strict singularities. 
It does not seem to affect the results, since analytical and numerical results match \aava{quite} well when using a \aava{power-law mean flow profile} and by varying the \aava{friction}. 
This conclusion is also shared by the work of \cite{AM2013} who studied corotation resonances for gravity waves propagating in stratified and vertically shear flows. 

Based on the analytical results, we have discussed in Sect. \ref{astrodi} possible applications to stellar and planetary interiors. We have estimated the rate of differential rotation in solar-like stars by the shear contrast (the rotation difference between the pole and the equator) and in giant gaseous planets through the local Rossby number (the ratio between the shear and the rotation frequency in cylindrical coordinates). We find that for K and G-type stars along their lifetime, and for Jupiter and Saturn at the present time, a regime where inertial waves are strongly damped is largely preferred in the convective envelope of these objects. Similar conclusions were found by \cite{AM2013} for internal gravity waves through critical levels in the core of solar-like stars.
\\

It is interesting to discuss different regimes of the wave transmission in terms of angular momentum transfer for cases of strong damping and  wave amplification. 
First, we have to underline that the theoretical analysis presented in Section 3 (using the Frobenius method) does not adequately characterise wave (over-)reflection as similarly observed in the numerical section \ref{secnum}. 
What we can access is the wave action flux on either side of the critical level. 
The analysis on changes in the wave  action flux across the critical level allows us to understand whether energy is deposited to or extracted from the mean flow, in line with the work carried out by \cite{M1961,BB1967,G1975,LB1985}. In the presence of a locally conical differential rotation, we have demonstrated that a valve effect can be found for critical levels other than the corotation, analogous to the results of \cite{A1972} and \cite{G1975}  for hydromagnetic and gravito-inertial waves with a vertical shear. For these peculiar critical levels, waves can be attenuated when going one direction, mainly featured by the sign of the rotation components in the box, or fully transmitted when going the other direction.
For cylindrical differential rotation, we have found a criterion analogous to the Miles-Howard theorem for stratified shear flows \citep{MH1964}, which for inertial waves as in this work can be formulated as:
\begin{equation}
R=\frac{k_z}{k_x}\frac{1-\Ro}{\Ro^2}
\left\}\begin{aligned}
&>1/4 \text{: wave attenuation;}\\
&<1/4 \text{: possible wave over- }\left|\begin{tabular}{l} reflection, \\ transmission. \end{tabular}\right.
\end{aligned}\right.
\label{resum}
\end{equation}
The above criterion depends on the shear Rossby number $\Ro=\Up/(2\Omega)$ and the vertical ($k_z$) and longitudinal ($k_x$) wavenumbers. \aava{This last point is an important difference from the Miles-Howard criterion, which does not involve wavenumbers. For this reason the analogy between Eq.(\ref{resum}) and the Miles-Howard stability criterion must be taken with care.}
We \aava{also} stress that \aava{Eq. (\ref{resum})} is very different from the Rayleigh's inflection point theorem for Rossby waves, which are a subclass of inertial waves when neglecting the vertical or the radial perturbed velocity \citep{B1966,LT1978}.
The Miles-Howard  criterion allows to disentangle between critical levels where strong wave attenuation is expected for $\mathrm{Ri}>1/4$ (where $\mathrm{Ri}$ is the Richardson number), and those where over-reflection and over-transmission can lead to potential shear instabilities for $\mathrm{Ri}<1/4$.
\citet{L1988} warns, however, that over-reflection and over-transmission are a necessary but not sufficient condition for shear instability. 
Such amplifications leading to the instability require peculiar conditions in a three-layer model, where the shear zone that features the critical level is surrounded by a region of incoming propagating waves, and a ``sink'' zone to force waves to cross the evanescent shear zone.
Special boundary conditions are necessary for the wave to return successively to the critical level and induce wave amplitude growth. Recent studies \citep[see e.g.][for a review]{CT2012} have revisited instabilities in stratified shear flows by studying multiple counter-propagating waves that can interact to grow in amplitude with time (with conditions as phase-locking). A parallel is drawn between over-reflection mechanisms and interacting counter-propagating waves by \citet{HH2007} to describe baroclinic instabilities for Rossby waves.
  
Contrary to what our results predict, \citet{BR2013} did not observe any instabilities of inertial waves when using cylindrical differential rotation. 
Several reasons can be put forward to explain this discrepancy, such as boundary conditions (as discussed in the previous paragraph), or the values of the shear and horizontal wavenumbers, since they may not be in the regime allowing instabilities according to the criterion Eq. (\ref{resum}), which further needs to be adapted to their global cylindrical geometry. 
We stress that when exploring different power-laws for the mean flow profiles, over-reflection was not retrieved for a non-linear mean flow in cylindrical differential rotation. With conical differential rotation, \cite{GB2016,GM2016} did observe instabilities, but only for sufficiently low viscosities, whereas our study showed little dependence on the \aava{friction}, and rather highlights possible over-transmission for non-linear flows. 
Lastly, we underline that a temporal analysis on the growth rate of perturbations should be undertaken to unravel instabilities, which has not been performed in this paper but in other separate papers \citep[][]{PPM2020,PPMB2020}.

This ab-initio analytical study is thus a first step to understand how inertial waves interact with a mean flow subject to latitudinal differential rotation at critical levels, in the context of tidal dissipation in differentially rotating stars and planets. Possible feedbacks of the perturbed wave on equilibrium quantities and the mean flow are not taken into account in this study, \aava{nor are non-linearities in the perturbed hydrodynamical wave equations}. Nonetheless, they should be considered in future studies since \cite{BO2010} and \cite{BR2013} suggested  important non-linear effects for inertial waves at corotation. Finally, magnetism may also play an important role to dissipate  or redistribute angular momentum at critical levels through  magnetic stresses \citep[e.g.][]{W2016, W2018,LO2018,AM2019}.
\begin{acknowledgements}
We would like to thank the anonymous referee, as well as A. Barker, for the helpful comments and suggestions regarding our work. A. Astoul, J. Park,  and S. Mathis acknowledge funding by the European Research Council through the ERC grant SPIRE 647383. The authors acknowledge the PLATO CNES funding at CEA/IRFU/DAp and IRAP.
This research has made use of NASA's Astrophysics Data System and of the software MATLAB  version R2018a.
Finally, we thank Quentin André and Jéremy Ahuir for fruitful discussions on the details of the analytical model.
\end{acknowledgements}

\bibliographystyle{aa} 
\bibliography{biblio} 

\begin{thebibliography}{108}
\expandafter\ifx\csname natexlab\endcsname\relax\def\natexlab#1{#1}\fi

\bibitem[{{Abramowitz} \& {Stegun}(1972)}]{AS1972}
{Abramowitz}, M. \& {Stegun}, I.~A. 1972, {Handbook of Mathematical Functions}

\bibitem[{{Acheson}(1972)}]{A1972}
{Acheson}, D.~J. 1972, Journal of Fluid Mechanics, 53, 401

\bibitem[{{Alvan} {et~al.}(2013){Alvan}, {Mathis}, \& {Decressin}}]{AM2013}
{Alvan}, L., {Mathis}, S., \& {Decressin}, T. 2013, \aap, 553, A86

\bibitem[{{Amard} {et~al.}(2019){Amard}, {Palacios}, {Charbonnel}, {Gallet},
  {Georgy}, {Lagarde}, \& {Siess}}]{AP2019}
{Amard}, L., {Palacios}, A., {Charbonnel}, C., {et~al.} 2019, arXiv e-prints,
  arXiv:1905.08516, {A\&A, in press, 10.1051/004-6361/201935160}

\bibitem[{{Andr{\'e}} {et~al.}(2017){Andr{\'e}}, {Barker}, \&
  {Mathis}}]{AB2017}
{Andr{\'e}}, Q., {Barker}, A.~J., \& {Mathis}, S. 2017, \aap, 605, A117

\bibitem[{{Andrews} \& {McIntyre}(1978)}]{AM1978}
{Andrews}, D.~G. \& {McIntyre}, M.~E. 1978, Journal of Fluid Mechanics, 89, 647

\bibitem[{{Astoul} {et~al.}(2019){Astoul}, {Mathis}, {Baruteau}, {Gallet},
  {Strugarek}, {Augustson}, {Brun}, \& {Bolmont}}]{AM2019}
{Astoul}, A., {Mathis}, S., {Baruteau}, C., {et~al.} 2019, \aap, 631, A111

\bibitem[{{Auclair-Desrotour} {et~al.}(2014){Auclair-Desrotour}, {Le
  Poncin-Lafitte}, \& {Mathis}}]{AL2014}
{Auclair-Desrotour}, P., {Le Poncin-Lafitte}, C., \& {Mathis}, S. 2014, in
  SF2A-2014: Proceedings of the Annual meeting of the French Society of
  Astronomy and Astrophysics, ed. J.~{Ballet}, F.~{Martins}, F.~{Bournaud},
  R.~{Monier}, \& C.~{Reyl{\'e}}, 199--203

\bibitem[{{Auclair Desrotour} {et~al.}(2015){Auclair Desrotour}, {Mathis}, \&
  {Le Poncin-Lafitte}}]{AM2015}
{Auclair Desrotour}, P., {Mathis}, S., \& {Le Poncin-Lafitte}, C. 2015, \aap,
  581, A118

\bibitem[{{Barker} \& {Ogilvie}(2009)}]{BO2009}
{Barker}, A.~J. \& {Ogilvie}, G.~I. 2009, \mnras, 395, 2268

\bibitem[{{Barker} \& {Ogilvie}(2010)}]{BO2010}
{Barker}, A.~J. \& {Ogilvie}, G.~I. 2010, \mnras, 404, 1849

\bibitem[{{Barnes} {et~al.}(2005){Barnes}, {Collier Cameron}, {Donati},
  {James}, {Marsden}, \& {Petit}}]{BCC2005}
{Barnes}, J.~R., {Collier Cameron}, A., {Donati}, J.~F., {et~al.} 2005, \mnras,
  357, L1

\bibitem[{{Barnes} {et~al.}(2017){Barnes}, {Jeffers}, {Haswell}, {Jones},
  {Shulyak}, {Pavlenko}, \& {Jenkins}}]{BJ2017}
{Barnes}, J.~R., {Jeffers}, S.~V., {Haswell}, C.~A., {et~al.} 2017, \mnras,
  471, 811

\bibitem[{{Baruteau} \& {Masset}(2008)}]{BM2008}
{Baruteau}, C. \& {Masset}, F. 2008, \apj, 672, 1054

\bibitem[{{Baruteau} \& {Rieutord}(2013)}]{BR2013}
{Baruteau}, C. \& {Rieutord}, M. 2013, Journal of Fluid Mechanics, 719, 47

\bibitem[{{Bazot} {et~al.}(2019){Bazot}, {Benomar}, {Christensen-Dalsgaard},
  {Gizon}, {Hanasoge}, {Nielsen}, {Petit}, \& {Sreenivasan}}]{BB2019}
{Bazot}, M., {Benomar}, O., {Christensen-Dalsgaard}, J., {et~al.} 2019, \aap,
  623, A125

\bibitem[{{Beaudoin} {et~al.}(2018){Beaudoin}, {Strugarek}, \&
  {Charbonneau}}]{BS2018}
{Beaudoin}, P., {Strugarek}, A., \& {Charbonneau}, P. 2018, \apj, 859, 61

\bibitem[{{Benbakoura} {et~al.}(2019){Benbakoura}, {R{\'e}ville}, {Brun}, {Le
  Poncin-Lafitte}, \& {Mathis}}]{BR2019}
{Benbakoura}, M., {R{\'e}ville}, V., {Brun}, A.~S., {Le Poncin-Lafitte}, C., \&
  {Mathis}, S. 2019, \aap, 621, A124

\bibitem[{{Benomar} {et~al.}(2018){Benomar}, {Bazot}, {Nielsen}, {Gizon},
  {Sekii}, {Takata}, {Hotta}, {Hanasoge}, {Sreenivasan}, \&
  {Christensen-Dalsgaard}}]{BB2018}
{Benomar}, O., {Bazot}, M., {Nielsen}, M.~B., {et~al.} 2018, Science, 361, 1231

\bibitem[{{Bolmont} {et~al.}(2017){Bolmont}, {Gallet}, {Mathis}, {Charbonnel},
  {Amard}, \& {Alibert}}]{BG2017}
{Bolmont}, E., {Gallet}, F., {Mathis}, S., {et~al.} 2017, \aap, 604, A113

\bibitem[{{Bolmont} \& {Mathis}(2016)}]{BM2016}
{Bolmont}, E. \& {Mathis}, S. 2016, Celestial Mechanics and Dynamical
  Astronomy, 126, 275

\bibitem[{{Booker} \& {Bretherton}(1967)}]{BB1967}
{Booker}, J.~R. \& {Bretherton}, F.~P. 1967, Journal of Fluid Mechanics, 27,
  513

\bibitem[{{Bretherton}(1966)}]{B1966}
{Bretherton}, F.~P. 1966, Quarterly Journal of the Royal Meteorological
  Society, 92, 325

\bibitem[{{Bretherton} \& {Garrett}(1968)}]{BG1968}
{Bretherton}, F.~P. \& {Garrett}, C.~J.~R. 1968, Proceedings of the Royal
  Society of London Series A, 302, 529

\bibitem[{{Broad}(1995)}]{B1995}
{Broad}, A.~S. 1995, Quarterly Journal of the Royal Meteorological Society,
  121, 1891

\bibitem[{{Brun} {et~al.}(2015){Brun}, {Garc{\'{\i}}a}, {Houdek}, {Nandy}, \&
  {Pinsonneault}}]{BG2015}
{Brun}, A.~S., {Garc{\'{\i}}a}, R.~A., {Houdek}, G., {Nandy}, D., \&
  {Pinsonneault}, M. 2015, \ssr, 196, 303

\bibitem[{{Brun} {et~al.}(2017){Brun}, {Strugarek}, {Varela}, {Matt},
  {Augustson}, {Emeriau}, {DoCao}, {Brown}, \& {Toomre}}]{BS2017}
{Brun}, A.~S., {Strugarek}, A., {Varela}, J., {et~al.} 2017, \apj, 836, 192

\bibitem[{{Bryan}(1889)}]{B1889}
{Bryan}, G.~H. 1889, Philosophical Transactions of the Royal Society of London
  Series A, 180, 187

\bibitem[{{Carpenter} {et~al.}(2012){Carpenter}, {Tedford}, {Heifetz}, \&
  {Lawrence}}]{CT2012}
{Carpenter}, J.~R., {Tedford}, E.~W., {Heifetz}, E., \& {Lawrence}, G.~A. 2012,
  Applied Mechanics Reviews, 64, 061001

\bibitem[{{Cartan}(1922)}]{C1922}
{Cartan}, E. 1922, Bull. Sci. Math., 46, 317

\bibitem[{{Damiani} \& {Mathis}(2018)}]{DM2018}
{Damiani}, C. \& {Mathis}, S. 2018, \aap, 618, A90

\bibitem[{{Debras} \& {Chabrier}(2019)}]{DC2019}
{Debras}, F. \& {Chabrier}, G. 2019, \apj, 872, 100

\bibitem[{{Decressin} {et~al.}(2009){Decressin}, {Mathis}, {Palacios}, {Siess},
  {Talon}, {Charbonnel}, \& {Zahn}}]{DM2009}
{Decressin}, T., {Mathis}, S., {Palacios}, A., {et~al.} 2009, \aap, 495, 271

\bibitem[{{Duguid} {et~al.}(2020){Duguid}, {Barker}, \& {Jones}}]{DB2020}
{Duguid}, C.~D., {Barker}, A.~J., \& {Jones}, C.~A. 2020, \mnras, 491, 923

\bibitem[{Eliassen \& Palm(1961)}]{EP1961}
Eliassen, A. \& Palm, E. 1961, On the Transfer of Energy in Stationary Mountain
  Waves, (Det Norske Videnskaps-Akademi i Oslo. Geofysiske publikasjoner) (I
  kommisjon hos Aschehoug)

\bibitem[{{Favier} {et~al.}(2014){Favier}, {Barker}, {Baruteau}, \&
  {Ogilvie}}]{FB2014}
{Favier}, B., {Barker}, A.~J., {Baruteau}, C., \& {Ogilvie}, G.~I. 2014,
  \mnras, 439, 845

\bibitem[{{Ford} \& {Rasio}(2006)}]{FR2006}
{Ford}, E.~B. \& {Rasio}, F.~A. 2006, \apjl, 638, L45

\bibitem[{{Fuller} {et~al.}(2016){Fuller}, {Luan}, \& {Quataert}}]{FL2016}
{Fuller}, J., {Luan}, J., \& {Quataert}, E. 2016, \mnras, 458, 3867

\bibitem[{{Galanti} {et~al.}(2019){Galanti}, {Kaspi}, {Miguel}, {Guillot},
  {Durante}, {Racioppa}, \& {Iess}}]{GK2019}
{Galanti}, E., {Kaspi}, Y., {Miguel}, Y., {et~al.} 2019, \grl, 46, 616

\bibitem[{{Gallet} {et~al.}(2018){Gallet}, {Bolmont}, {Bouvier}, {Mathis}, \&
  {Charbonnel}}]{GB2018}
{Gallet}, F., {Bolmont}, E., {Bouvier}, J., {Mathis}, S., \& {Charbonnel}, C.
  2018, \aap, 619, A80

\bibitem[{{Gallet} {et~al.}(2017){Gallet}, {Bolmont}, {Mathis}, {Charbonnel},
  \& {Amard}}]{GB2017}
{Gallet}, F., {Bolmont}, E., {Mathis}, S., {Charbonnel}, C., \& {Amard}, L.
  2017, \aap, 604, A112

\bibitem[{{Garc{\'{\i}}a} {et~al.}(2007){Garc{\'{\i}}a}, {Turck-Chi{\`e}ze},
  {Jim{\'e}nez-Reyes}, {Ballot}, {Pall{\'e}}, {Eff-Darwich}, {Mathur}, \&
  {Provost}}]{GT2007}
{Garc{\'{\i}}a}, R.~A., {Turck-Chi{\`e}ze}, S., {Jim{\'e}nez-Reyes}, S.~J.,
  {et~al.} 2007, Science, 316, 1591

\bibitem[{{Gastine} {et~al.}(2013){Gastine}, {Wicht}, \& {Aurnou}}]{GW2013}
{Gastine}, T., {Wicht}, J., \& {Aurnou}, J.~M. 2013, \icarus, 225, 156

\bibitem[{{Gerkema} {et~al.}(2008){Gerkema}, {Zimmerman}, {Maas}, \& {van
  Haren}}]{GZ2008}
{Gerkema}, T., {Zimmerman}, J.~T.~F., {Maas}, L.~R.~M., \& {van Haren}, H.
  2008, Reviews of Geophysics, 46, RG2004

\bibitem[{{Gliatto} \& {Held}(2020)}]{GH2020}
{Gliatto}, M.~T. \& {Held}, I.~M. 2020, Journal of Atmospheric Sciences, 77,
  859

\bibitem[{{Goldreich} \& {Nicholson}(1989)}]{GN1989}
{Goldreich}, P. \& {Nicholson}, P.~D. 1989, \apj, 342, 1075

\bibitem[{{Goldreich} \& {Tremaine}(1979)}]{GT1979}
{Goldreich}, P. \& {Tremaine}, S. 1979, \apj, 233, 857

\bibitem[{{Greenspan}(1969)}]{G1969}
{Greenspan}, H.~P. 1969, Cambridge University Press.

\bibitem[{{Grimshaw}(1979)}]{G1979}
{Grimshaw}, R. 1979, Geophysical and Astrophysical Fluid Dynamics, 14, 303

\bibitem[{{Grimshaw}(1975{\natexlab{a}})}]{G1975b}
{Grimshaw}, R.~H.~J. 1975{\natexlab{a}}, Tellus Series A, 27, 351

\bibitem[{{Grimshaw}(1975{\natexlab{b}})}]{G1975}
{Grimshaw}, R.~H.~J. 1975{\natexlab{b}}, Journal of Fluid Mechanics, 70, 287

\bibitem[{{Guenel} {et~al.}(2016{\natexlab{a}}){Guenel}, {Baruteau}, {Mathis},
  \& {Rieutord}}]{GB2016}
{Guenel}, M., {Baruteau}, C., {Mathis}, S., \& {Rieutord}, M.
  2016{\natexlab{a}}, \aap, 589, A22

\bibitem[{{Guenel} {et~al.}(2016{\natexlab{b}}){Guenel}, {Mathis}, {Baruteau},
  \& {Rieutord}}]{GM2016}
{Guenel}, M., {Mathis}, S., {Baruteau}, C., \& {Rieutord}, M.
  2016{\natexlab{b}}, ArXiv e-prints, arXiv:1612.05071

\bibitem[{{Harnik} \& {Heifetz}(2007)}]{HH2007}
{Harnik}, N. \& {Heifetz}, E. 2007, Journal of Atmospheric Sciences, 64, 2238

\bibitem[{{Hut}(1980)}]{H1980}
{Hut}, P. 1980, \aap, 92, 167

\bibitem[{{Jackson} {et~al.}(2008){Jackson}, {Greenberg}, \& {Barnes}}]{JG2008}
{Jackson}, B., {Greenberg}, R., \& {Barnes}, R. 2008, \apj, 678, 1396

\bibitem[{{Jones}(1967)}]{J1967}
{Jones}, W.~L. 1967, 30, 439

\bibitem[{{Jones}(1968)}]{J1968}
{Jones}, W.~L. 1968, Journal of Fluid Mechanics, 34, 609

\bibitem[{{Jouve} \& {Ogilvie}(2014)}]{JO2014}
{Jouve}, L. \& {Ogilvie}, G.~I. 2014, Journal of Fluid Mechanics, 745, 223

\bibitem[{{Kaspi} {et~al.}(2017){Kaspi}, {Guillot}, {Galanti}, {Miguel},
  {Helled}, {Hubbard}, {Militzer}, {Wahl}, {Levin}, {Connerney}, \&
  {Bolton}}]{KG2017}
{Kaspi}, Y., {Guillot}, T., {Galanti}, E., {et~al.} 2017, \grl, 44, 5960

\bibitem[{{Kippenhahn} {et~al.}(2012){Kippenhahn}, {Weigert}, \&
  {Weiss}}]{K2012}
{Kippenhahn}, R., {Weigert}, A., \& {Weiss}, A. 2012, {Stellar Structure and
  Evolution}

\bibitem[{{Lai}(2012)}]{L2012}
{Lai}, D. 2012, \mnras, 423, 486

\bibitem[{{Lainey} {et~al.}(2020){Lainey}, {Casajus}, {Fuller}, {Zannoni},
  {Tortora}, {Cooper}, {Murray}, {Modenini}, {Park}, {Robert}, \&
  {Zhang}}]{LC2020}
{Lainey}, V., {Casajus}, L.~G., {Fuller}, J., {et~al.} 2020, Nature Astronomy,
  TBD, TBD

\bibitem[{{Lainey} {et~al.}(2017){Lainey}, {Jacobson}, {Tajeddine}, {Cooper},
  {Murray}, {Robert}, {Tobie}, {Guillot}, {Mathis}, {Remus}, {Desmars},
  {Arlot}, {De Cuyper}, {Dehant}, {Pascu}, {Thuillot}, {Le Poncin-Lafitte}, \&
  {Zahn}}]{LJ2017}
{Lainey}, V., {Jacobson}, R.~A., {Tajeddine}, R., {et~al.} 2017, \icarus, 281,
  286

\bibitem[{{Latter} \& {Balbus}(2009)}]{LB2009}
{Latter}, H.~N. \& {Balbus}, S.~A. 2009, \mnras, 399, 1058

\bibitem[{{Lin} \& {Ogilvie}(2018)}]{LO2018}
{Lin}, Y. \& {Ogilvie}, G.~I. 2018, \mnras, 474, 1644

\bibitem[{{Lindzen}(1988)}]{L1988}
{Lindzen}, R.~S. 1988, Pure and Applied Geophysics, 126, 103

\bibitem[{{Lindzen} \& {Barker}(1985)}]{LB1985}
{Lindzen}, R.~S. \& {Barker}, J.~W. 1985, Journal of Fluid Mechanics, 151, 189

\bibitem[{{Lindzen} \& {Tung}(1978)}]{LT1978}
{Lindzen}, R.~S. \& {Tung}, K.~K. 1978, Journal of Atmospheric Sciences, 35,
  1626

\bibitem[{{Luan} {et~al.}(2018){Luan}, {Fuller}, \& {Quataert}}]{LF2018}
{Luan}, J., {Fuller}, J., \& {Quataert}, E. 2018, \mnras, 473, 5002

\bibitem[{{Maas} \& {Lam}(1995)}]{ML1995}
{Maas}, L.~R.~M. \& {Lam}, F. P.~A. 1995, Journal of Fluid Mechanics, 300, 1

\bibitem[{{Maslowe}(1986)}]{M1986}
{Maslowe}, S.~A. 1986, Annual Review of Fluid Mechanics, 18, 405

\bibitem[{{Mathis}(2009)}]{M2009}
{Mathis}, S. 2009, \aap, 506, 811

\bibitem[{{Mathis}(2015)}]{M2015}
{Mathis}, S. 2015, \aap, 580, L3

\bibitem[{{Mathis}(2019)}]{M2019}
{Mathis}, S. 2019, in EAS Publications Series, Vol.~82, EAS Publications
  Series, 5--33

\bibitem[{{Mathis} \& {de Brye}(2012)}]{MB2012}
{Mathis}, S. \& {de Brye}, N. 2012, \aap, 540, A37

\bibitem[{{Mathis} {et~al.}(2004){Mathis}, {Palacios}, \& {Zahn}}]{MP2004}
{Mathis}, S., {Palacios}, A., \& {Zahn}, J.~P. 2004, \aap, 425, 243

\bibitem[{{Mathis} {et~al.}(2018){Mathis}, {Prat}, {Amard}, {Charbonnel},
  {Palacios}, {Lagarde}, \& {Eggenberger}}]{MP2018}
{Mathis}, S., {Prat}, V., {Amard}, L., {et~al.} 2018, \aap, 620, A22

\bibitem[{{Miles}(1961)}]{M1961}
{Miles}, J.~W. 1961, Journal of Fluid Mechanics, 10, 496

\bibitem[{{Miles} \& {Howard}(1964)}]{MH1964}
{Miles}, J.~W. \& {Howard}, L.~N. 1964, Journal of Fluid Mechanics, 20, 331

\bibitem[{{Militzer} {et~al.}(2019){Militzer}, {Wahl}, \& {Hubbard}}]{MW2019}
{Militzer}, B., {Wahl}, S., \& {Hubbard}, W.~B. 2019, \apj, 879, 78

\bibitem[{{Morse} \& {Feshbach}(1953)}]{MF1953}
{Morse}, P.~M. \& {Feshbach}, H. 1953, {Methods of theoretical physics}

\bibitem[{{Ogilvie}(2009)}]{O2009}
{Ogilvie}, G.~I. 2009, \mnras, 396, 794

\bibitem[{{Ogilvie}(2014)}]{O2014}
{Ogilvie}, G.~I. 2014, \araa, 52, 171

\bibitem[{{Ogilvie} \& {Lesur}(2012)}]{OL2012}
{Ogilvie}, G.~I. \& {Lesur}, G. 2012, \mnras, 422, 1975

\bibitem[{{Ogilvie} \& {Lin}(2004)}]{OL2004}
{Ogilvie}, G.~I. \& {Lin}, D.~N.~C. 2004, \apj, 610, 477

\bibitem[{{Ogilvie} \& {Lin}(2007)}]{OL2007}
{Ogilvie}, G.~I. \& {Lin}, D.~N.~C. 2007, \apj, 661, 1180

\bibitem[{{Park, J.} {et~al.}(2020{\natexlab{a}}){Park, J.}, {Prat, V.}, \&
  {Mathis, S.}}]{PPM2020}
{Park, J.}, {Prat, V.}, \& {Mathis, S.} 2020{\natexlab{a}}, A\&A, 635, A133

\bibitem[{{Park, J.} {et~al.}(2020{\natexlab{b}}){Park, J.}, {Prat, V.},
  {Mathis, S.}, \& {Bugnet, L.}}]{PPMB2020}
{Park, J.}, {Prat, V.}, {Mathis, S.}, \& {Bugnet, L.} 2020{\natexlab{b}}, to be
  submitted to A\&A

\bibitem[{{Press}(1981)}]{P1981}
{Press}, W.~H. 1981, \apj, 245, 286

\bibitem[{{Rieutord}(2015)}]{R2015}
{Rieutord}, M. 2015, {Fluid Dynamics: An Introduction}

\bibitem[{{Rieutord} {et~al.}(2001){Rieutord}, {Georgeot}, \&
  {Valdettaro}}]{RG2001}
{Rieutord}, M., {Georgeot}, B., \& {Valdettaro}, L. 2001, Journal of Fluid
  Mechanics, 435, 103

\bibitem[{{Rieutord} \& {Valdettaro}(1997)}]{RV1997}
{Rieutord}, M. \& {Valdettaro}, L. 1997, Journal of Fluid Mechanics, 341, 77

\bibitem[{{Rieutord} \& {Valdettaro}(2010)}]{RV2010}
{Rieutord}, M. \& {Valdettaro}, L. 2010, Journal of Fluid Mechanics, 643, 363

\bibitem[{Ringot(1998)}]{R1998}
Ringot, O. 1998, PhD thesis, thèse de doctorat dirigée par Schatzman, Évry
  Terre, océan, espace Paris 7 1998

\bibitem[{{Schmid} {et~al.}(2002){Schmid}, {Henningson}, \&
  {Jankowski}}]{SH2002}
{Schmid}, P., {Henningson}, D., \& {Jankowski}, D. 2002, Applied Mechanics
  Reviews, 55, B57

\bibitem[{{Schou} {et~al.}(1998){Schou}, {Antia}, {Basu}, {Bogart}, {Bush},
  {Chitre}, {Christensen-Dalsgaard}, {Di Mauro}, {Dziembowski}, {Eff-Darwich},
  {Gough}, {Haber}, {Hoeksema}, {Howe}, {Korzennik}, {Kosovichev}, {Larsen},
  {Pijpers}, {Scherrer}, {Sekii}, {Tarbell}, {Title}, {Thompson}, \&
  {Toomre}}]{SA1998}
{Schou}, J., {Antia}, H.~M., {Basu}, S., {et~al.} 1998, \apj, 505, 390

\bibitem[{{Shampine} \& {Reichelt}(1997)}]{SR1997}
{Shampine}, L.~F. \& {Reichelt}, M.~W. 1997, SIAM J. Sci. Comp., 18, 1

\bibitem[{{Thompson} {et~al.}(2003){Thompson}, {Christensen-Dalsgaard},
  {Miesch}, \& {Toomre}}]{TC2003}
{Thompson}, M.~J., {Christensen-Dalsgaard}, J., {Miesch}, M.~S., \& {Toomre},
  J. 2003, \araa, 41, 599

\bibitem[{{Tsang} \& {Lai}(2009)}]{TL2009}
{Tsang}, D. \& {Lai}, D. 2009, \mnras, 400, 470

\bibitem[{{Varela} {et~al.}(2016){Varela}, {Strugarek}, \& {Brun}}]{VS2016}
{Varela}, J., {Strugarek}, A., \& {Brun}, A.~S. 2016, Advances in Space
  Research, 58, 1507

\bibitem[{{Watts} {et~al.}(2004){Watts}, {Andersson}, \& {Williams}}]{WA2004}
{Watts}, A.~L., {Andersson}, N., \& {Williams}, R.~L. 2004, \mnras, 350, 927

\bibitem[{{Wei}(2016)}]{W2016}
{Wei}, X. 2016, \apj, 828, 30

\bibitem[{{Wei}(2018)}]{W2018}
{Wei}, X. 2018, \apj, 854, 34

\bibitem[{Yamanaka \& Tanaka(1984)}]{YT1984}
Yamanaka, M. \& Tanaka, H. 1984, Journal of the Meteorological Society of
  Japan, 62, 1

\bibitem[{{Zahn}(1966)}]{Z1966a}
{Zahn}, J.~P. 1966, Annales d'Astrophysique, 29, 313

\bibitem[{{Zahn}(1977)}]{Z1977}
{Zahn}, J.~P. 1977, \aap, 500, 121

\bibitem[{{Zahn} {et~al.}(1997){Zahn}, {Talon}, \& {Matias}}]{ZT1997}
{Zahn}, J.~P., {Talon}, S., \& {Matias}, J. 1997, \aap, 322, 320

\end{thebibliography}

\begin{appendix}
\section{Derivation of the \aava{wave propagation} equation}
\label{derPc}
We detail in this section the derivation of the \aava{ODE} (cf Eqs. (\ref{Pc}) and (\ref{coefs})) in a dissipative medium and for forced inertial waves.
 Writing the perturbation variables as wave-like functions in $x$ and $z$-directions, the hydrodynamic equations (\ref{pert:1}), (\ref{pert:2}), and (\ref{rhop})
are respectively:
\begin{align}
-is\bo u +v\Up\bo e_x+\bO\times\bo u=-\frac{\bo\nabla p}{\rom}-\frac{\rho}{\rho_0}\bo e_z+\bo f,\label{pert_mom}\\
ik_xu+v'+ik_zw=0, \label{continuity}\\
-i\sigma\rho +v\ft U'=0, \label{thermo}
\end{align}
with $s=\sigma+i\sf$ a complex frequency that includes the Rayleigh friction frequency $\sf$. We have removed the symbol $'$ in each perturbed quantity, and it now refers to the derivative with respect to $y$.
The projection of the curl of the perturbed momentum equation (\ref{pert_mom}) on the $(\bo e_x,\bo e_y,\bo e_z)$ basis is:
\begin{eqnarray}
&is\,(ik_zv-w')-is'w-(\bO\cdot\bn)u=-\rho'/\rho_0+(\bn\times\bo f)_x\label{curl1}&\\
&is\,(ik_xw-ik_zu)+ik_zvU'-(\bO\cdot\bn)v=ik_x\rho/\rho_0+(\bn\times\bo f)_y \label{curl2}&\\
&is\,(u'-ik_xv)+is'u-(vU')'-(\bO\cdot\bn)w=(\bn\times\bo f)_z. \label{curl3}&
\end{eqnarray}
Note that the term $-\bn\times(\bn p/\rom)=(\bn\rom\times\bn p)/\rom^2$
is of order $\epsilon$ thus neglected. This enables us to define and use the quantity $\Pi=p/\rho_0$ that we call reduced pressure in the present study.
Using the continuity equation, the linear combination $ik_z(\ref{curl1})-ik_x(\ref{curl3})$ reads:
\begin{equation}
\begin{aligned}
&is(v''-k_\perp^2v)+is'v'+ik_x(vU')'+(\bO\cdot\bn)(-ik_zu+ik_xw)\\=
&-ik_z\rho'/\rho_0+[\bn\times(\bn\times\bo f)]_y,
\end{aligned}\label{curl4}
\end{equation}
where $k_\perp^2=k_x^2+k_z^2$.
Thus, $is\left[(\bO\cdot\bn)\cfrac{(\ref{curl2})}{is}-(\ref{curl4})\right]$ gives a second-order ordinary differential equation on the latitudinal velocity $v$, with a source term $S$:
\begin{equation}
Av''+Bv'+Cv=S.
\label{PcS}
\end{equation}
Coefficients are written as:
\begin{equation}
\begin{aligned}
A=&s^2-\ft^2,\\
B=&ik_z\ft U'\left(1-\frac{s}{\sigma}\right)-k_x\ft^2\frac{\Up}{\sigma}\left(1+\frac{\sigma}{s}\right)-2ik_zf\ft,\\
C=&-k_\perp^2s^2+k_z^2f(f-U')-ik_xk_z\ft f\frac{\Up}{\sigma}\left(1+\frac{\sigma}{s}\right)\\
&-k_x^2\ft^2\frac{U'^2}{\sigma^2}\left(1+\frac{\sigma}{s}\right)-ik_xk_z\ft\frac{\Um'^2}{\sigma^2}s\left(1-\frac{\sigma^2}{s^2}\right)\\
&+U''\left[\frac{k_x}{\sigma}\left(s\sigma-\ft^2\right)+ik_z\ft\left(1-\frac{s}{\sigma}\right)\right],\\
S=&-is[\bn\times(\bn\times\bo f)]_y+\left(k_x\frac{U'}{s}\ft+\bO\cdot\bn\right)(\bn\times\bo f)_y.
\end{aligned}
\label{Pcvis}
\end{equation}
\section{Derivation of the polarisation relations}
\label{derPi}
The Navier-Stokes equation with Rayleigh friction, without forcing, and projected onto the Cartesian basis reads:
\begin{eqnarray}
-isu + \left( U' - f\right)v +\widetilde{f}w & = & - i k_x \Pi 
\label{(1)},\\ 
-is v +fu & = & -\Pi'   \label{(2)},\\
-is w -\widetilde{f}u +v \ft\frac{\Um'}{i\sigma}& = & -ik_z\Pi,  \label{(3)}
\end{eqnarray}
where we have used Eq. (\ref{thermo})  to replace the pertubed density.
In order to obtain an equation for the reduced pressure and the latitudinal velocity perturbations only, we apply the following linear combination:
\begin{equation}
\partial_z\left[\widetilde{f}(\ref{(1)})-is(\ref{(3)})\right]-\partial_x\left[is(\ref{(1)})+\widetilde{f}(\ref{(3)})\right],
\end{equation}
which yields, using the continuity equation (\ref{continuity}):
\begin{equation}
\Pi =\frac{1}{sk^2_\perp}\left\{iAv'+\left[(\Um'-f)\left(ik_xs-k_z\widetilde{f}\right)+\ft\frac{\Um'}{\sigma}\left(k_zs-ik_x\ft\right)\right]v\right\}. \\
\label{(9)}
\end{equation}
To get the perturbed vertical velocity, one can do the linear combination $-\ft(\ref{(1)})+is(\ref{(3)})$ which yields, after some algebra:
\begin{equation}
w=\frac{1}{sk_\perp^2}\left\{\left(isk_z-k_x\ft\right)v'+k_x\left[ik_z(\Um'-f)-k_x\ft\frac{\Um'}{\sigma}\right]v\right\}.
\label{w} 
\end{equation}
Furthermore, $is(\ref{(1)}+\ft(\ref{(3)})$ gives the perturbed longitudinal velocity: 
\begin{equation}
u=\frac{1}{sk_\perp^2}\left\{\left(isk_x+k_z\ft\right)v'+k_z\left[ik_z(f-\Um')+k_x\ft\frac{\Um'}{\sigma}\right]v\right\}.
\label{u}
\end{equation}
\section{The WKBJ approximation}
\label{wkbj}
Rapidly oscillating solutions are often studied within the WKBJ approximation \citep{P1981}.
At the pole, the solution of the ODE (\ref{Pcpole}) in this approximation takes the form:
\begin{equation}
\Psi=\frac{1}{\sqrt{\kappa(y)}}\left(A\e^{i\int \kappa(y')\,\mathrm{d}y'}+B\e^{-i\int \kappa(y')\,\mathrm{d}y'}\right),
\end{equation}
where $A$ and $B$ are the complex amplitudes of the wave function $\Psi$ and $\kappa(y)$ is the complex potential associated \aava{with} the ODE. We can determine the validity domain of this approximation like \citet{AM2013} did.
In the WKBJ approximation, $\Psi$ satisfies
\begin{equation}
\frac{\mathrm{d}^2\Psi(y)}{\mathrm{d}y^2}=f(y)\Psi(y),
\label{schro4}
\end{equation}
where $f(y)=-\kappa(y)^2\approx-R/(y-\yo)^2$ for a constant shear. The WKBJ approximation is valid provided that: (i) $\Ro<1$ so that $R>0$, and (ii) $k_\perp^2$ is negligible in front of $-R/(y-\yo)^2$.
Then, we introduce the Liouville transformation:
\begin{equation}
W(y)=f^{1/4}\Psi\ \text{ and }\ \xi(y)=\int^y f^{1/2}\,\mathrm{d}y' .
\end{equation}
We deduce 
\begin{align}\begin{aligned}
\frac{\mathrm{d}W}{\mathrm{d}\xi}&=\frac{1}{4}f^{-5/4}f'\Psi+f^{-1/4}\Psi', \\
\frac{\mathrm{d}^2W}{\mathrm{d}\xi^2}&=\frac{-5}{16}f^{-11/4}f'^2\Psi+\frac{1}.{4}f^{-7/4}f''\Psi+f^{-3/4}\Psi''.
\end{aligned}\end{align}
Equation (\ref{schro4}) thus becomes:
\begin{equation}
\frac{\mathrm{d}^2W}{\mathrm{d}\xi^2}=[1+\Phi(y)]W\ \ \text{ with }\ \ \Phi=\frac{4ff''-5f'^2}{16f^3}.
\end{equation}
The WKBJ approximation states that $|\Phi| \ll1$. Given the definition of $f$, this leads to the validity condition $|R|\gg1/4$.
\section{Analytical properties in the no-shear regions (zones I and III)}
\subsection{Wave-like solutions}
\label{solI_III}
The ordinary differential equation without shear can be written:
\begin{equation}
\left(s^{2}-\tilde{f}^{2}\right)\frac{d^{2}v}{dy^{2}}-2ik_{z}f\tilde{f}\frac{dv}{dy}+\left[k_{z}^{2}\left(f^{2}-s^{2}\right)-k_{x}^{2}s^{2}\right]v=0.
\label{Pcuni}
\end{equation}
For the following, we introduce $s_1=\omega+i\sf$ in zone I, and $s_3=\omega-k_x\Lambda+i\sf$ in zone III. 
In the no-flow region I, the analytic solution of Eq. (\ref{Pcuni}) can be written in the form: 
\begin{equation}
v(y)=A_\mathrm{I}\exp\left(ik_\mathrm{I}y\right)+A_\mathrm{R}\exp\left(ik_\mathrm{R}y\right),
\label{v_IR}
\end{equation}
which is the sum of an incident and a reflected waves of amplitude $A_\mathrm{I}$ and $A_\mathrm{R}$ and wavenumbers:
\begin{equation}
\begin{aligned}
k_1=\frac{-k_{z}f\tilde{f}+\sqrt{s_1^2\left[k_z^2f^2+k_\perp^2\left(\tilde{f}^{2}-s_1^{2}\right)\right]}}{\tilde{f}^{2}-s_1^{2}},\\
k_\mathrm{R}=\frac{-k_{z}f\tilde{f}-\sqrt{s_1^2\left[k_z^2f^2+k_\perp^2\left(\tilde{f}^{2}-s_1^{2}\right)\right]}}{\tilde{f}^{2}-s_1^{2}},
\end{aligned}
\end{equation}
respectively. These relationships are given by the dispersion relation satisfied by the incident and reflected waves:
\begin{equation}
s^2=\frac{\bO\cdot\bm k}{\bm k^2}=\frac{\left(k_\mathrm{I,R}\tilde{f}+k_{z}f\right)^{2}}{k_{x}^{2}+k_\mathrm{I,R}^{2}+k_{z}^{2}}.
\end{equation}

Similarly, we write the latitudinal velocity in the uniform mean flow region III as:
\begin{equation}
v(y)=A_\mathrm{T}\exp\left(ik_\mathrm{T}y\right),
\label{v_T}
\end{equation}
where $A_\mathrm{T}$ is the amplitude of the transmitted wave and 
\begin{equation}
k_\mathrm{T}=\frac{-k_{z}f\tilde{f}\pm\sqrt{s_3^2\left[k_z^2f^2+k_\perp^2\left(\tilde{f}^{2}-s_3^{2}\right)\right]}}{\tilde{f}^{2}-s_3^{2}}
\label{kT}
\end{equation}
is its wavenumber, also given by the dispersion relation of the transmitted wave.
It is important to stress that all varying parameters in the model, like $\Lambda$, $k_x$, $k_z$ and $\omega$, have been chosen such that $k_\mathrm{I}$, $k_\mathrm{R}$ and $k_\mathrm{T}$ are not complex without \aava{friction}. It prevents widely diverging waves in zones I or III.
\subsection{Wave action flux}
\label{wAI_III}
As we did in Sect. \ref{Co}, one can also determine the wave action flux (Eq. (\ref{A})) in region I:
\begin{equation}
\mA_\mathrm{I-R}/\rom=\frac{\sqrt{\omega^2\left[k_z^2f^2+k_\perp^2\left(\ft^2-\omega^{2}\right)\right]}}{2k_\perp^2\omega^2}\left(\left|A_\mathrm{I}\right|^{2}-\left|A_\mathrm{R}\right|^{2}\right),~
\end{equation}
and in region III:
\begin{equation}
\mA_\mathrm{T}/\rom=\pm\frac{\sqrt{\sn^2\left[k_z^2f^2+k_\perp^2\left(\ft^2-\sn^{2}\right)\right]}}{2k_\perp^2\sn^2}\left|A_\mathrm{T}\right|^{2},~
\end{equation}
with $\sn=\omega-k_x\Lambda$, and where we have used the reduced pressure for all three waves:
\begin{equation}
\Pi=\frac{1}{sk_\perp^2}\left[(\ft^2-s^2)k_\mathrm{I,R,T}+f\left(k_z\ft-ik_xs\right)\right]A_\mathrm{I,R,T}\e^{ik_\mathrm{I,R,T}y}
\label{Pivis}.
\end{equation}
It is noteworthy that in the no-shear regions the latitudinal flux of energy $\overline{pv}$ is preserved \citep[see also][]{EP1961}, and so is the wave action flux. Indeed, the characteristic frequencies ($\omega$ and $\sn$) in both zones (I and  III, respectively) are constants. Furthermore, as we stated in Sec. \ref{Co} the direction of the group velocity is given by $\sign{(\overline{pv})}=\sign{(\sigma\mA)}$. By consequence, the direction of energy propagation is constrained by the sign of the Doppler-shifted frequency $\sn$ in zone III, and by the sign of inertial frequency $\omega$ in zone I. 
Considering $\omega>0$, the incident wave is properly named since the group velocity is positive, likewise the  reflective wave which has negative group velocity and is so moving downward. It is a little more complicated for the so-called transmitted wave. The sign of $\sn$ is directly related to the presence or the absence of a critical level inside zone II.

The table below summarises which frequency has to be exited so that waves can meet one or several critical points in the shear region of range $0<y<1$, for a linear mean flow profile.\\

{\centering
\begin{tabular}{c||cc}
critical point \rule[-1ex]{0pt}{4ex}&$\Lambda>0$&$\Lambda<0$\\
\hline
\rule[-1ex]{0pt}{4ex} $\sigma=0$&$0<\omega<k_x\Lambda$&$k_x\Lambda<\omega<0$\\
\rule[-1ex]{0pt}{4ex}$\sigma=\ft$&$\ft<\omega<k_x\Lambda+\ft$&$k_x\Lambda+\ft<\omega<\ft$\\
\rule[-1ex]{0pt}{4ex}$\sigma=-\ft$&$-\ft<\omega<k_x\Lambda-\ft$&$k_x\Lambda-\ft<\omega<-\ft$\\
\end{tabular}\vspace{0.5cm}}
Thus, if corotation is met in zone II with $\Lambda>0$, we automatically have a negative Doppler-shifted frequency in zone III, i.e. $\sn<0$. Therefore, one must choose the $-$ sign in the expression of $k_\mathrm{T}$ (Eq. (\ref{kT})) in order to construct a wave that moves away from the critical level. The same reasoning applies for the critical level $\sigma=-\ft$.  It is trickier for the singularity $\sigma=\ft$ because $\ft-k_x\Lambda<\sn<\ft$, so $\sn$ can be either positive or negative in the interval $[0,\pi/2]$ depending on the value of $\ft$ and $k_x\Lambda$.
\end{appendix}
\end{document}